\documentclass[sts]{imsart}

\RequirePackage{amsthm,amsmath,amsfonts,amssymb}
\RequirePackage[authoryear]{natbib}
\RequirePackage[colorlinks,citecolor=blue,urlcolor=blue]{hyperref}
\RequirePackage{graphicx, placeins, tabularx, booktabs, longtable}

\startlocaldefs
\theoremstyle{plain}

\theoremstyle{remark}

\endlocaldefs

\emergencystretch=\maxdimen
\begin{document}

\begin{frontmatter}
\title{Area-based epigraph and hypograph indices for functional outlier detection.}

\runtitle{Area-based epigraph and hypograph indices for functional outlier detection.}

\begin{aug}
\author[A]{\fnms{Belén}~\snm{Pulido}\thanks{}\ead[label=e1]{belen.pulido@uc3m.es}\orcid{0000-0003-2105-959X}},
\author[B]{\fnms{Alba M.}~\snm{Franco-Pereira}\ead[label=e2]{albfranc@ucm.es}},
\author[C]{\fnms{Rosa E.}~\snm{Lillo}\ead[label=e3]{rosaelvira.lillo@uc3m.es}}
\and
\author[D]{\fnms{Fabian}~\snm{Scheipl}\ead[label=e4]{fabian.scheipl@stat.uni-muenchen.de}}

\address[A]{Belén Pulido is postdoctoral researcher at uc3m-Santander Big Data Institute,
Universidad Carlos III de Madrid, Getafe, Madrid\printead[presep={\ }]{e1}.}

\address[B]{Alba M. Franco-Pereira is Associate Professor at the Department of Statistics and O.R., Universidad Complutense de Madrid, Madrid, Spain and she also belongs to Instituto de Matemática Interdisciplinar (IMI) at the same university \printead[presep={\ }]{e2}.}

\address[C]{Rosa E. Lillo is Full Professor at the Department of Statistics, Universidad Carlos III de Madrid, Getafe, Spain and and she is the director of uc3m-Santander Big Data Institute at the same university \printead[presep={\ }]{e3}.}

\address[D]{Fabian Scheipl is Lecturer at the Department of Statistics, Ludwig-Maximilians-Universität München, Munich, Germany and Principal Investigator at the Munich Center for Machine Learning (MCML)\printead[presep={\ }]{e4}.}

\end{aug}

\begin{abstract}
Detecting outliers in Functional Data Analysis  is challenging because curves can stray from the majority in many different ways.  The Modified Epigraph Index (MEI) and Modified Hypograph Index (MHI) rank functions by the fraction of the domain on which one curve lies above or below another. While effective for spotting shape anomalies, their construction limits their ability to flag magnitude outliers. This paper introduces two new metrics, the Area-Based Epigraph Index (ABEI) and Area-Based Hypograph Index (ABHI) that quantify the  area between curves, enabling simultaneous sensitivity to both magnitude and shape deviations. Building on these indices, we present EHyOut, a robust procedure that recasts functional outlier detection as a multivariate problem: for every curve, and for its first and second derivatives, we compute ABEI and ABHI and then apply multivariate outlier-detection techniques to the resulting feature vectors. Extensive simulations show that EHyOut remains stable across a wide range of contamination settings and often outperforms established benchmark methods. Moreover,
applications to Spanish weather data and United Nations world population
data further illustrate the practical utility and meaningfulness of this methodology.
\end{abstract}

\begin{keyword}
\kwd{Epigraph}
\kwd{Hypograph}
\kwd{Outliers}
\kwd{Functional data}

\end{keyword}

\end{frontmatter}

\section{Introduction}

The analysis of data where observations can be considered as functions, has become a cornerstone of modern statistics. This field of study is called Functional Data Analysis (FDA) and provides tools to explore patterns and variability in phenomena that evolve over a continuum. A complete overview of FDA can be found in \cite{ramsay}, and \cite{ferraty}, while some interesting reviews of functional data can be found in \cite{horvath}, \cite{hsing}, and \cite{wang}. This paradigm has become important in several areas, such as medicine, finance and environmental sciences, where grasping the structures and tendencies in data leads to more informed decisions and innovative advancements.

Despite extensive research, there is no universally accepted definition of what an outlier is. \cite{hawkins1980identification} defined an outlier as ``an observation which deviates so much from the other observations as to arouse suspicions that it was generated by a different mechanism.'' \cite{barnett1994outliers} described an outlier as ``an observation (or subset of observations) which appears to be inconsistent with the remainder of that set of data''. These definitions highlight that outliers are often identified by their inconsistency with the rest of the data, which may not always be precise or mathematically defined.

Defining and identifying outliers in FDA continues to be a significant problem. If undetected, such outliers can greatly distort statistical conclusions and how well models work. Conversely, they might also highlight specific data of interest that require closer examination.

Functional outliers can manifest in multiple ways, including deviations in magnitude, shape, phase, or a combination of all of these types. This variety presents a significant challenge to the development of general-purpose outlier detection methods. See the discussion in \cite{herrmann_geometric_2021}.

A significant body of research on functional outlier detection leverages the concept of statistical depth, which provides a center-outward ordering of observations \citep[see][e.g.]{vardi, fraiman, lop2009}. For example, the functional boxplot \citep{sun2011genton}  applies band depth notions to flag outlying curves. Moreover, depth measures continue to be an active area of investigation for outlier detection with recent works such as \cite{jimenez-varon_pointwise_2023}.

A different approach for ordering functions from top to bottom, or vice-versa is using the the epigraph and hypograph indices, first proposed by \cite{franc2011}. They were proposed to characterize extremality of functions but 
have been also considered for outlier detection in the functional data literature. Some examples are the outliergram \citep{arribas} and the functional boxplot based on epigraphs and hypographs \citep{martin2016}. 
However, the utility of these indices for general-purpose outlier detection is limited by their design, relying on the Lebesgue measure,  which quantifies the proportion of the domain of dominance but remains unaffected by the magnitude of the vertical distance between curves.

To address this limitation, this paper introduces improved versions of these indices: the Area-Based Epigraph and Hypograph indices, ABEI and ABHI, respectively. These new approaches take into account the area between different curves. The paper then defines and evaluates EHyOut, a novel approach for outlier detection in functional data which aims to identify outlying curves regardless of outlier type, using ABEI and ABHI. 

This paper is organized as follows. Section~\ref{indices} reviews the definition of the epigraph and hypograph indices, and proposes the alternative definitions based on areas. In Section~\ref{s_ehyout}, the EHyOut methodology for outlier detection in functional data based on the proposed indices is defined. The section also describes a large-scale simulation set-up used for deciding algorithm details based on simulations and for evaluating and comparing the performance of our proposal.  
Section \ref{simul} evaluates EHyOut by comparing its performance with benchmark methodologies on simulated datasets, and Section \ref{real} applies EHyOut to two real-world datasets: the Spanish weather dataset and the World population dataset from the United Nations.
Finally, Section~\ref{conc} concludes with final remarks and avenues for future research.

\section{Epigraph and hypograph indices} \label{indices}

The epigraph and hypograph indices were first introduced by \cite{franc2011} as a measure of the ``extremality'' of a given curve with respect to a reference set of curves. These indices provide an ordering of the curves in the sample, ranking them from top to bottom or vice-versa as an alternative to statistical functional depth functions, which provide a center-outward ranking of curves. Various proposals for statistical (functional) depths can be found in \cite{vardi}, \cite{fraiman}, \cite{cuevas}, \cite{cuesta},  \cite{lop2009}, \cite{lop2011}, and \cite{sguera2014}.

The epigraph and hypograph indices have been applied in different contexts, including the outliergram proposed in \cite{arribas}, the functional boxplot in \cite{martin2016}, the homogeneity test from \cite{franc2020}, or the clustering methodology introduced in \cite{pulido2023, pulido2024}.

To formally define these indices, consider the space of real continuous functions $C(\mathcal{I}, \mathbb{R})$  over a compact interval $\mathcal{I}$ to $\mathbb{R}$. Let $X \colon \mathcal{I} \longrightarrow \mathbb{R} $ be a stochastic process with probability distribution $P_{X}$. The graph of a function $x$ in $C(\mathcal{I},\mathbb{R})$ is denoted as $G(x) = \{(t,x(t)): \ t \in \mathcal{I} \}.$ 

Given a set of curves $\{x_1(t),...,x_n(t)\}$, the modified epigraph and the hypograph indices of a curve $x$, denoted as $\mathrm{MEI}_n(x)$ and $\mathrm{MHI}_n(x)$ respectively, are defined as follows:
\begin{equation} \label{def_mei}
    \mathrm{MEI}_n(x) =1- \sum_{i=1}^n{ \frac{ \lambda (\{t \in \mathcal{I} : x_i(t) \geq x(t)\})}{n \lambda(\mathcal{I})}},
\end{equation}
 where $\lambda$ stands for Lebesgue's measure on $\mathbb{R}$. The modified hypograph index (MHI) is obtained by considering the inequality in the opposite direction:
\begin{equation} \label{def_mhi}
    \mathrm{MHI}_n(x) =\sum_{i=1}^n{ \frac{ \lambda (\{t \in \mathcal{I} : x_i(t) \leq x(t)\})}{n \lambda(\mathcal{I})}}.
\end{equation}  

These definitions offer an interpretation of the indices as the proportion of time (when $\mathcal{I}$ is treated as a time interval) the sample curves are above/below the given curve $x$. Specifically, when $\text{MEI}_n(x) = 1$, it means that the given curve $x$ is entirely above all other sample curves throughout the domain, indicating it is the ``highest'' curve in the dataset. Conversely, when $\text{MHI}_n(x) = 0$, the curve $x$ is entirely below all other sample curves, making it the ``lowest'' in the dataset. Moreover, these indices verify three important characteristics specifically remarked in \cite{pulido2023}:
\begin{itemize}
    \item MEI and MHI ranges between 0 and 1.
    \item A linear relation exists between MEI and MHI. 
    \item MEI and MHI can be expressed in terms of each other as follows: \\ $\text{MHI}_{n,X}(x(t)) = 1-\text{MEI}_{n,-X}(-x(t)),$ \\ where $\text{MHI}_{n,X}(x(t))$ represents MEI on the sample given by the process $X$, and analogous for MEI, and $-X$ represents the ``inverted'', sign-flipped process.
\end{itemize}

\subsection{Area based indices}

The previously defined indices given by Equations~\ref{def_mei} and \ref{def_mhi}, have proven useful for specific tasks like constructing functional boxplots for shape outliers \citep{martin2016}. However, MEI and MHI where used to create functional quartiles by providing a top-to-bottom ordering. Their definition rely on the Lebesgue measure of the sets $\{t \in \mathcal{I} : x_i(t) \leq x(t)\}$ and $\{t \in \mathcal{I} : x_i(t) \geq x(t)\}$, respectively. These definitions ignore the magnitude of the changes between curves, making the indices unable to distinguish between marginal and extreme magnitude deviations. Moreover, a curve that is truly a shape outlier but stay below another curve for a small interval can be hard to detect.  See Fig.~\ref{dgp12_ind} for a clear example. These limitations are particularly relevant in real-world applications, where the nature of outliers is often unknown, and even a single curve may exhibit multiple types of anomalies. This challenge motivates the need for a new definition of these indices, enabling them to capture a broader range of outliers effectively.

While the period of time that a set of curves lies above or below a given curve provides useful information, it does not capture the magnitude of these deviations. To address this limitation, we propose new indices that take into account the area between curves, thus reflecting not only the ``time'' that separates curves, but also the distance between them. These new indices are called area-based indices, and include the area-based epigraph index (ABEI) and the area-based hypograph index, defined as follows:
\begin{align} \label{def_abei}
    \text{ABEI}_n(x) &=\sum_{i=1}^n \int_{\mathcal{I}} (x_i(t)-x(t))_+ dt, \ \text{and} \\
  \label{def_abhi}
    \text{ABHI}_n(x) &=\sum_{i=1}^n \int_{\mathcal{I}} (x(t)-x_i(t))_+ dt,
\end{align}
where $(\centerdot)_+$ represents the positive part of $(\centerdot)$.

A curve can remain above another for an extended period with only minor variations in shape. By focusing on the accumulated deviation across the entire domain, rather than just the proportion of time one curve dominates the other, these definitions better capture the magnitude and structure of differences between curves. ABEI quantifies the total area where the sample curves exceed a given curve $x$, while ABHI captures the total area where the sample curves fall below $x$. Specifically, when $\text{ABEI}_n(x)=0$, it indicates that $x$ is never exceeded by any sample curve, making it the ``highest'' curve in terms of area dominance. Conversely, when $\text{ABHI}_n(x) = 0$, the curve $x$ always lies above all sample curves, designating it as the ``lowest'' in this framework.

ABEI and ABHI can be compared to MEI and MHI (Equations~(\ref{def_mei}) and (\ref{def_mhi})) in terms of the three characteristics previously noted about the later:
\begin{itemize}
    \item ABEI and ABHI have a lower bound equal to 0, but they are not upper bounded.
    \item While there is no linear relationship between ABEI and ABHI, the following relation holds: $$\text{ABEI}_n(x)+\text{ABHI}_n(x) = \sum_{i=1}^n \int |x_i(t)-x(t)| dt.$$
    \item ABHI can be expressed in terms of ABEI as follows: $\text{ABHI}_{n,X}(x)=\text{ABEI}_{n,-X}(-x)$. This equation states that ABHI of a curve $x(t)$ with respect to a sample from process $X$ is equivalent to the ABEI of the curve $-x(t)$ with respect to a sample from process $-X$.
\end{itemize}

The lack of a linear relationship between ABEI and ABHI allows both indices to be used together for comparisons, unlike MEI and MHI, where such comparisons were less meaningful.

\begin{figure*}[htbp]
\centering
\includegraphics[width=\textwidth]{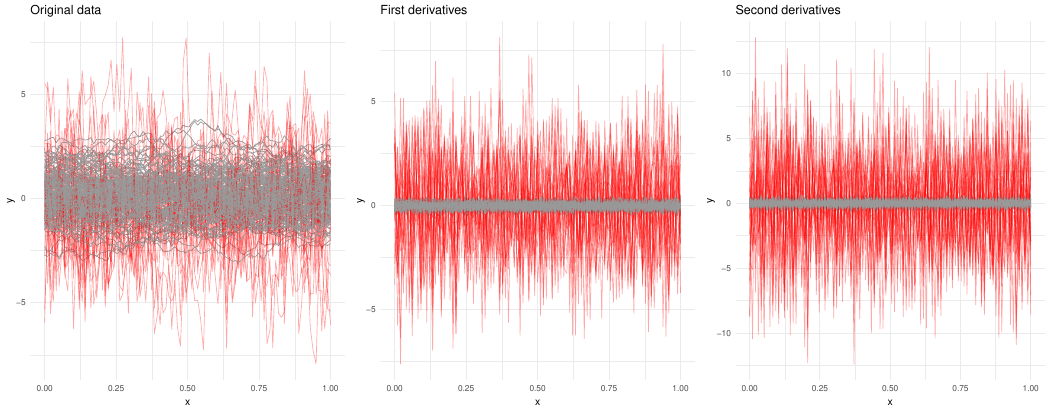}
\caption{Sample of 200 simulated curves with 10\% of outliers (red). The three panels display original data (left), first derivatives (middle) and second derivatives (right). \label{dgp12}}
\end{figure*}

\begin{figure}[htbp]
\centering
\includegraphics[width=0.47\textwidth]{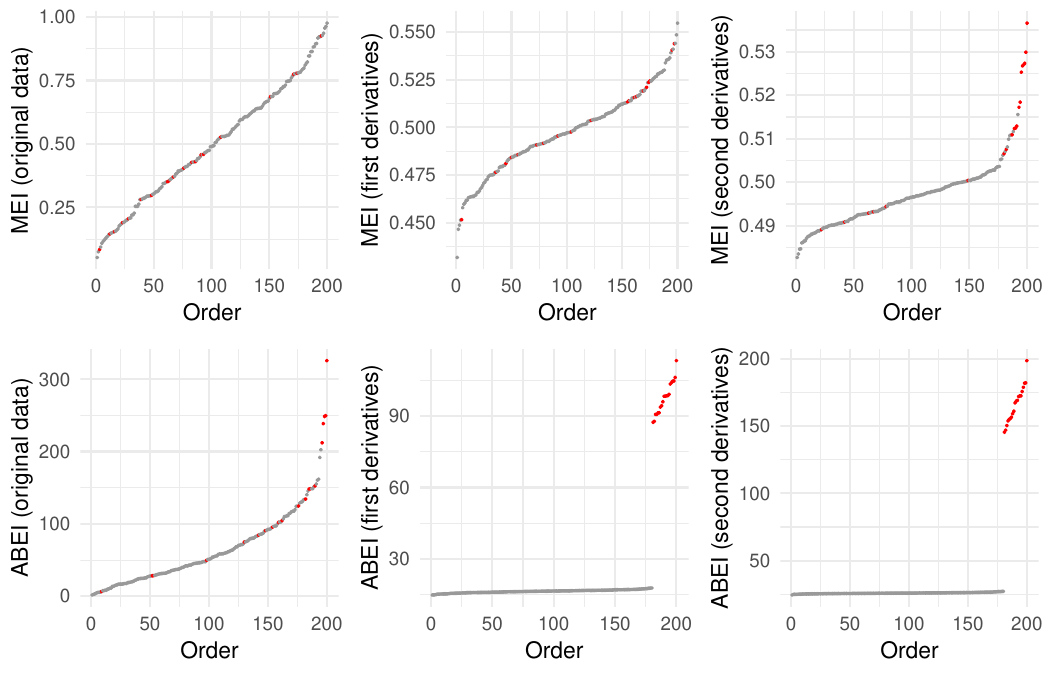}
\caption{Ordering of the curves from Fig. \ref{dgp12} based on different indices. The first row corresponds to MEI, while the second row shows the results for ABEI. The columns represent the original data (left), first derivatives (middle), and second derivatives (right).\label{dgp12_ind}}
\end{figure}

To illustrate the behavior of these indices, we consider an example dataset of 200 curves, 10\% of which are shape outliers (Fig. \ref{dgp12}). This figure presents the original data (left panel) along with its first and second derivatives (middle and right panels).

Fig. \ref{dgp12_ind} provides a detailed comparison, where the first row illustrates MEI applied to the data and its derivatives, while the second row shows the results for ABEI. The findings indicate that MEI fails to effectively isolate outliers in any of the datasets, whereas ABEI performs significantly better, particularly when applied to derivatives.

\begin{figure}[htbp]
\centering
\includegraphics[width=0.47\textwidth]{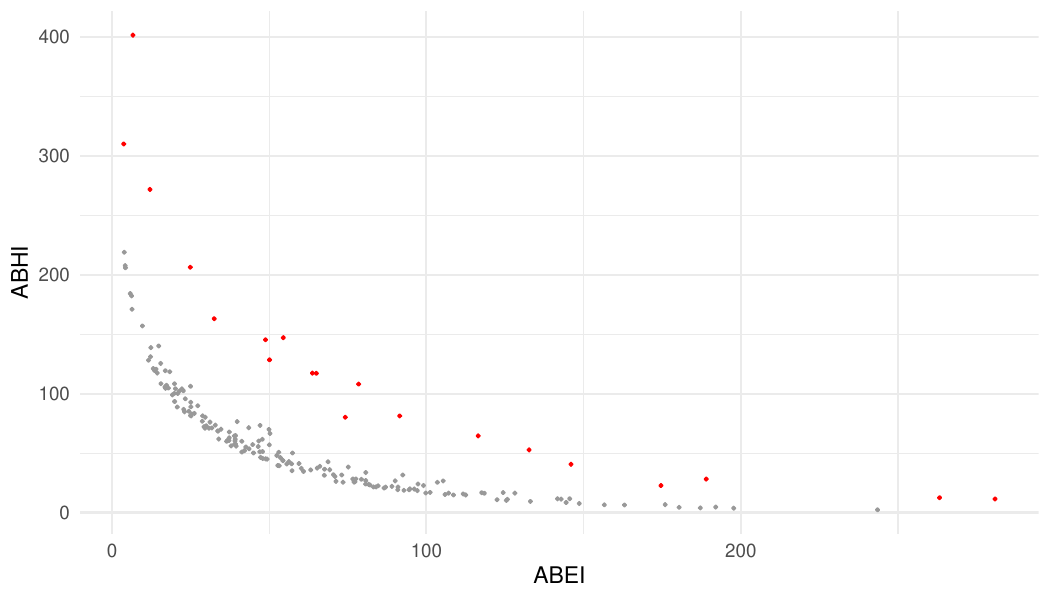}
\caption{Two-dimensional representation of ABEI (horizontal axis) vs ABHI (vertical axis), both computed on the curves in Fig. \ref{dgp12}. \label{dgp12_AB}}
\end{figure}

Moreover, ABEI can be combined with ABHI, as these indices do not exhibit the linear dependence observed between MEI and MHI  \citep{pulido2023}. This is illustrated in Fig. \ref{dgp12_AB}, which highlights how the combination of these indices provides an effective approach for detecting outliers.

\section{Outlier detection with area-based indices} \label{s_ehyout}

This section introduces EHyOut, a methodology for outlier detection in functional data based on the ABEI and ABHI  indices. This methodology aims to distinguish between outlier and non-outlier curves without requiring prior knowledge of the outlier type or any other information about the data.

EHyOut is based on applying multivariate methods to a dataset in a reduced dimension, with the main innovation being a specific transformation of functional data into a multivariate format. This approach mirrors the structure of EHyClus, a clustering method for functional data presented in \cite{pulido2023, pulido2024}. It involves the following three-step process:
\begin{enumerate}
     \item \textbf{Prepare the functional data.} Represent the functional data in terms of a twice-differentiable spline basis. In this case, cubic spline interpolation is considered to obtain the first and second derivatives of the data.
    \item \textbf{Apply indices to the data.} The ABEI and ABHI indices defined in Equations~(\ref{def_abei}) and (\ref{def_abhi}) are applied to the cubic spline representation of the data and their derivatives, obtaining a dataset of size $n \times 6$, where $n$ is the original number of curves (sample size).
    \item \textbf{Apply outlier detection techniques for multivariate data.} The well-known multivariate outlier detection method, comedian method \citep[COM,][]{sajesh2012outlier}, is applied to the dataset obtained in Step 2, achieving a final classification in outlying and non outlying curves. 
\end{enumerate}

EHyOut involves two key decisions that require empirical justification. The first choice is related to variable selection (step 2), while the second choice implies the selection of the most effective multivariate outlier detection method (step 3). A comprehensive simulation study was designed to first guide these internal choices and then to facilitate a fair comparison against established benchmark methods.

\subsection{Data generation processes} \label{dgps}

To ensure a comprehensive and rigorous evaluation, a diverse set of 19 distinct data generation processes (DGPs) is considered. These DGPs are drawn from a wide range of simulation models available in the functional outlier detection literature, including those from \cite{ojo2022} (DGP1-DGP10),  \cite{hernandez_kernel_2016} (DGP11), \cite{jimenez-varon_pointwise_2023} (DGP12-DGP16), \cite{nagy_depth-based_2017} (DGP17-DGP18), and \cite{herrmann_geometric_2021} (DGP19). This last DGP is a modification of those proposed in that work, adapted to include spherical patterns. 

These 19 DGPs were specifically chosen to generate a variety of outlier types, distinguished by variations in magnitude, amplitude, shape, or a combination thereof. Magnitude outliers maintain the common data shape but are shifted. Amplitude outliers also preserve this shape but differ in scale. Shape outliers, in contrast, exhibit a fundamentally different form from the majority of the data. Each DGP consists of 200 curves defined over the interval $[0,1]$ with $\alpha$ representing the proportion of outliers in the sample.

The full mathematical formulation for each of the 19 DGPs is provided in Appendix~\ref{appA}. This appendix also includes a graphical example of each DGP with 10\% of outliers to visually illustrate the nature of the generated inliers and outliers.

\subsection{Algorithmic specification of EHyOut}

Steps 2 and 3 of EHyOut require a decision-making process. Step 2 involves choosing the indices-based variables to considered, while Step 3 focuses on choosing the most suitable multivariate outlier detection methodology.

The purpose of Step 2 is to determine the optimal combination of indices (ABEI and ABHI for the original data and/or its 1st, 2nd derivatives) for a given dataset. This is done by evaluating the discriminative power of different combinations using the Area Under the Curve (AUC) as the evaluation metric. 

\begin{figure}[ht]
\centering
\includegraphics[width=0.47\textwidth]{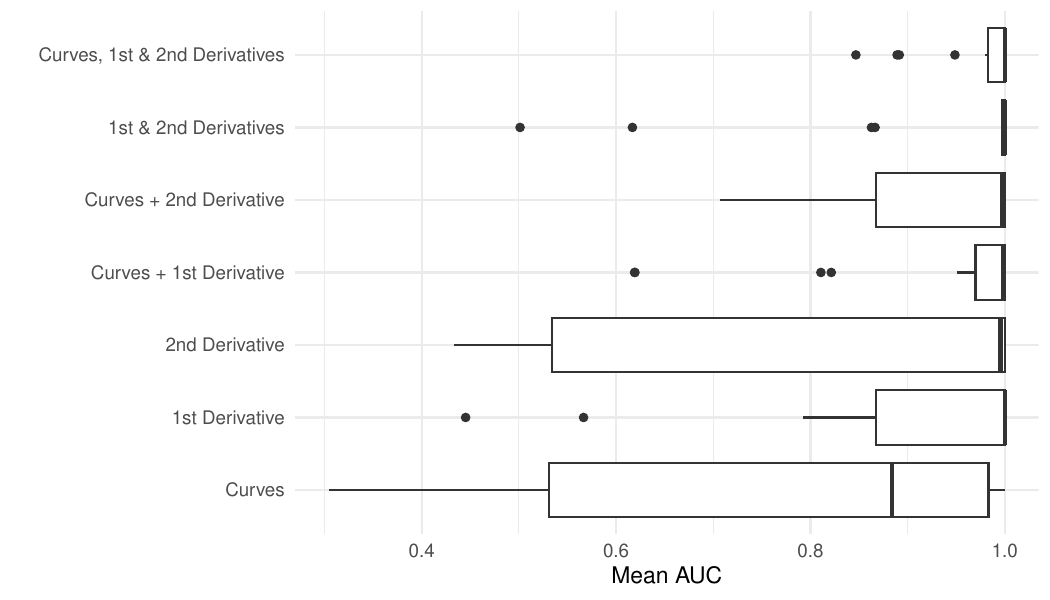}
\caption{Boxplots of the mean AUC for a robust Mahalanobis distance from 100 simulations across 19 different DGPs. The methods are distinguished by the type of information used: the smoothed curves, their derivatives, or combinations thereof.  \label{comb_ind}}
\end{figure}

A total of seven different combinations of original data, first derivatives, and second derivatives with indices, have been considered. 
To determine the best combination, we compared their performance by applying a robust Mahalanobis distance and calculating the mean AUC across 100 simulations for each of 19 DGPs described in Section~\ref{dgps} are compared (Fig.~\ref{comb_ind}). Relying on the first and second derivative indices produces highly variable outcomes. When the smoothed curves are included, the variability is reduced and every AUC stays above 0.8, whereas the derivative-only option falls to roughly 0.5 in at least one case. Therefore, for the next steps we will compute ABEI and ABHI on the original curves and their first two derivatives.

A similar approach is followed to select the outlier detection methodology in Step 3. In the multivariate case, traditional univariate methods may fail since an observation does not need to be extreme across all variables to be an outlier. Instead, both the distance from the data centroid and the overall data shape must be considered. The Mahalanobis distance \citep{mahalanobis1936generalized} is a common metric for this purpose, with robust Mahalanobis distance (RMD) variants being particularly useful. To mitigate the impact of outliers, robust statistical methods are employed to estimate the central location and the covariance matrix for the Mahalanobis distance.

The following five robust methodologies have been considered. The process of selecting a threshold for the identification of outliers is a difficult task that is data dependent and can have dramatic effects on the final result of the method.
Note that the thresholds used in this work are those proposed in the original works or defined in their publicly available implementations.

\begin{itemize}
    \item \textbf{FASTMCD}: Computes an RMD using the minimum covariance determinant (MCD) in its fast implementation \citep{rousseeuw1999fast}. The algorithm iterates over many random subsets, which may be computationally expensive for large datasets. The threshold is set at the 97.5 percentile of the corresponding $\chi^2$-distribution.
    
    \item \textbf{FASTMCD-Adj}: The same RMD as FASTMCD, but with an adjusted threshold as proposed by \cite{filzmoser2005multivariate}.
    
    \item \textbf{OGK}: The orthogonalized Gnanadesikan-Kettenring (OGK) estimator, a robust location and scatter estimator based on a modified version of the Gnanadesikan–Kettenring method \citep{maronna2002robust,gnanadesikan1972robust}. The threshold is set at the 97.5 percentile of the corresponding $\chi^2$-distribution.
    
    \item \textbf{COM}: The comedian method, introduced by \cite{sajesh2012outlier}, based on the robust median absolute deviation (MAD) estimator \citep{falk1997mad}, is employed to identify multivariate outliers. The method computes robust estimates of location and scatter, which are then used to calculate a robust Mahalanobis distance ($RD_i$) for each observation $x_i$. An observation is flagged as an outlier if its distance exceeds a predefined cutoff value defined using a boxplot-type criterion. 

    \item \textbf{RMDsh}: An RMD estimator incorporating shrinkage, as proposed by \cite{cabana2021multivariate}. Shrinkage estimators provide a balance between variance and bias. The threshold is set at the 97.5 percentile of the corresponding $\chi^2$- distribution.
\end{itemize}

The implementations for FASTMCD and OGK are sourced from the 
\verb|rrcov| R package \citep{rrcovRpackage}, while COM relies on the \verb|robustbase| R package \citep{robustbaseRpackage}. The RMDsh method is implemented in R based on a MATLAB version given by the authors.

\begin{figure}[htbp]
\centering
\includegraphics[width=0.47\textwidth]{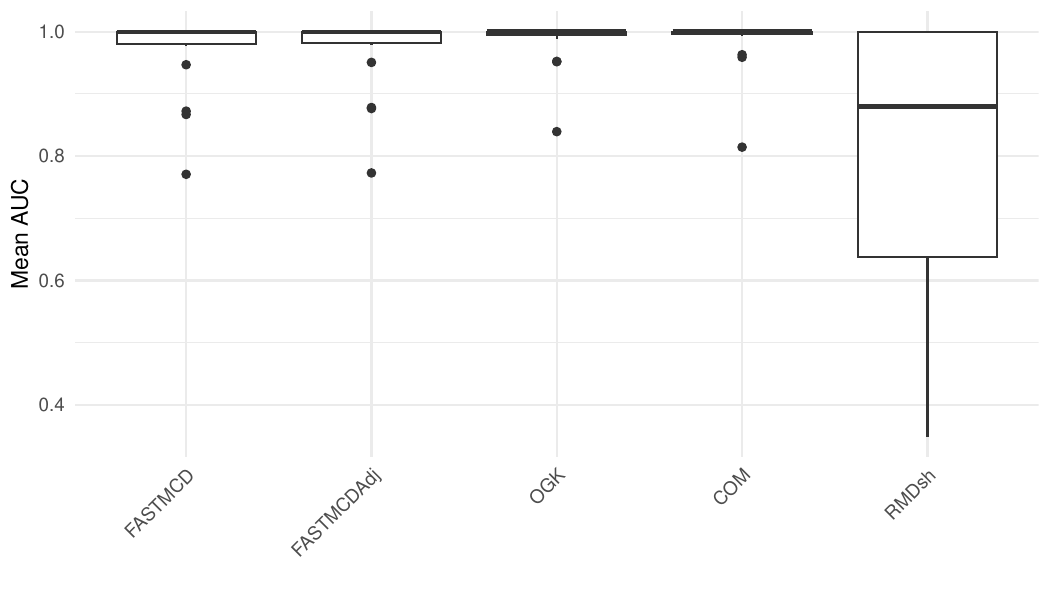}
\caption{Boxplots of the mean AUC for different multivariate outlier detection methods over 100 simulation runs of 19 DGPs.  \label{indM_met}}
\end{figure}

Fig.~\ref{indM_met} displays boxplots of the mean AUC (from 100 simulation runs per method across the 19 DGPs detailed in Section~\ref{dgps}), aiding in the selection of the most effective methodology. Although OGK demonstrates the lowest dispersion, the COM method achieves the highest mean AUC. Crucially, the difference in dispersion between COM (the second-best in this regard) and OGK is very small. Therefore, considering its leading mean AUC and this small difference in variability, COM is selected for the subsequent steps of EHyOut.

In this way, EHyOut follows a process that begins with data smoothing, followed by the computation of the ABEI and ABHI for the smoothed curves as well as their first and second derivatives. Finally, the Comedian method is applied to identify the outlying observations.

\section{Benchmarking EHyOut against alternative approaches} \label{simul}

EHyOut, as introduced in Section~\ref{s_ehyout}, is applied to each DGP described in Section~\ref{dgps} and compared against a range of state-of-the-art functional data approaches for outlier detection, each representing a distinct strategy for identifying outliers in functional datasets. This comprehensive comparison allows us to evaluate the strengths and limitations of EHyOut, offering valuable insights into its performance and robustness across diverse scenarios.

The objective of this section is to identify an outlier detection method for functional data that can reliably detect outliers of different types across a range of diverse functional datasets. To this end, methodologies designed for specific outlier types have also been considered. For instance, methods like the Outliergram, which are tailored for shape outliers, may fail when shape outliers are absent. However, these methods also sometimes fail in scenarios where shape outliers are present.\\

We evaluate the following methods: 
\begin{itemize}

    \item \textbf{FASTMUOD} identifies outliers using the method proposed by \cite{ojo2022}, which provides three indices that quantify its magnitude, shape, and amplitude outlyingness. A classical boxplot procedure is then applied to these indices to determine appropriate cutoffs for outlier detection.

    \item \textbf{OG} denotes the outliers detected by the Outliergram introduced in \cite{arribas}. This method primarily targets shape outliers by constructing a scatterplot of the modified epigraph index (MEI) against the modified band depth (MBD). Outliers are identified as points that fall below the parabola formed in the $(\text{MEI}, \text{MBD})$ plot, using a boxplot to flag distant observations. Additionally, the functional boxplot is employed to detect magnitude outliers.

    \item \textbf{MSPLOT} refers to the magnitude-shape (MS) plot based on directional outlyingness for multivariate functional data, as developed by \cite{dai2018multivariate}. It decomposes the overall outlyingness of a curve into magnitude outlyingness (mean directional outlyingness, MO) and shape outlyingness (variation of directional outlyingness, VO). These are visualized in a scatterplot of $(MO, VO)^T$, and outliers are identified using the squared robust Mahalanobis distance, estimated via the minimum covariance determinant (MCD) algorithm \citep{rousseeuw1999fast}. The distribution of these distances is approximated using an $F$-distribution, following the approach in \cite{hardin2005robust}. Curves with robust distances exceeding a threshold from the upper tails of the Snedecor-F distribution are flagged as outliers.

    \item \textbf{TVD} is a two stage procedure introduced by \cite{huang2019decomposition} which uses the Total Variation Depth (TVD) and the Modified Shape Similarity Index (MSS). First, a classical boxplot is applied to the MSS index values to detect shape outliers. After removing these, a functional boxplot based on TVD is used on the remaining data to detect magnitude outliers. For the MSS boxplot, we use the inflation factor of Fc = 3.

    \item \textbf{MBD} employs the functional boxplot proposed by \cite{sun2011genton}. This method uses the Modified Band Depth (MBD) to order curves from center to outwards, definig a 50\% central region based on the deepest curves. Outliers are then identified as any curves that lie completely outside the whiskers, which are defined by a 1.5-fold inflation of this central region's range.

    \item \textbf{MDS-LOF} applies metric multidimensional scaling (MDS) to reduce the dimensionality of the functional data to five components, followed by the application of the Local Outlier Factor (LOF) algorithm for outlier scoring, as described in \cite{herrmann_geometric_2021}.  In this case 5 dimension embedding is considered, and it is denoted as MDS5LOF. An observation is flagged as an outlier if its LOF score exceeded the 90th percentile of all computed LOF scores.

    \item \textbf{BP-PWD} is a two steps method described in \cite{jimenez-varon_pointwise_2023} designed to handle shape and magnitude outliers separately. It first identify shape outliers by constructing a boxplot based on the pointwise data depth (PWD). After removing shape outliers, a functional boxplot \citep{sun2011genton} is used to identify magnitude outliers.

\end{itemize}

The performance of EHyOut is evaluated against the outlined benchmark methodologies using two primary criteria: classification effectiveness and computational efficiency. For classification effectiveness, the Matthews Correlation Coefficient (MCC) \citep{matthews1975comparison} is employed using the thresholds proposed by the original authors for each method. To provide a threshold-independent evaluation, the AUC of those methods using a unique score to identify outliers is also computed. Finally, computational efficiency is assessed by measuring the execution time (ET) of each algorithm in seconds.

The MCC formulation is:
\begin{equation*}
\textstyle
MCC = \frac{ TP \times TN - FP \times FN }{ \sqrt{ (TP + FP)(TP + FN)(TN + FP)(TN + FN) } },
\end{equation*}
where TP, TN, FP, and FN represent True Positives, True Negatives, False Positives, and False Negatives, respectively, from the confusion matrix.

A key advantage of MCC, particularly well-suited for assessing outlier detection, is its inherent robustness to class imbalance, a common characteristic of outlier detection datasets where outliers are significantly outnumbered by inliers. While common metrics like accuracy and F1-score are widely adopted, they can be misleading in such imbalanced settings. For instance, accuracy can be high even if no outliers are detected. \cite{chicco2020advantages} have shown MCC to be a more truthful and informative metric than accuracy and F1-score under these conditions. Furthermore, while precision and recall highlight specific aspects such as minimizing false alarms (high precision) or maximizing detection rate (high recall), MCC synthesizes performance across all four quadrants of the confusion matrix into a single one. This provides a more balanced and comprehensive assessment of overall classification quality in scenarios like outlier detection, where both missing an outlier (FN) and falsely flagging an inlier (FP) can have significant consequences.

More generally, evaluation measures of outlier detection based on continuous scores of "outlyingness" are preferable to evaluation metrics that rely on a dichotomization of data points into outliers and inliers for methods that return such continuous scores, since results will be highly sensitive to the more or less arbitrarily chosen thresholds used for this dichotomization. See Appendix~\ref{appC} for tables containing the corresponding AUC values for those  methods that return continuous outlyingness scores (EHyOut, OG, MDB, MDS5LOF).

\subsection{Simulation results}

The performance of EHyOut and the proposed benchmarking methodologies were evaluated on the 19 DGPs described in Section~\ref{dgps}, considering outlier proportions of $\alpha = 0.01, 0.05, \text{ and } 0.1$. Detailed results for each of these $\alpha$ values, including True Positive Rate (TPR), False Positive Rate (FPR) and AUC when available, from 100 simulations per DGP, are presented in Appendix~\ref{appC}.

The results in Appendix~\ref{appC} when comparing MCC for different $\alpha$ values indicate that the behavior of the eight methodologies remained substantially stable in nearly all cases. Consequently, for clarity and conciseness in the main text, the subsequent discussion and presentation of results will primarily focus on $\alpha=0.1$. Furthermore, our analysis reveals that high AUC values do not always coincide with perfect TPR and FPR. This suggests that while many methods effectively rank outliers, their default thresholds could be improved. Fine-tuning these thresholds is a complex, data-dependent challenge that is beyond the scope of this work. Instead of undertaking such case-by-case optimization, our study provides a robust evaluation based on the thresholds originally proposed by the respective authors. Even for EHyOut implementation, the threshold used is the one given by the authors of the Comedian Method. While better partitions are theoretically possible with custom thresholds, this work provides a crucial and practical benchmark of the effectiveness of these methods.

\begin{figure*}[htbp]
\centering
\includegraphics[width=\textwidth]{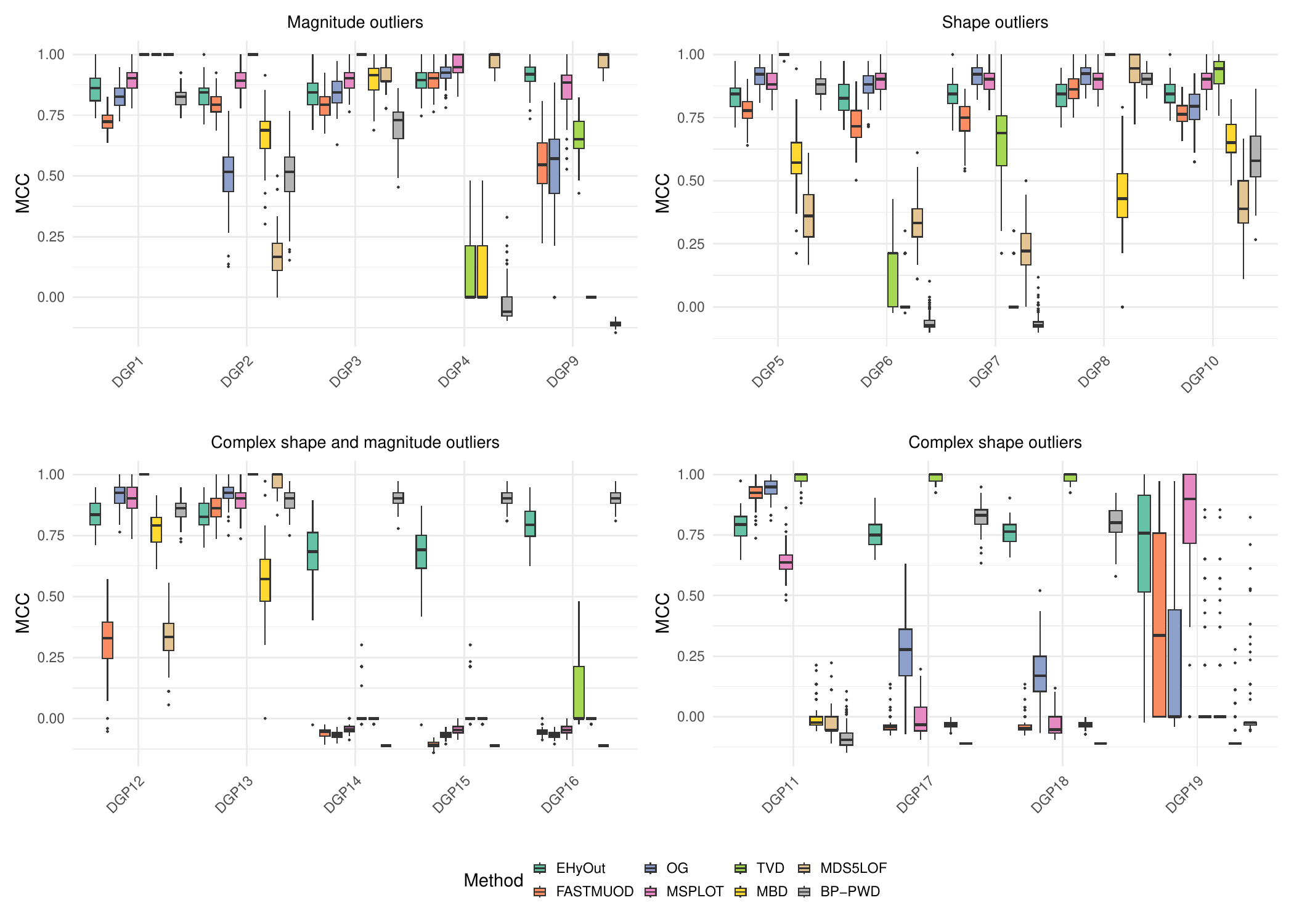}
\caption{Comparison of MCC performance for eight methodologies across 19 DGPs. Boxplots show the distribution of MCC scores from 100 simulations with $\alpha=0.1$. Each of the four subplots presents a group of DGPs based on their outlier type, allowing for comparisons of the color-coded methodologies for each specific scenario. \label{dgp_mcc}}
\end{figure*}

Fig.~\ref{dgp_mcc} presents a detailed comparison of the methodologies performance for each DGP. It
displays boxplots of the MCC derived from 100 simulations using $\alpha=0.1$, where DGPs are grouped by outlier type into four subplots.  Within each subplot, the DGPs are organized along the x axis, allowing for easy comparisons of the eight methodologies for each specific scenario.

\begin{figure*}[htbp]
\centering
\includegraphics[width=\textwidth]{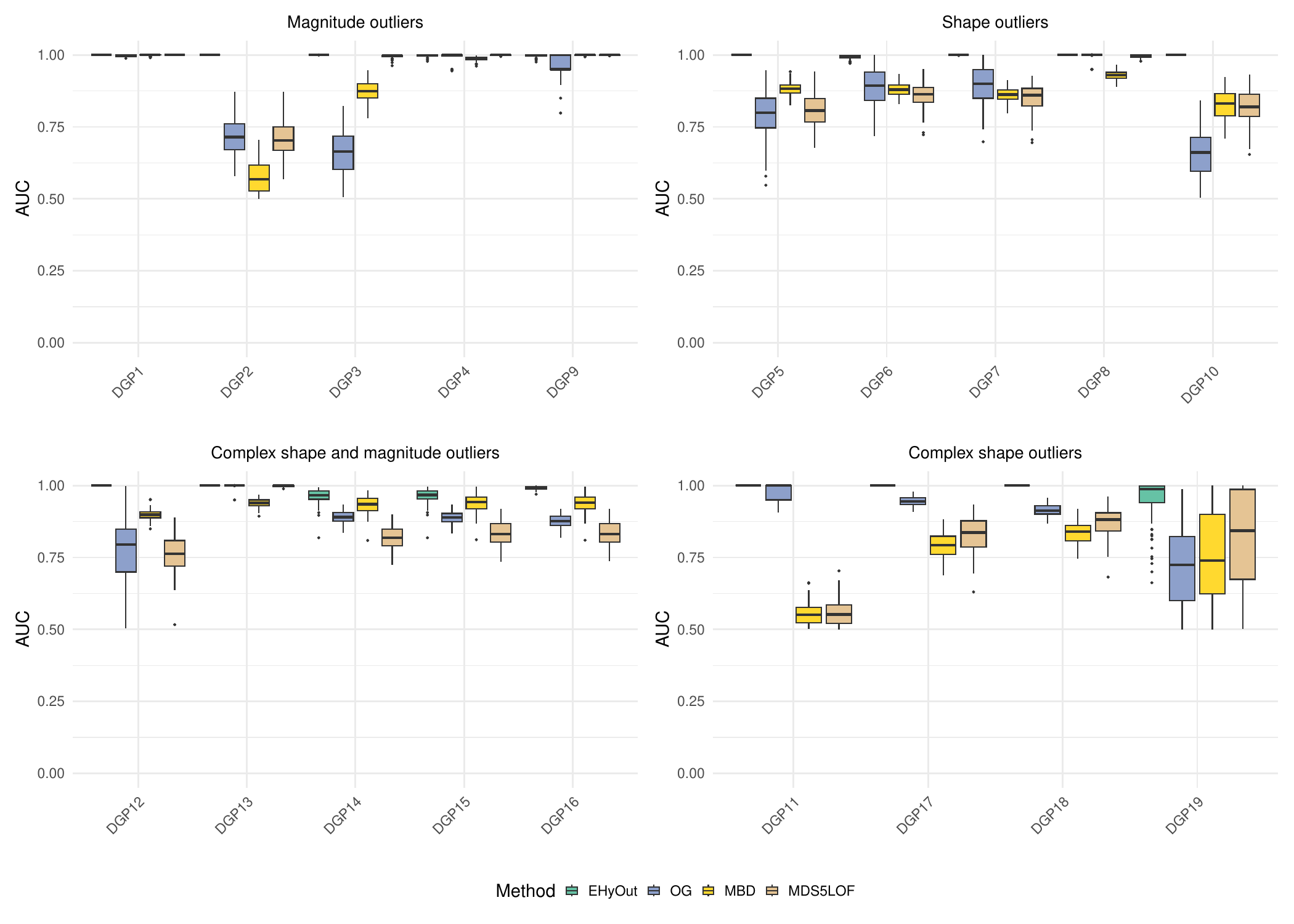}
\caption{Comparison of AUC performance for eight methodologies across 19 DGPs. Boxplots show the distribution of AUC scores from 100 simulations with $\alpha=0.1$. Each of the four subplots presents a group of DGPs based on their outlier type, allowing for comparisons of the color-coded methodologies for each specific scenario. \label{dgp_auc}}
\end{figure*}

In the same way, Fig~\ref{dgp_auc} present the same comparisons in terms of AUC. Note that FASTMUOD, MSPLOT, TVD, and BP-PWD AUC values are not available because they do not use a single score to flag outliers.

For magnitude outliers (DGPs 1, 2, 3, 4, and 9), EHyOut, FASTMUOD, and MSPLOT consistently achieved high MCC scores.  Nevertheless, while TVD and MDS5LOF demonstrated strong performance on some of these DGPs, they underperformed on others. Regarding AUCs, EHyOut and MSPLOT always obtain AUCs close to 1. The case of DGP9 is particularly insightful, as the gap between its high AUCs and not that high MCC scores suggests that a different decision threshold could yield a more accurate classification outcome.  In the case of shape outliers (DGPs 5, 6, 7, 8, and 10), EHyOut, MSPLOT, OG, and FASTMUOD showed consistent results across these DGPs. Conversely, the remaining methodologies, despite achieving good results for certain DGPs in this category, struggled with outlier detection in others. 

DGPs 12 to 16 feature more complex shape and magnitude outliers. Given that these DGPs originate from \cite{jimenez-varon_pointwise_2023}, the superior performance of the therein-proposed BP-PWD methodology (when applied to its native DGPs) is expected. Nevertheless, among the benchmarked methodologies, EHyOut was the only one to achieve median MCC values above 0.5 for all these challenging sets. For instance, OG, TVD, and MSPLOT performed well on DGP12 and DGP13 but struggled with DGPs 14 and 15. A comparison of the MCC results with the threshold-independent AUC results show that there are some DGPs obtaining small MCC values, while hihg AUCs. This is particulary clear for example, in DGP14, where all methods except EHyOut and BP-PWD, obtain MCC values close to 0. Finally, for DGPs 11, 17, 18, and 19, which represent diverse shape outliers, EHyOut again was the sole method achieving median MCCs greater than 0.5 across all these cases. TVD emerged as a strong competitor, outperforming EHyOut on DGPs 11, 17, and 18, but performed poorly on DGP19. Although MCC may achieve higher AUC scores on certain DGPs, EHyOut is the only method that demonstrates consistently superior performance across all scenarios.

This detailed DGP-by-DGP analysis highlights EHyOut's consistent and robust performance. While the benchmarked methodologies occasionally surpass EHyOut on specific DGPs, they often lacked consistency, struggling with outlier detection in other scenarios.
Moreover, comparing results in terms of MCC (Figure~\ref{dgp_mcc}) and AUC (Figure~\ref{dgp_auc}) reveals that AUC values are generally much higher than MCC ones. This fact highlights that many methods are limited by their default thresholds rather than their inherent discriminative ability. Nonetheless, the overall performance pattern mirrors our MCC findings: EHyOut is the only methodology to achieve consistently high performance. This suggests that performances of some of the other methods could be enhanced by using an adaptive threshold based on the outlier scores themselves. While this is a valuable direction for future research, it is not tackled here, as our focus remains on evaluating the methods as proposed by their authors.

\begin{figure}[htbp]
\centering
\includegraphics[width=0.47\textwidth]{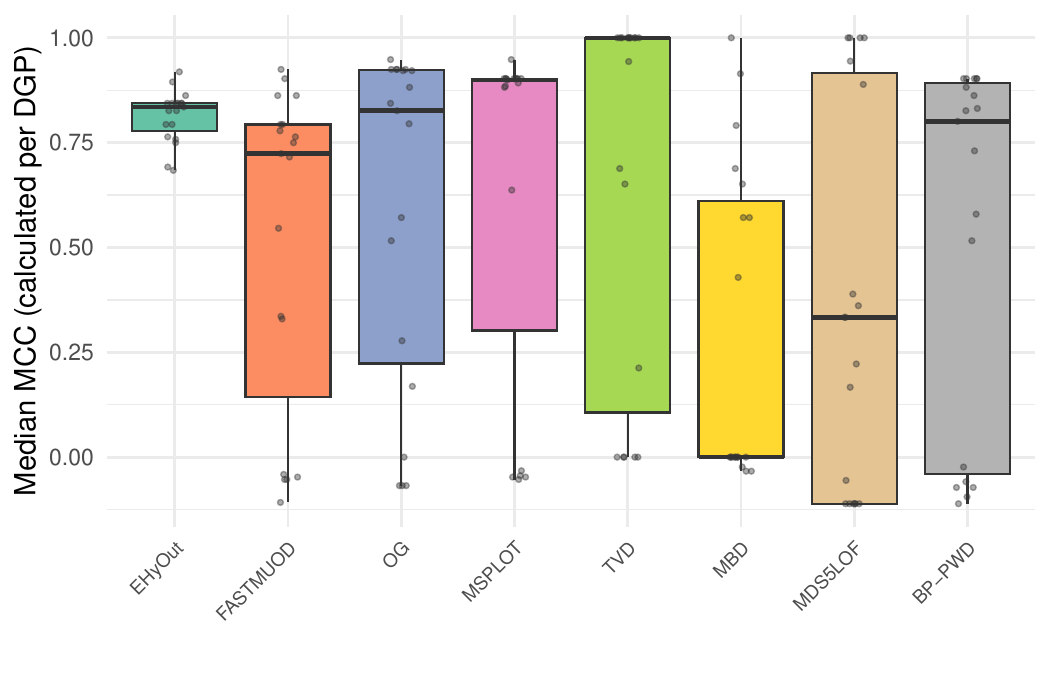}
\caption{Distribution of median MCC scores for various outlier detection methodologies across 19 DGPs. Each boxplot summarizes the per-DGP median MCCs for a given methodology. All individual per-DGP median MCC values are overlaid as jittered grey dots, illustrating the full spread of performance for each method.
\label{bp_mcc}}
\end{figure}

To provide an aggregate view, Fig.~\ref{bp_mcc} displays the distributions of median MCCs across all 19 DGPs for each outlier detection methodology. This summary reinforces the performance trends observed in the detailed per-DGP results. While some benchmarked methodologies reach higher median MCCs on individual DGPs, EHyOut is distinguished by its very stable and strong overall performance. Notably, EHyOut is the only methodology whose median MCC score (per DGP) remained above 0.5 for all 19 DGPs. Furthermore, the significantly smaller interquartile range of EHyOut's MCC distribution across DGPs underscores its consistency and reliability across a diverse range of data with different outlier types.

\begin{figure}[htbp]
\centering
\includegraphics[width=0.47\textwidth]{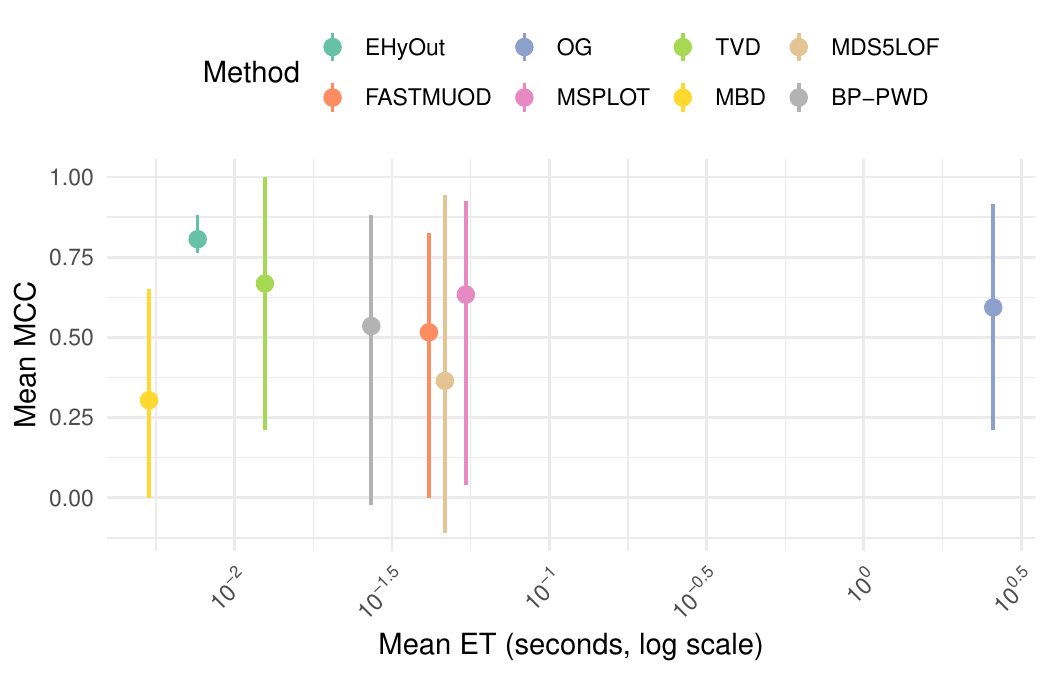}
\caption{Mean MCC versus Mean ET (log scale, in seconds) for different methods, aggregated across all 19 DGPs. Points represent the mean values. Vertical lines indicate the Interquartile Range (IQR) of MCC for each methodology.
\label{time_mcc}}
\end{figure}

Finally, Fig.~\ref{time_mcc} illustrates the trade-off between mean execution time (ET) and mean MCC for each methodology, aggregated across all simulations for the 19 DGPs with an outlier proportion of $\alpha=0.1$. A logarithmic scale is employed for ET (in seconds) to enhance visualization. This plot highlights EHyOut as offering a favorable balance: it achieves the highest mean MCC with an extremely small interquartile range (IQR) for its MCC scores and the second fastest mean ET. Although TVD reaches MCC values equal to 1 in some instances, its overall MCC distribution exhibits considerably larger variability. Conversely, while MBD is the fastest methodology in terms of ET, its MCC performance is not competitive with the leading methods. Table~\ref{t_mean} provides the precise mean ET values (on the original, non-logarithmic scale) and other summary statistics used for plotting Fig.~\ref{time_mcc}.

\begin{table}[htbp]
\tabcolsep=0pt
\caption{Performance comparison of different methods. Metrics include Mean Time (seconds), Mean MCC, and the first (Q1) and third (Q3) quartiles of MCC, considering values aggregated from all 19 DGPs. Bold values indicate the best performance in each column, and underlined values highlight the second-best result.}
\label{t_mean}
\begin{tabular*}{\columnwidth}{@{\extracolsep{\fill}}lrrrr@{}}
\hline
 Method & Mean Time & Mean MCC & MCC Q1 & MCC Q3 \\ 
  \hline
    EHyOut & \underline{0.008} & \textbf{0.806} & \textbf{0.764} & 0.882 \\ 
    FASTMUOD & 0.041 & 0.516 & 0.000 & 0.826 \\ 
    OG & 2.578 & 0.594 & \underline{0.213} & 0.916 \\ 
    MSPLOT & 0.054 & 0.633 & 0.039 & 0.925 \\ 
    TVD & 0.012 & \underline{0.668} & \underline{0.213} & \textbf{1.000} \\ 
    MBD & \textbf{0.005} & 0.304 & 0.000 & 0.651 \\ 
    MDS5LOF & 0.047 & 0.365 & -0.111 & \underline{0.944} \\ 
    BP-PWD & 0.027 & 0.535 & -0.024 & 0.882 \\
   \hline
\end{tabular*}
\end{table}

In conclusion, for applications requiring an outlier detection methodology for functional data that performs effectively without prior specification of outlier types, EHyOut emerges as the most robust choice. This is supported by its superior performance in both threshold-dependent classification (MCC, Figs.~\ref{dgp_mcc} and \ref{bp_mcc}) and threshold-independent evaluation (AUC, Fig~\ref{dgp_auc}).  Furthermore, regarding ET, EHyOut is consistently one of the fastest algorithms, outperformed only by MBD, as detailed in Fig.~\ref{time_mcc}. However, MBD's advantage in speed is offset by its comparatively weaker performance on MCC. Therefore, EHyOut distinguishes itself as a highly efficient and reliable methodology for outlier detection in functional data analysis, specifically when the outlier typology is unknown.

\section{Application to real data} \label{real}

In this section, EHyOut is applied to two different real scenarios widely studied in the literature for detecting outliers: Spanish weather data and population growth patterns in different countries.

\subsection{Spanish weather data}

The Spanish weather data, collected by the Agencia Estatal de Meteorología (AEMET), includes daily average temperature, precipitation, and wind speed from 73 weather stations across Spain for the period 1980–2009. This dataset also provides information about the regions in which these stations are located. These data are available in the \verb|fda.usc| R package \citep{febrero2012statistical} and have been analyzed in the FDA literature, such as in \cite{dai2018multivariate}, \cite{herrmann_geometric_2021} and \cite{ojo2022}. 

\begin{figure}[htbp]
\begin{minipage}[b]{0.49\linewidth} 
\centering
\includegraphics[width=\linewidth]{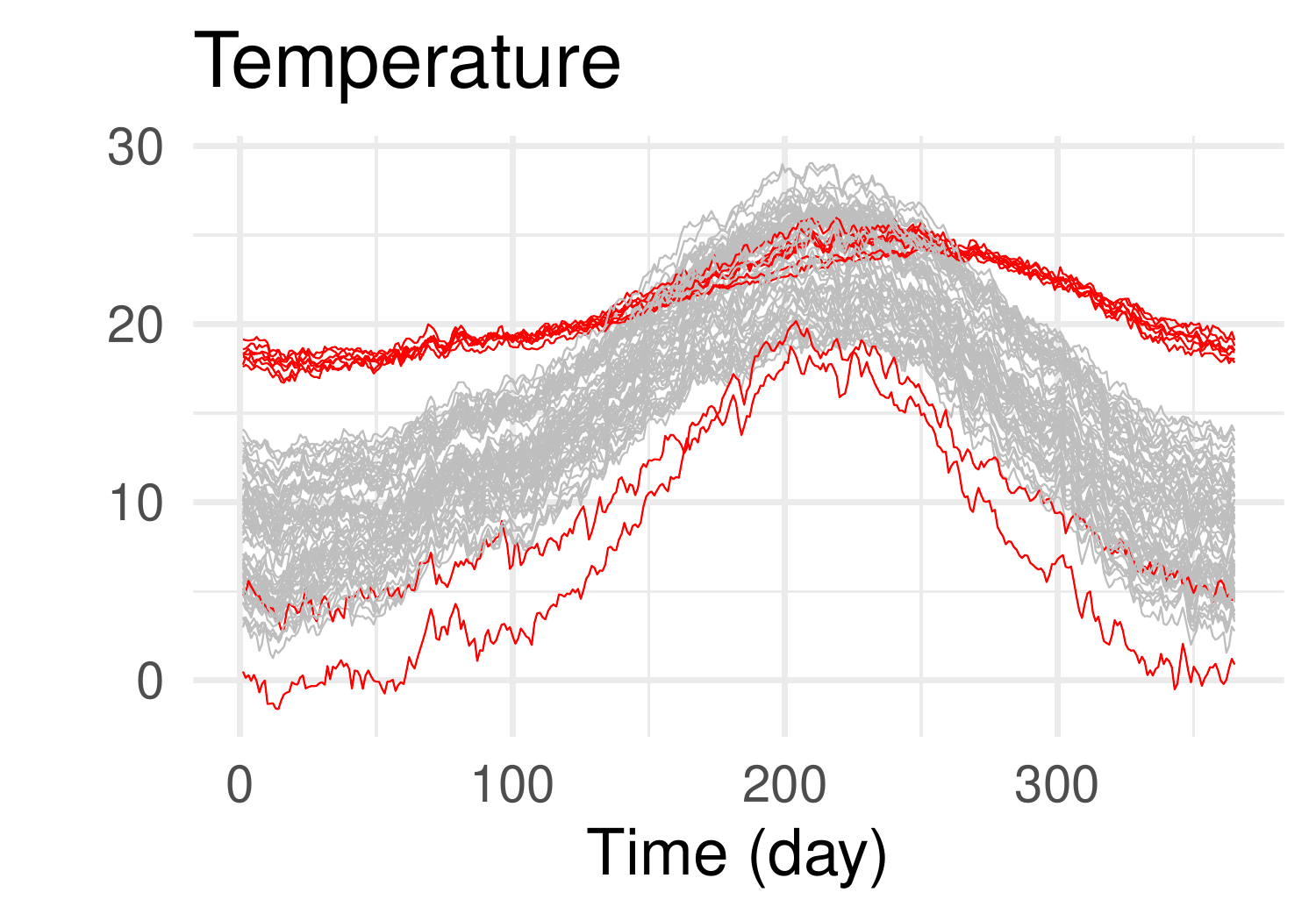}
\end{minipage}
\begin{minipage}[b]{0.49\linewidth}
\centering
\includegraphics[width=\linewidth]{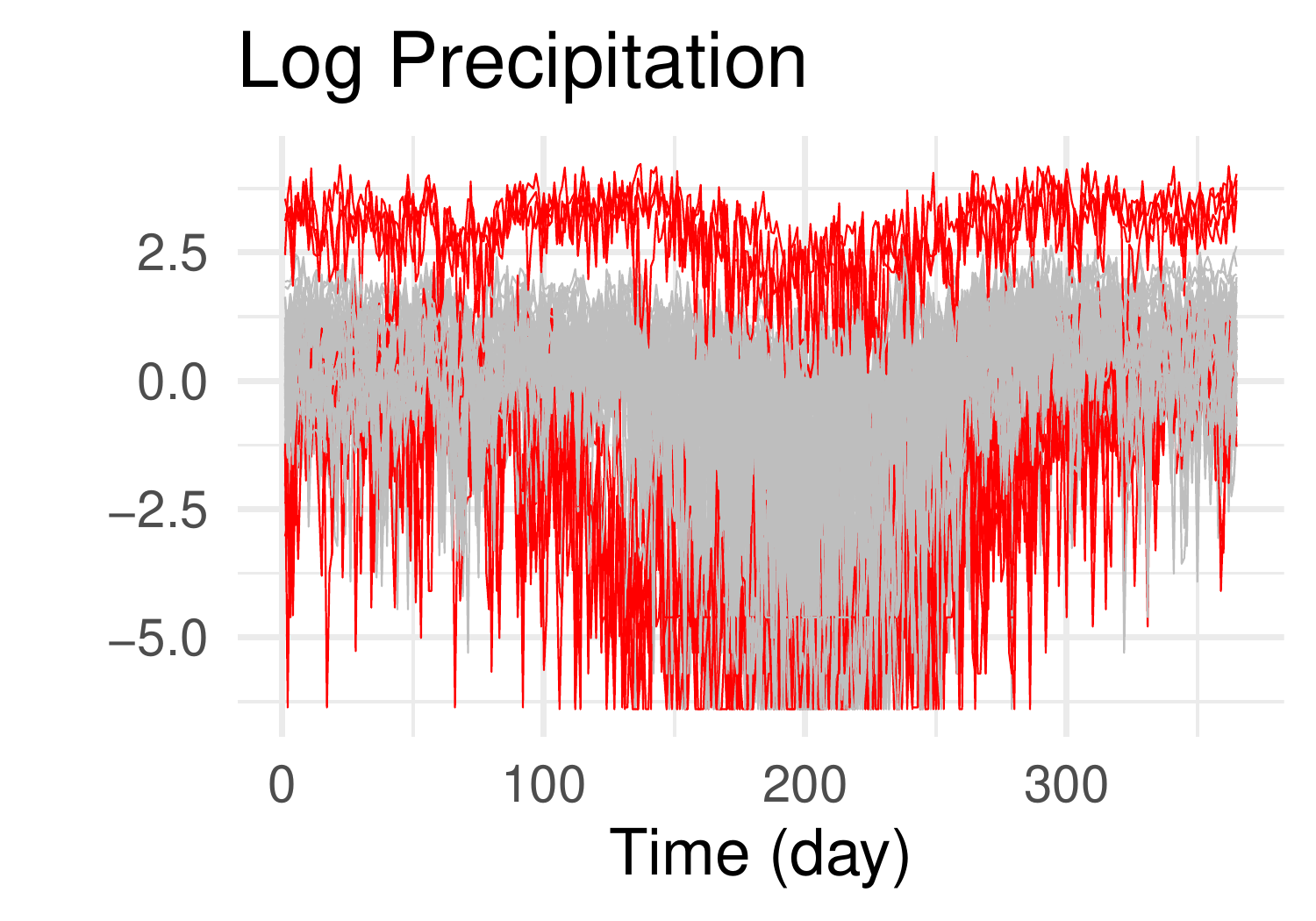}
\end{minipage}
\caption{Outlying curves identified using EHyOut. Outliers are highlighted in red. Temperature curves are displayed on the left, and log precipitation curves on the right.} 
\label{temp_prec}
\end{figure}

\begin{figure}[htbp]
    \centering
    \includegraphics[width=0.45\textwidth]{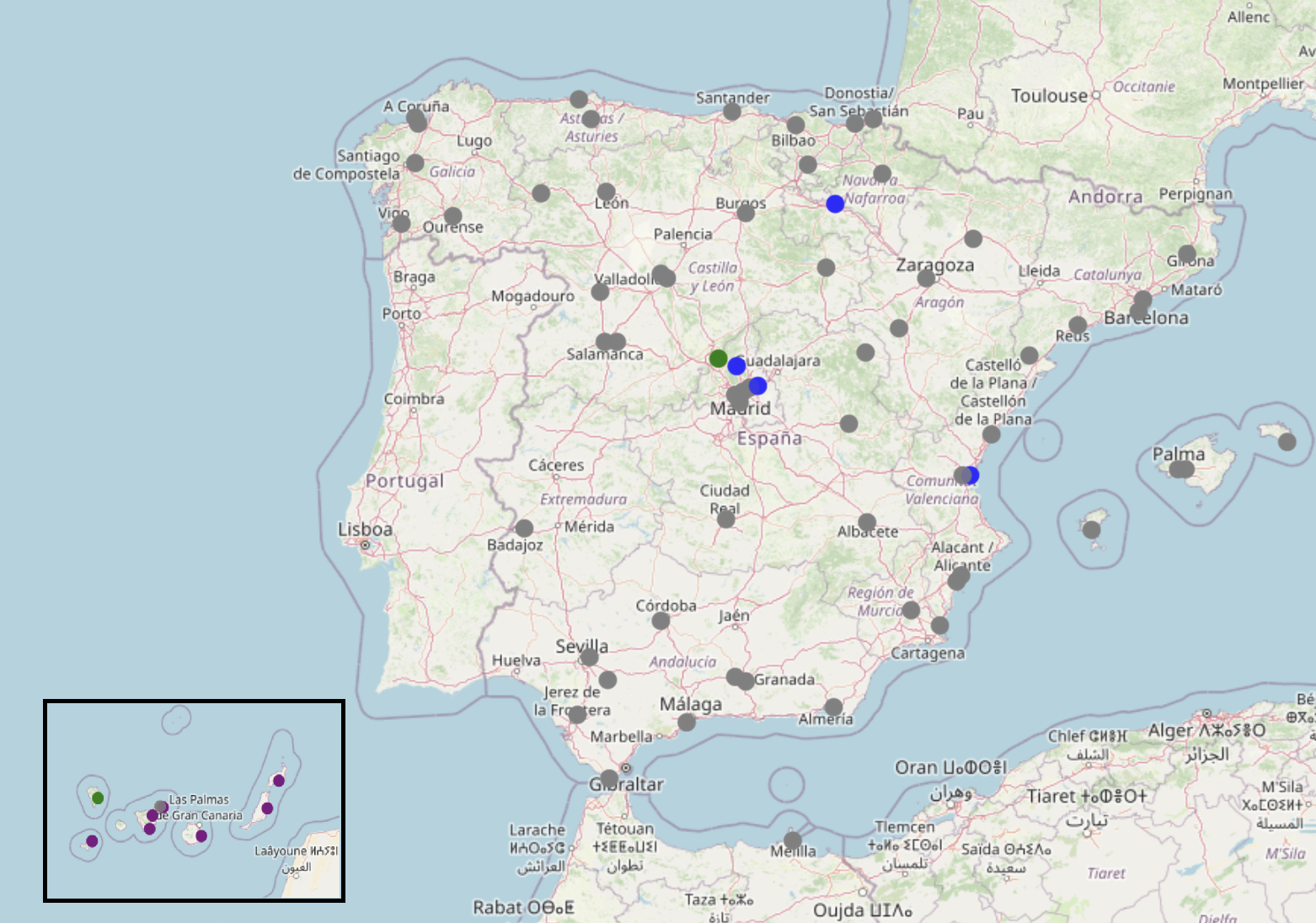}
    \caption{Weather stations identified as outliers using EHyOut. Stations outlying in both temperature and log precipitation are shown in purple, those outlying only in temperature are in blue, and those outlying only in log precipitation are in green.}
    \label{temp_prec_fig}
\end{figure}

For our analysis, we use the temperature and log precipitation data separately. In this way, we obtain curves that are outliers only for precipitation or only for temperature. However, in this analysis we compare both results and obtain those curves that outline in both procedures. Fig.~\ref{temp_prec} illustrates the temperatures on the left and log precipitation on the right, with the outliers obtained with EHyOut flagged in red. Table~\ref{tem_prec_names} and Fig.~\ref{temp_prec_fig} help to identify the geographic locations of these outliers. The map uses purple points to mark stations that coincide to be outliers for both temperature and log precipitation, blue points for stations that are outliers only for precipitation, and green points for those outlying only for temperature.

\begin{table}[htbp]
  \caption{Weather stations identified as outliers using EHyOut. Stations with outlying temperatures are on the left, while those with outlying log precipitation are on the right.}
  \label{tem_prec_names}
  \centering 
  \small 
  \setlength{\tabcolsep}{3pt} 
  \begin{tabularx}{\linewidth}{@{} >{\raggedright\arraybackslash}X | >{\raggedright\arraybackslash}X @{}}
    \toprule 
    Temperature & Log Precipitation \\
    \midrule 
    Fuerteventura (aeropuerto), Lanzarote (aeropuerto), Las Palmas de Gran Canaria (Gando), Hierro (aeropuerto), Tenerife (sur), Sta.Cruz de Tenerife, Izana, La Palma (aeropuerto), Navacerrada (puerto) & Fuerteventura (aeropuerto), Lanzarote (aeropuerto), Las Palmas de Gran Canaria (Gando), Hierro (aeropuerto), Tenerife (sur), Sta.Cruz de Tenerife, Izana, Logroño (Agoncillo), Valencia,  Madrid (Torrejón), Colmenar Viejo (FAMET) \\
    \bottomrule 
  \end{tabularx}
\end{table}

EHyOut identified 9 stations with outlying behavior in precipitation and 11 for temperature, 7 of which were the same. The stations flagged as outliers in both precipitation and temperature were located in the Canary Islands, which are far from the other stations and exhibit distinct weather patterns. 
These findings are consistent with the results from the MS-plot in \cite{dai2018multivariate} and Fast-MUOD in \cite{ojo2022}, where all Canary Islands are identified as temperature outliers. Additionally, the station in Navacerrada is detected as a temperature outlier, which is also noted by Fast-MUOD due to its consistently low temperatures. 
Regarding precipitation, EHyOut aligns with the MS-plot in identifying almost all Canary Islands as outliers, and it agrees with both strategies in marking the remaining four stations on the peninsula as outliers due to their higher precipitation volumes.

\subsection{Population growth in different countries}

In this section we consider the world population dataset from the United Nations, analyzed by \cite{nagy_depth-based_2017}, \cite{dai_functional_2020} and \cite{ojo2022}. This comprehensive dataset includes yearly estimates of the total population for 233 countries and autonomous regions, recorded every July from 1950 to 2010, except from \cite{ojo2022} where it ranges until 2015. Consistent with the methodologies employed by \cite{nagy_depth-based_2017}, \cite{dai_functional_2020} and \cite{ojo2022}, we filter the dataset to include only those countries with populations ranging between one million and fifteen million as of July 1, 1980. This criterion results in a subset of 105 countries out of the initial 233. The resulting functional data for our analysis thus comprise 105 population curves, each observed at 65 time points spanning from 1950 to 2010.

Applying EHyOut to this dataset, 38 countries appear as outliers, distributed in continents as follows: 
\begin{itemize}
    \item \textbf{Africa}:
Madagascar, Malawi, Mozambique, Rwanda, Somalia, South Sudan, Uganda, Zimbabwe, Angola, Cameroon, Sudan, Burkina Faso, Cote d'Ivoire, Ghana, Niger

\item \textbf{Asia}:
Kazakhstan, Afghanistan, Nepal, Cambodia, Malaysia, Iraq, Saudi Arabia, Syrian Arab Republic, United Arab Emirates, Yemen

\item \textbf{Europe}:
Belarus, Czech Republic, Hungary, Bosnia and Herzegovina, Greece, Portugal, Serbia, Belgium, Netherlands

\item \textbf{America}:
Cuba, Chile, Ecuador
\item \textbf{Oceania}: Australia
\end{itemize}

\begin{figure}[htbp]
\centering
\includegraphics[width=0.47\textwidth]{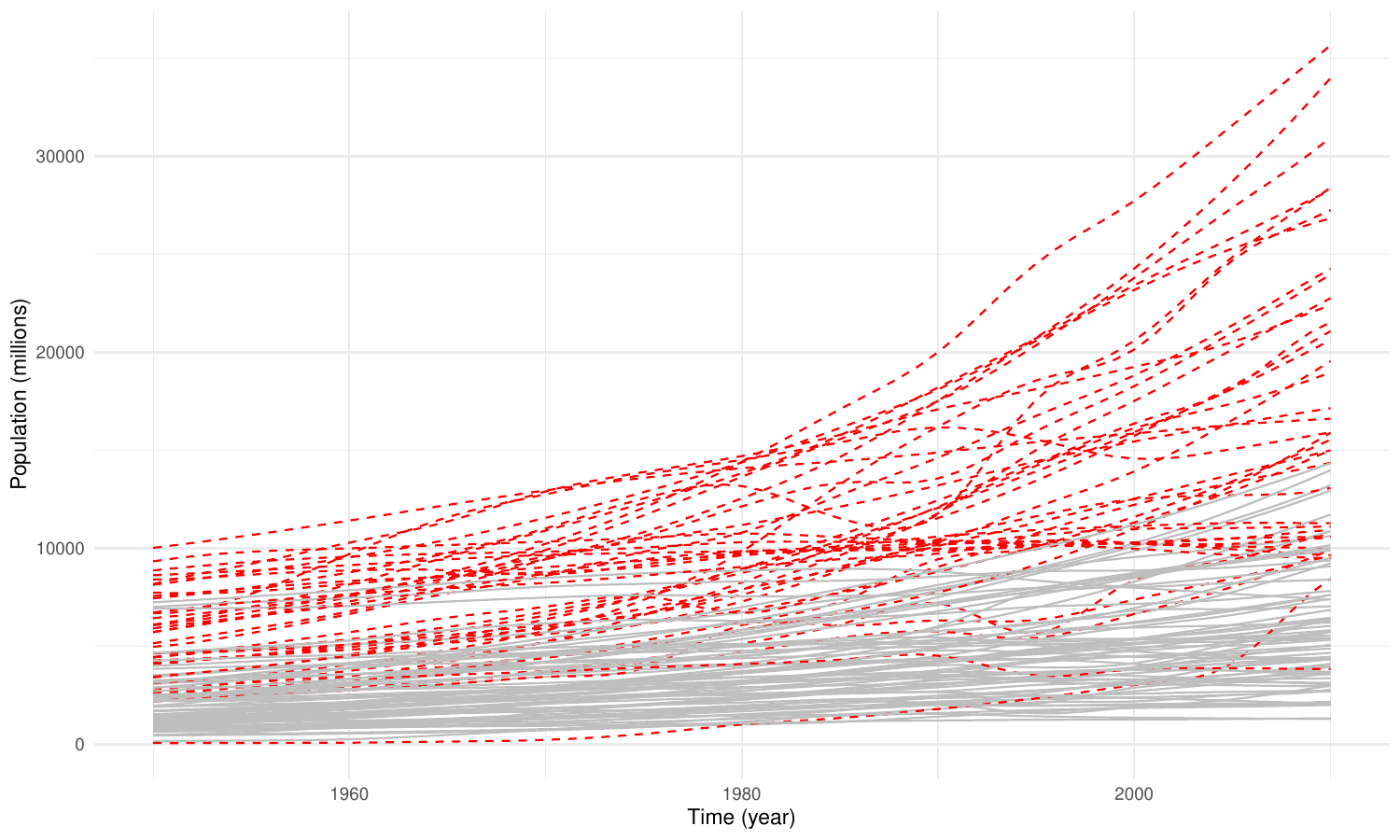}
\caption{Countries flagged as outliers, highlighted in red for EHyOut.} \label{countries}
\end{figure}

Moreover, Fig.~\ref{countries} displays population growth curves for the 105 selected countries, with outliers flagged by EHyOut highlighted in red.

EHyOut obtains a classification in outliers and non-outlier that have differences and similarities compared to those previously achieved in previous works. For instance, countries like Saudi Arabia, Iraq, Afganistan, Malasya and Uganda exhibited the highest population values by the end of the studied period, highlighting their intense growth during the period under review. In the case of Fast-MUOD from \cite{ojo2022},  which considered data up to 2015, Sudan was also identified as experiencing rapid growth in recent years. Nevertheless, the grouping proposed by \cite{nagy_depth-based_2017} neither classified Sudan as outlier when considering the end of the period in 2010.  Other nations, such as Cameroon or Yemen also flagged due to their significant growth patterns. In general, this pattern is particularly prevalent in countries across Africa and the Middle East. Conversely, some countries in Eastern Europe, such as Hungary and Czech Republic, were also flagged as outliers due to their moderate yet stable demographic behavior over time. A notable distinction of EHyOut compared to classifications by \cite{nagy_depth-based_2017}, \cite{dai_functional_2020} or \cite{ojo2022} is its inclusion of countries consistently ranked highly over time, such as Chile, Australia, and Cuba, alongside other European nations with shared behaviors like the Netherlands, Greece, and Portugal. In contrast, other strategies tend to flag curves with low flat behavior. Despite that, the classification given by EHyOut seems consistent, classifying curves with consistently high or significantly differing population trends as outliers, while defining those with flatter trends as within the norm.

\section{Conclusion} \label{conc}

The epigraph and hypograph indices as introduced by \cite{franc2011} have previously been applied for detecting shape outliers, but further theoretical analysis as well as empirical results obtained by us showed it did not consistently
 perform well if diverse types of outliers were present in the data. This work introduces the Area-Based Epigraph and Hypograph Indices (ABEI and ABHI) to face this problem. These indices explicitly incorporate the magnitude of deviations by considering the area between curves. This fundamental distinction allows ABEI and ABHI to be more sensitive to a broader range of outlying behaviors, particularly those not solely defined by shape.

EHyOut is a methodology for outlier detection in functional data  that is built upon these new indices, leveraging the power of ABEI and ABHI applied not only to the original functional data but also to its first and second derivatives. This strategy transforms a functional outlier detection problem into a multivariate outlier detection problem to which a robust multivariate outlier detection technique, the Comedian method, can be applied. An extensive simulation study, evaluated with both threshold-dependent (MCC) and threshold-independent (AUC) metrics, demonstrated that EHyOut is the most robust and reliable methodology among those benchmarked for detecting diverse functional outlier types (magnitude, amplitude, shape, and combinations thereof), all while maintaining highly competitive computational efficiency.

However, in real world applications where the ground truth is unknown, performance metrics like MCC and AUC cannot be computed. Practitioners must instead rely on a  decision threshold to classify observations as inliers or outliers, setting suitable threshold values depending on the context and domain knowledge. A primary objective of this study was to conduct a fair and reproducible benchmark using the thresholds proposed by the original authors of each method. Our analysis revealed that a method's theoretical potential (high AUC) does not always translate to optimal practical performance (lower MCC), indicating that default thresholds are not universally ideal. This highlights a clear and valuable avenue for future research: the development of a data-driven, adaptive thresholding strategy tailored specifically for EHyOut, potentially based on the distribution of its underlying ABEI and ABHI scores. Investigating this strategy was outside the scope of the present work, whose purpose was the comparative evaluation of existing methods. EHyOut was also applied to two real datasets, obtaining meaningful results.

Future research will also explore the adaptation of these area-based indices to the multivariate functional data setting, similarly to the work of \cite{pulido2024}. Moreover, area-based indices in this work have been defined from a pragmatic sampling perspective, leaving their probabilistic characterization for future work. Finally, the code for implementing EHyOut is currently available in the \href{https://github.com/bpulidob/ehyout}{ehyout} GitHub repository and including the methodology presented here in the \verb|ehymet| R package \citep{ehymetpackageR2024} constitutes a near future task.

\newpage

\begin{appendix}
\section{Data Generation Processes. Formulation and graphical examples}\label{appA}

This appendix provides the detailed mathematical formulation and a graphical example for each of the 19 Data Generation Processes (DGPs) used in the simulation study. Each DGP consists of 200 curves defined over the interval  $[0,1]$ with $\alpha$ of outliers. The figures below show an example for each DGP with $\alpha = 0.1$.

\begin{itemize}
    \item[DGP 1.] \textit{Main model}: $X_i(t) = 4t + e_i(t)$, for $i = 1,\ldots,n$. $e_i(t)$ is a Gaussian process with zero mean and covariance function $\gamma(s, t) = \exp\left(-|t - s|\right),$ where  $s, t \in \mathcal{I}.$ 
    
    \textit{Outlier model}: $X_i(t) = 4t + 8k_i + e_i(t)$, with \\ $k_i \in \{-1, 1\}$, and $P(k_i = -1) = P(k_i = 1) = 0.5.$ 
    
    \textit{Type of outliers:} \textbf{magnitude outliers} shifted from the main model.

    \begin{figure}[htbp]
    \centering
    \includegraphics[width=.5\textwidth]{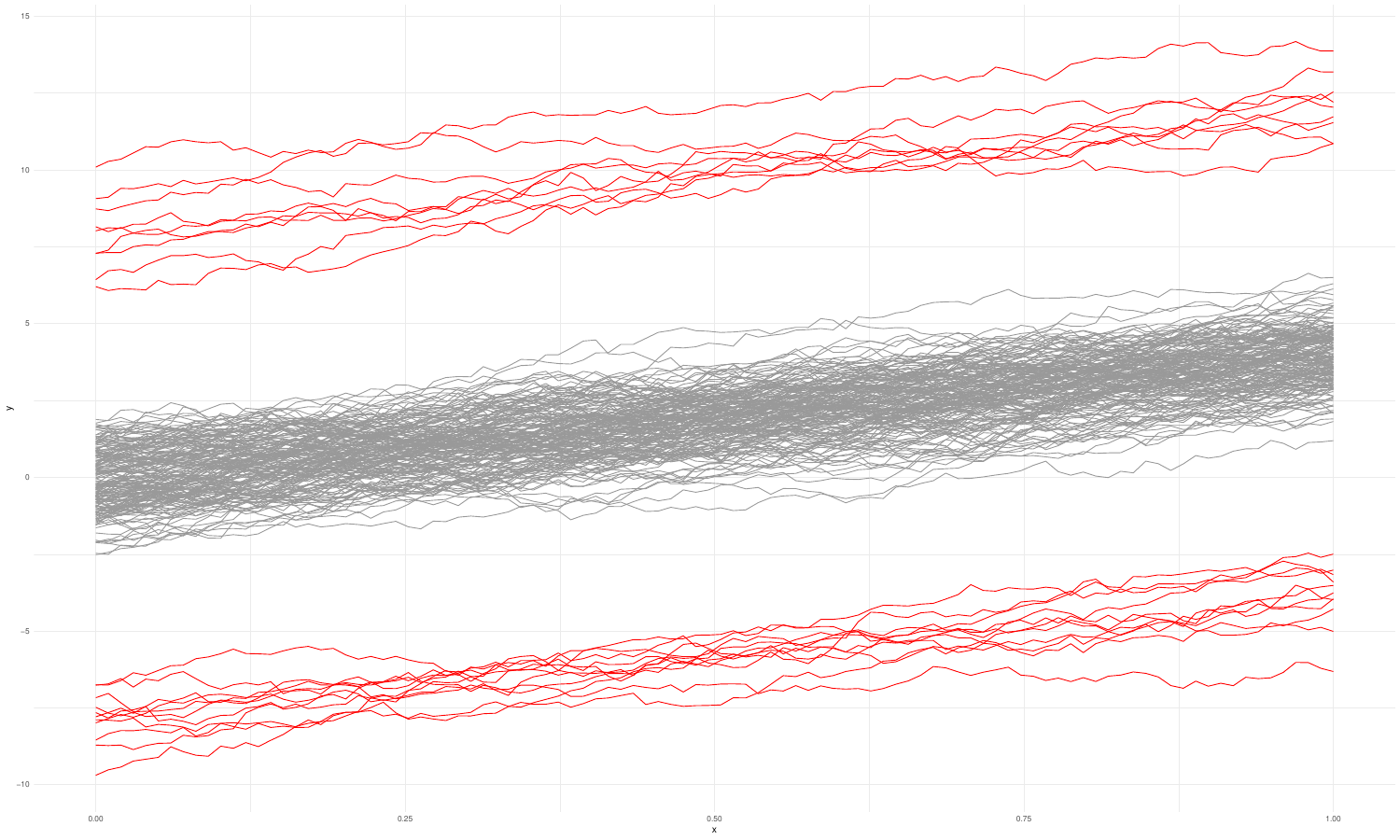}
    \caption{DGP1. A sample of 200 simulated curves with 10\% of outliers (red).  }
    \end{figure}

    \item[DGP 2.] \textit{Main model}: same as DGP1.
    
    \textit{Outlier model}: $X_i(t) = 4t + 8k_i I_{[T_i \leq t \leq T_i + 0.05]} + e_i(t)$ with $T_i \sim \mathcal{U}(0.1, 0.9)$ and $I$ an indicator function. 
    
    \textit{Type of outliers:} \textbf{magnitude outliers} for a subinterval of the domain, which produce spikes along the domain.

    \begin{figure}[htbp]
    \centering
    \includegraphics[width=.5\textwidth]{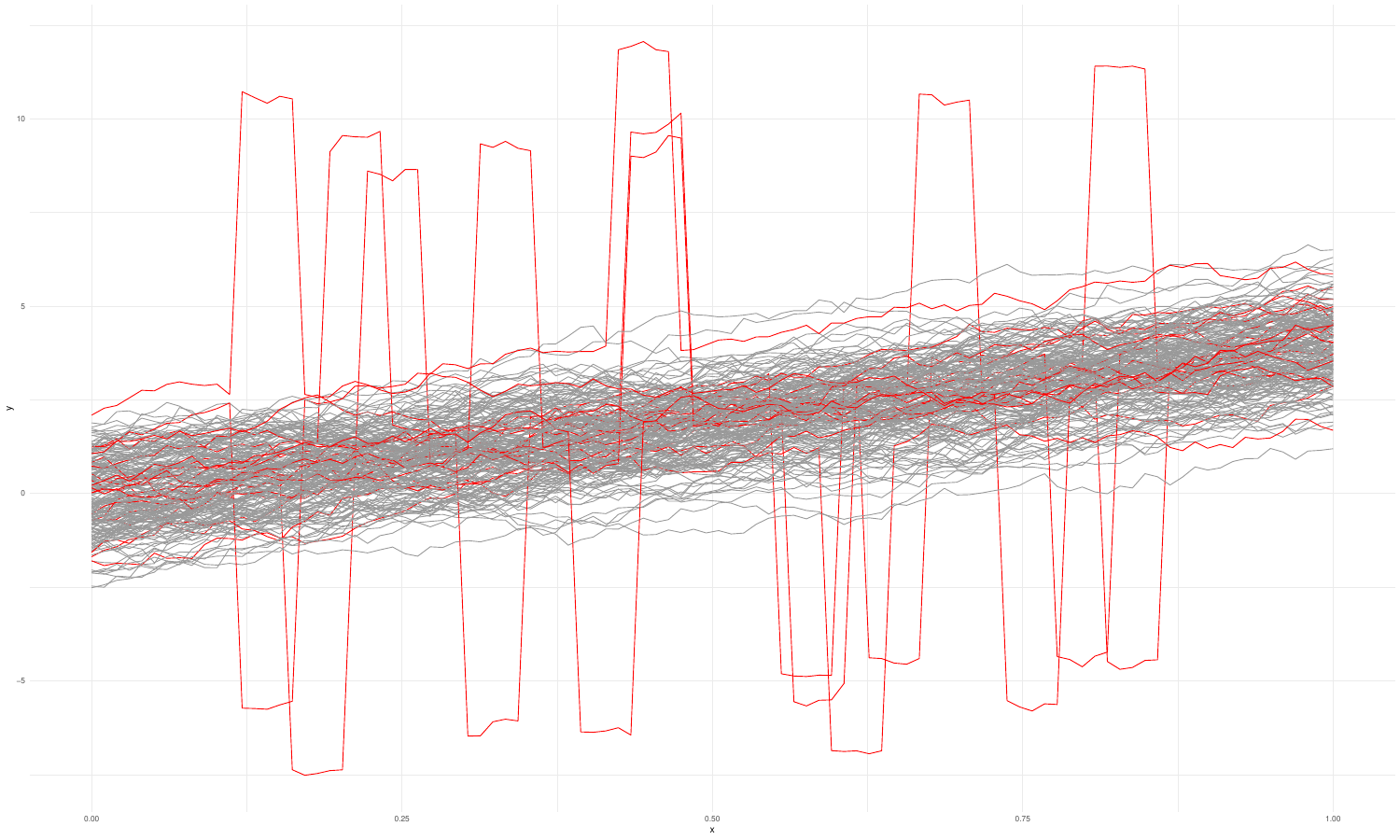}
    \caption{DGP2. A sample of 200 simulated curves with 10\% of outliers (red).  }
    \end{figure}

    \item[DGP 3.] \textit{Main model}: same as DGP1.
    
    \textit{Outlier model}: $X_i(t) = 4t + 8k_i I_{[T_i \leq t]} + e_i(t)$. 
    
    \textit{Type of outliers:} \textbf{magnitude outliers} for a subinterval of the domain.

    \begin{figure}[htbp]
    \centering
    \includegraphics[width=.5\textwidth]{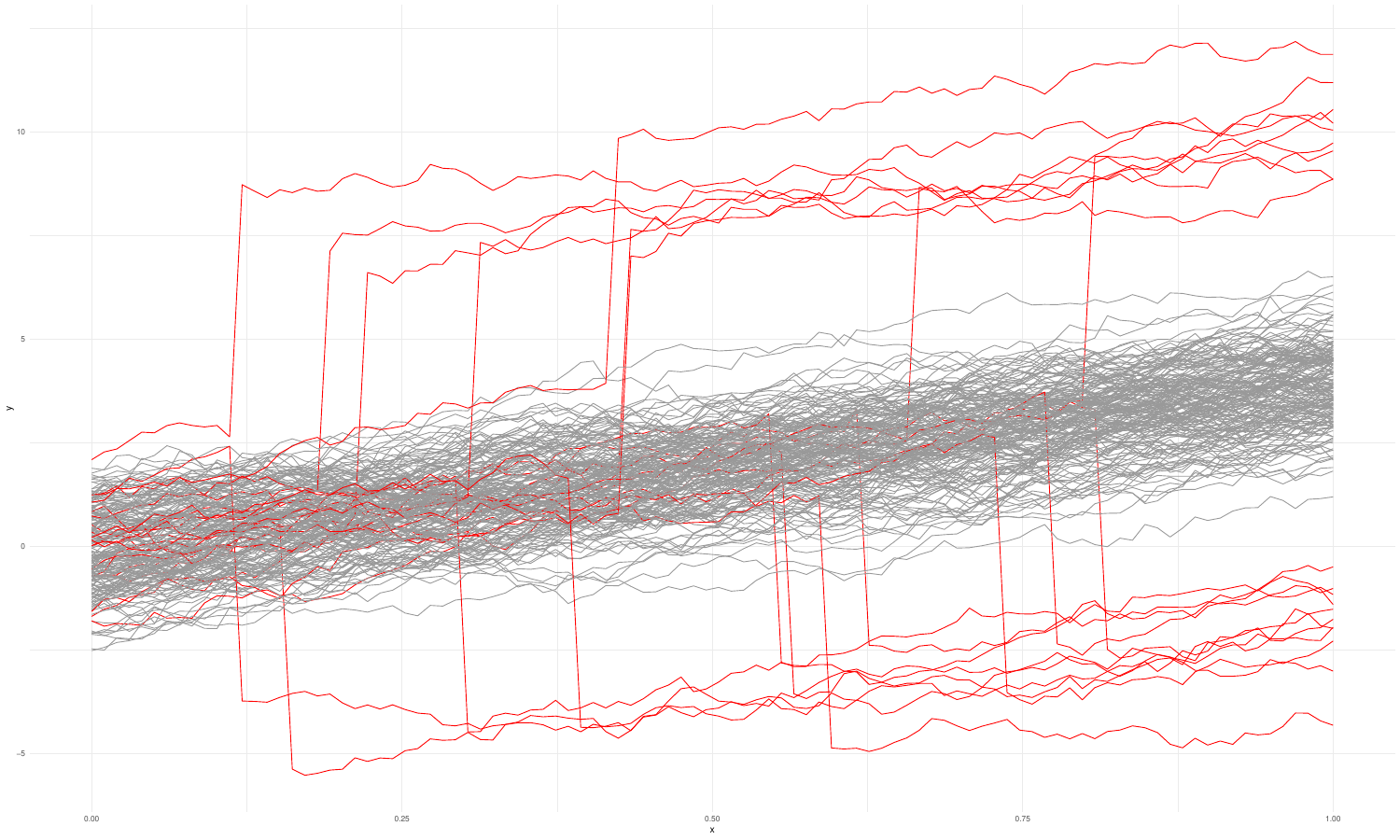}
    \caption{DGP3. A sample of 200 simulated curves with 10\% of outliers (red).  }
    \end{figure}

    \item[DGP 4.] \textit{Main model}: $X_i(t) = \mu t(1 - t)^{3/2} + \bar{e}_i(t)$, where $\bar{e}_i(t)$ is a Gaussian process with zero mean and covariance function $\bar{\gamma}(s, t) = 0.3 \exp\left(-|t - s|/0.3\right)$, where $s, t \in \mathcal{I}$.
    
    \textit{Outlier model}: $X_i(t) = 30 (1 - t) t^{3/2} + \bar{e}_i(t)$.  

    \textit{Type of outliers:} \textbf{magnitude outliers} (in phase), similar to the main curves, but slightly shifted horizontally and reversed.

    \begin{figure}[htbp]
    \centering
    \includegraphics[width=.5\textwidth]{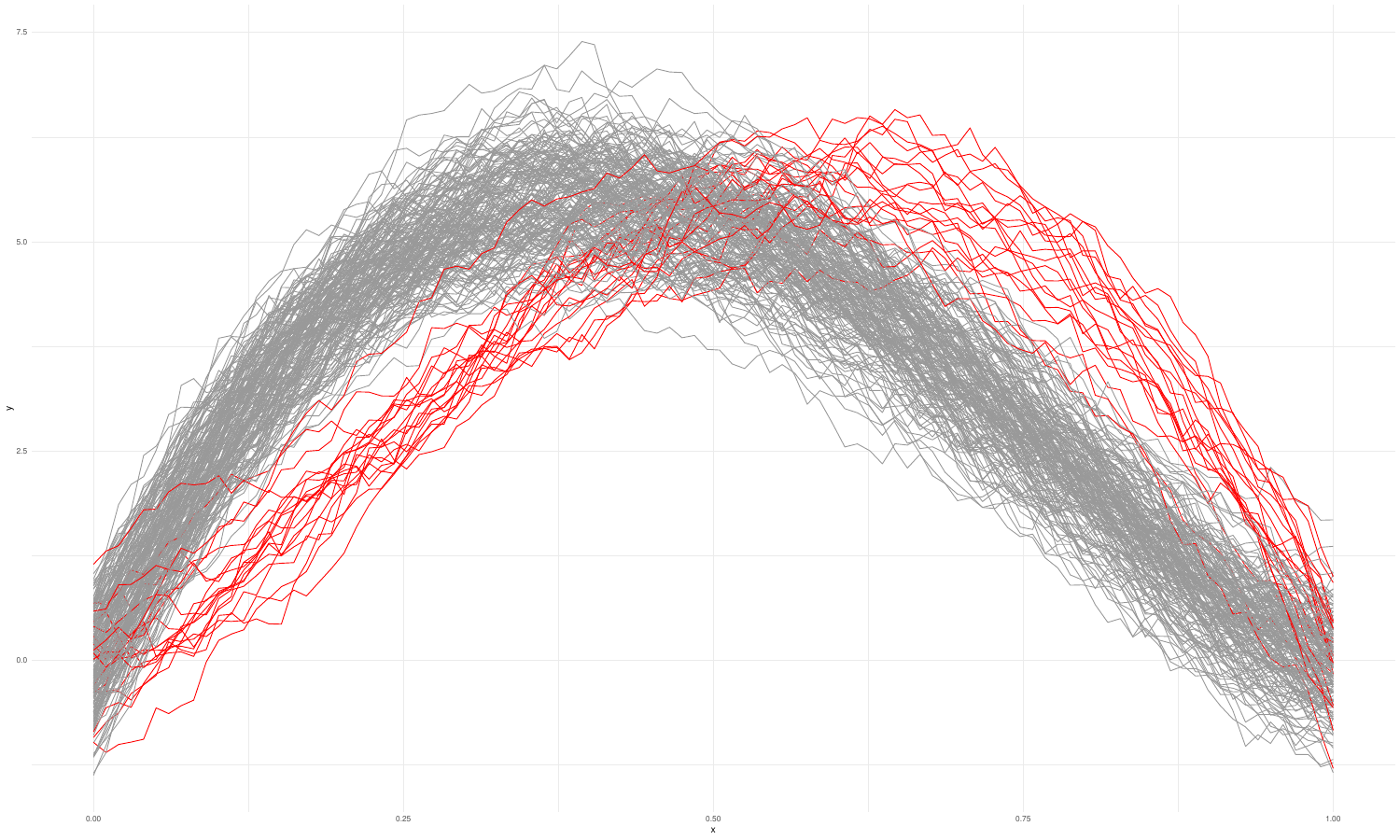}
    \caption{DGP4. A sample of 200 simulated curves with 10\% of outliers (red).  }
    \end{figure}

    \item[DGP 5.] \textit{Main model}: same as DGP1.
    
    \textit{Outlier model}: $X_i(t) = 4t + {e_2}_i(t)$. ${e_2}_i(t)$ is a Gaussian process with zero mean and covariance function $\gamma_2(s, t) = 5 \exp\left(-2|t - s|^{0.5}\right)$, where \\ $s, t \in \mathcal{I}$. 
    
    \textit{Type of outliers:}  \textbf{shape outliers} with a different covariance structure.

    \begin{figure}[htbp]
    \centering
    \includegraphics[width=.5\textwidth]{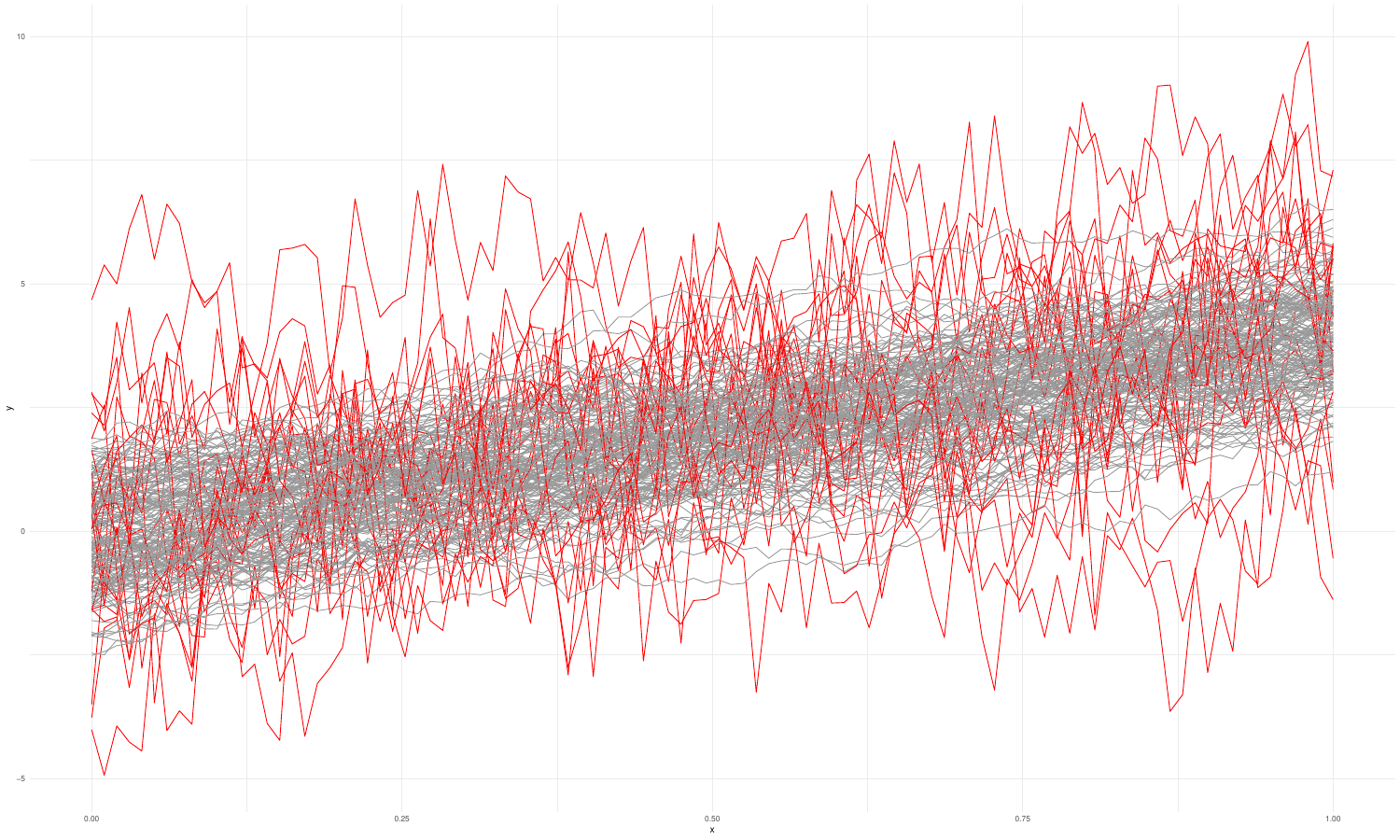}
    \caption{DGP5. A sample of 200 simulated curves with 10\% of outliers (red).  }
    \end{figure}

    \item[DGP 6.] \textit{Main model}: same as DGP1.
    
    \textit{Outlier model}:  $Xi(t) = 4 t + 1.8 (-1)^u + \\ (-1)^{1-u}\left(\frac{1}{\sqrt{0.02\pi}}\right) \exp\left(-50(t - v)^2\right) + e_i(t)$,\\ where $u \sim Ber(0.5)$, and $v \sim \mathcal{U}(0.25, 0.75)$.
    
    \textit{Type of outliers:} \textbf{shape outliers} in a subinterval of the domain.

    \begin{figure}[htbp]
    \centering
    \includegraphics[width=.5\textwidth]{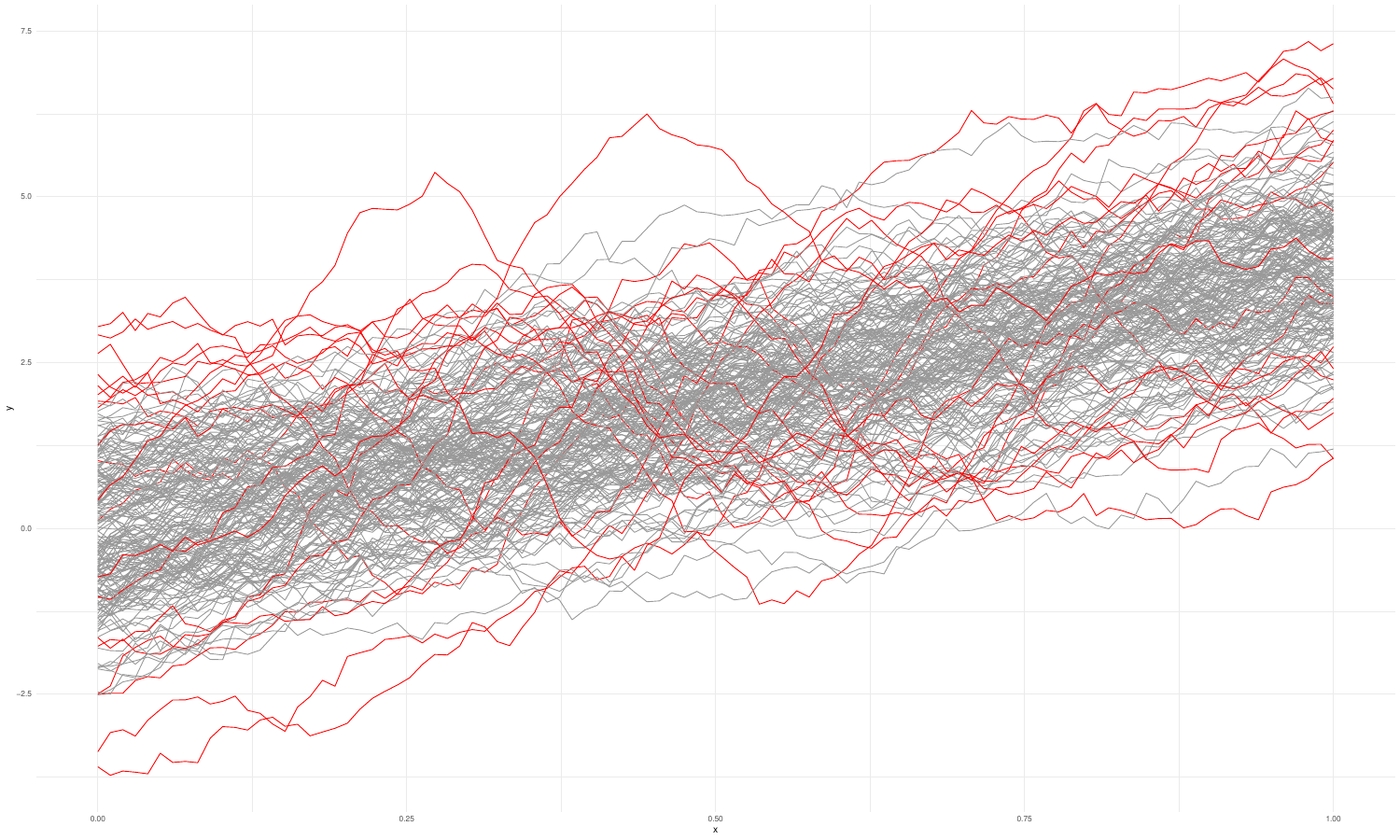}
    \caption{DGP6. A sample of 200 simulated curves with 10\% of outliers (red).  }
    \end{figure}

    \item[DGP 7.] \textit{Main model}: same as DGP1.
    
    \textit{Outlier model}:\newline  $X_i(t)=4t+2\sin\left(4\pi \left(t+\theta\right)\right)+e_i(t),$ with $\theta \sim \mathcal{U}(0.25,0.75)$.
    
    \textit{Type of outliers:} \textbf{shape outliers} that are periodic.

    \begin{figure}[htbp]
    \centering
    \includegraphics[width=.5\textwidth]{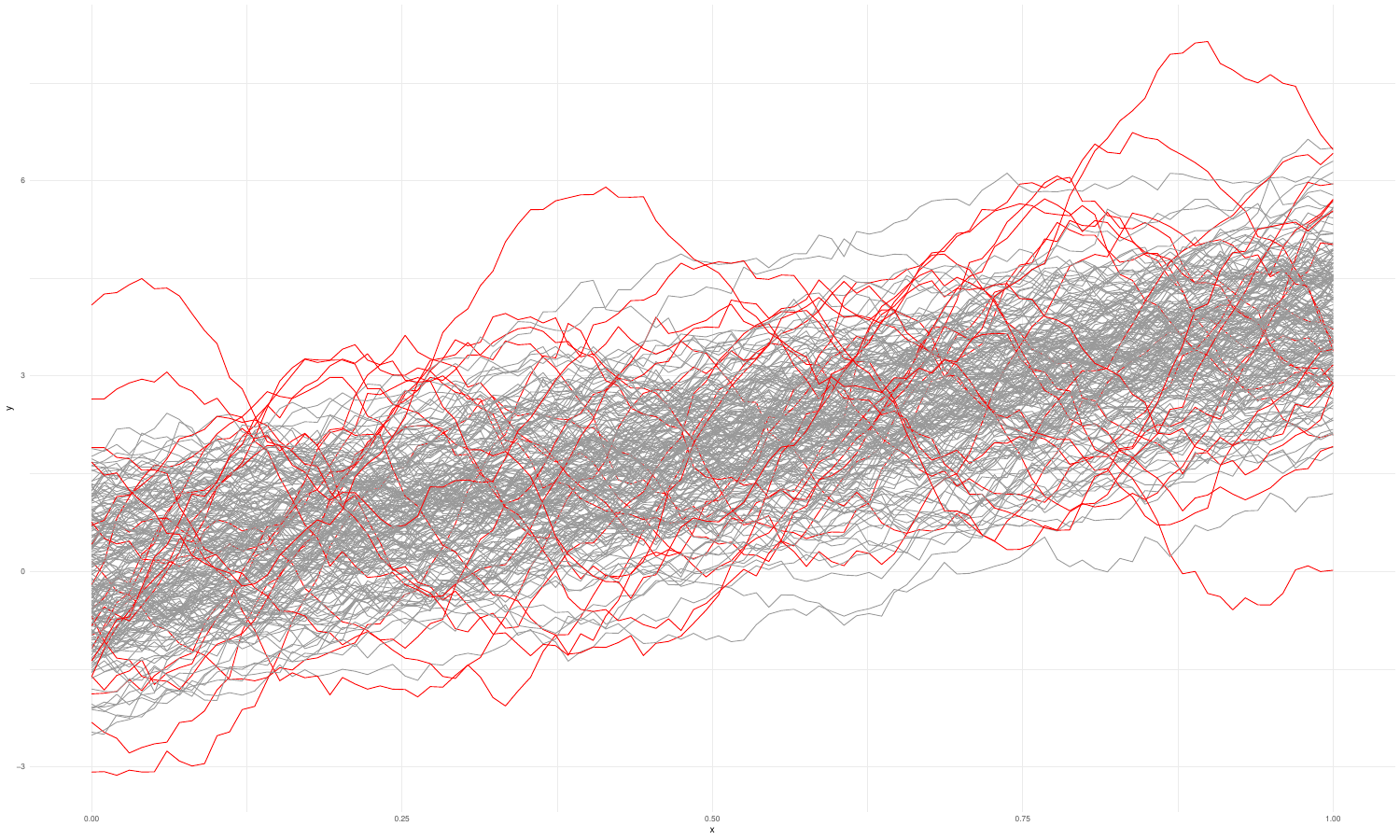}
    \caption{DGP7. A sample of 200 simulated curves with 10\% of outliers (red).  }
    \end{figure}

    \item[DGP 8.] \textit{Main model}: $X_i(t)=2\sin\left(15\pi t\right) + e_i(t).$ 
    
    \textit{Outlier model}: $X_i(t)=2\sin\left(15\pi t + 2\right) + e_i(t).$
    
    \textit{Type of outliers:} \textbf{shape outliers} that are periodic.

    \begin{figure}[htbp]
    \centering
    \includegraphics[width=.5\textwidth]{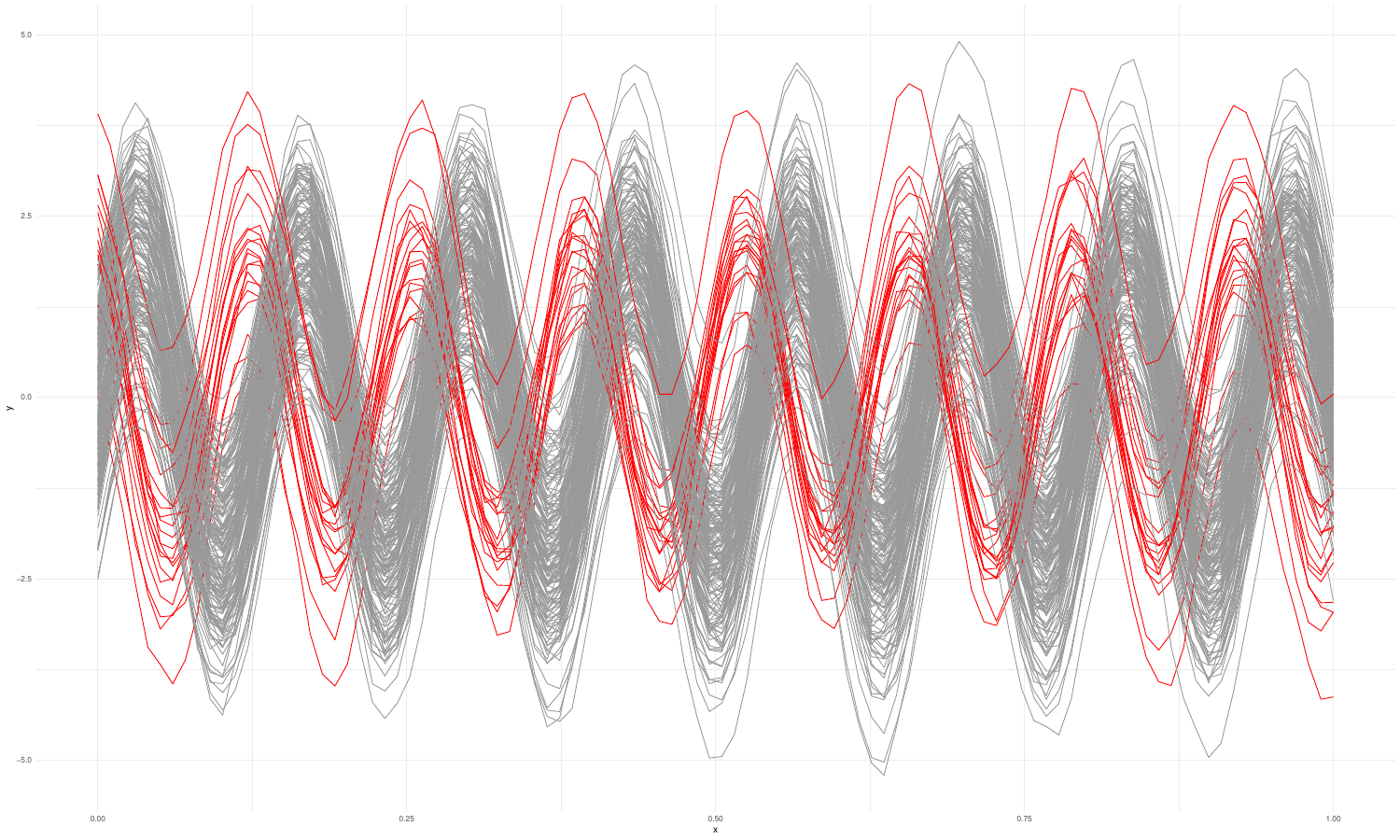}
    \caption{DGP8. A sample of 200 simulated curves with 10\% of outliers (red).  }
    \end{figure}

    \item[DGP 9.] \textit{Main model}: $X_i(t)= a_{1i}\sin\theta+a_{2i}\cos\theta+e_i(t).$ 
    
    \textit{Outlier model}: \\ $X_i(t)=\left(b_{1i}\sin\theta+b_{2i}\cos\theta\right)\left(1-u_i\right)+\\ \left(c_{1i}\sin\theta+c_{2i}\cos\theta\right)u_i+e_i(t).$ $t \in [0,1], \\ \theta \in [0,2\pi]$, $a_{1i}, a_{2i} \sim \mathcal{U}(3,8)$, \\ $b_{1i}, \ b_{2i}  \sim \mathcal{U}(1.5,2.5)$, $c_{1i}, \ c_{2i} \sim \mathcal{U}(9,10.5)$, \\ and $u_i\sim Ber(0.5)$.
    
    \textit{Type of outliers:} \textbf{magnitude outliers}.

    \begin{figure}[htbp]
    \centering
    \includegraphics[width=.5\textwidth]{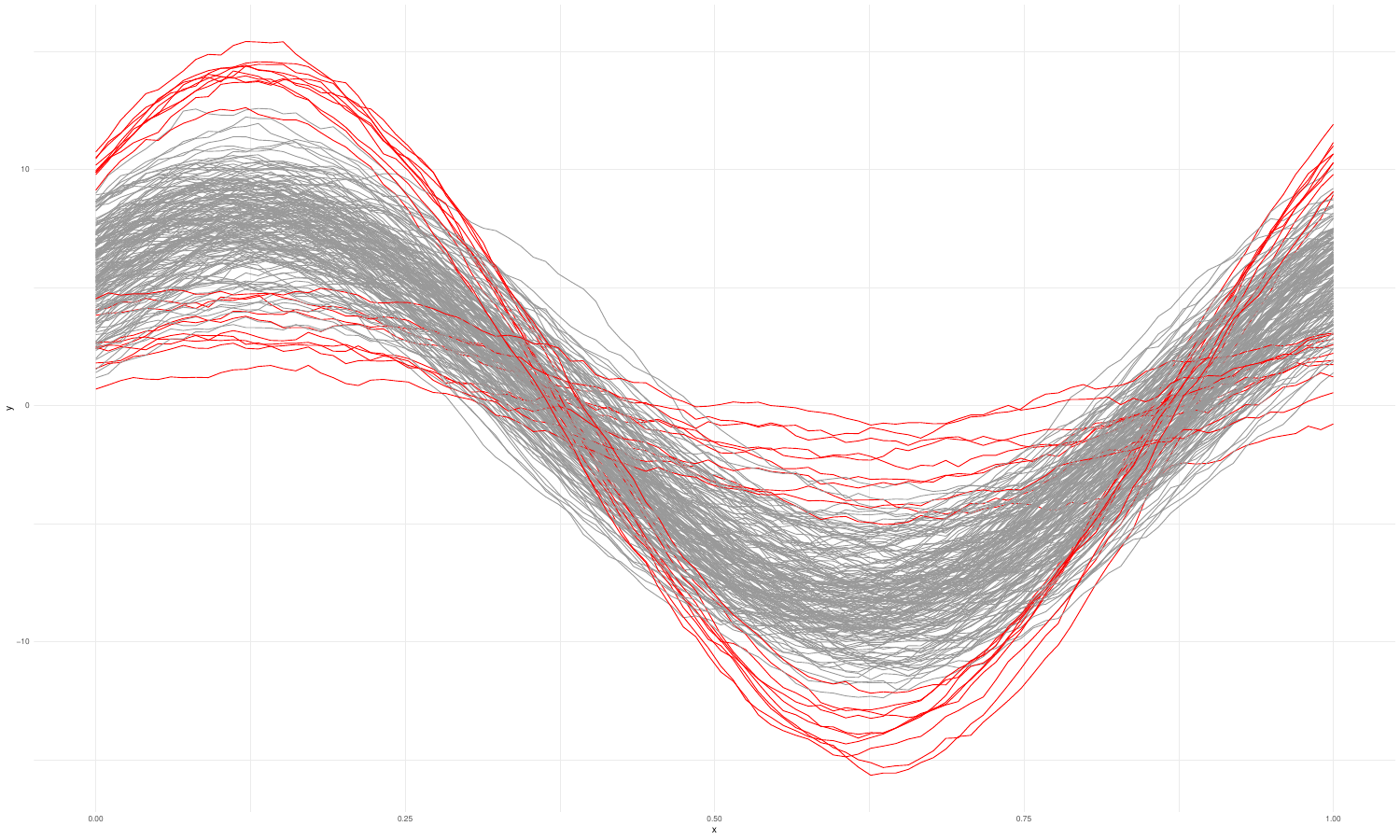}
    \caption{DGP9. A sample of 200 simulated curves with 10\% of outliers (red).  }
    \end{figure}

    \item[DGP 10.] \textit{Main model}: same as DGP1.
    
    \textit{Outlier model}: same as DGPs 1, 2, 5, 7 (with the same probability). 
    
    \textit{Type of outliers:} both \textbf{magnitude and shape outliers}. 

    \begin{figure}[htbp]
    \centering
    \includegraphics[width=.5\textwidth]{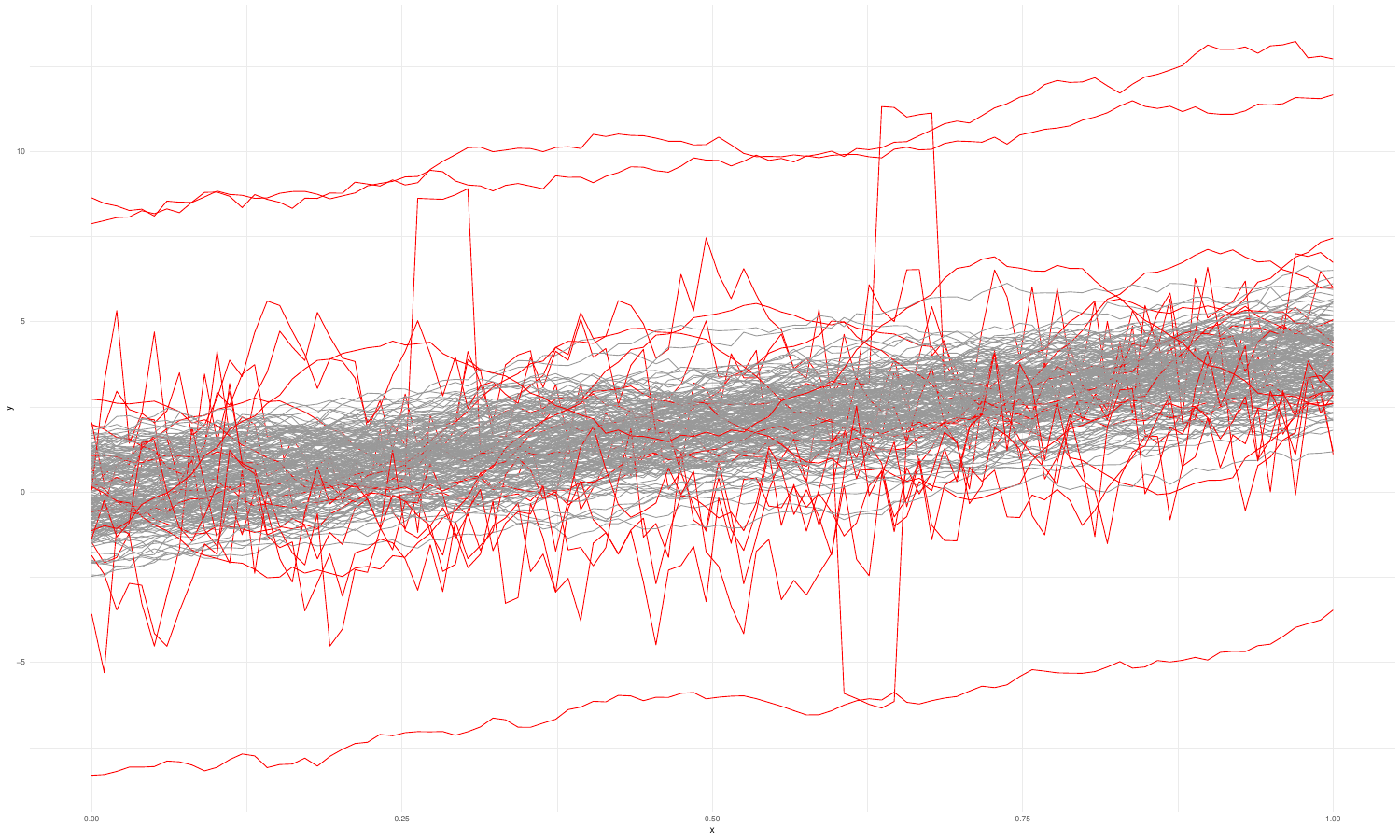}
    \caption{DGP10. A sample of 200 simulated curves with 10\% of outliers (red).  }
    \end{figure}

    \item[DGP 11.] \textit{Main model}: $X_i(t)=a_i+0.05t+\sin \left(\pi x_t^2\right)$.
    
    \textit{Outlier model}: $X_i(t)=b_i+0.05t+\cos \left(20 x_t\right)$ where $x_t=\frac{t}{500}\in [0,1], \ t = 1,\ldots, 500, \\ \ a_i \sim \mathcal{N}(\mu_a = 5, \sigma_a = 4), \ b_i \sim \mathcal{N}(\mu_b = 5, \sigma_b = 3).$ 
    
    \textit{Type of outliers:} \textbf{shape outliers}.

    \begin{figure}[htbp]
    \centering
    \includegraphics[width=.5\textwidth]{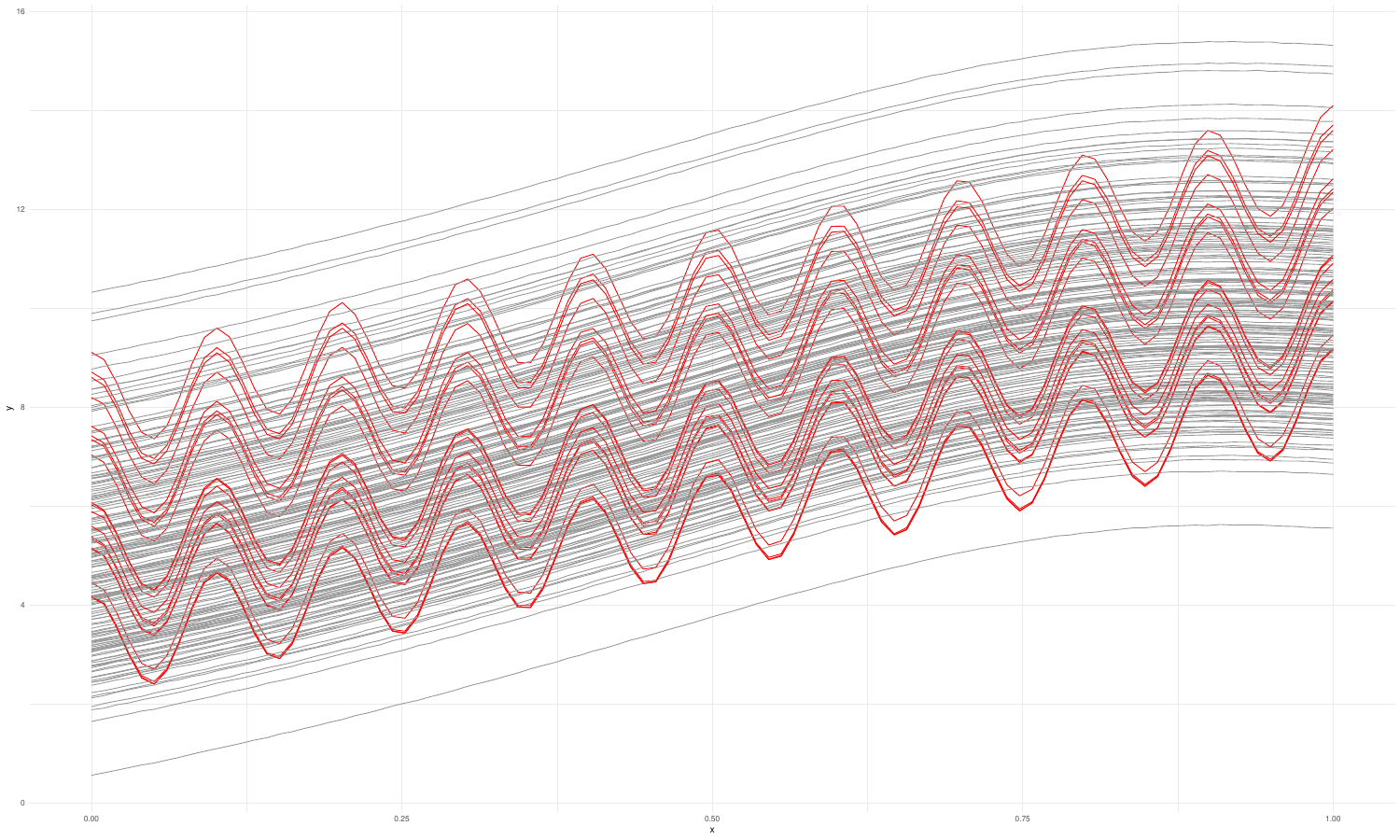}
    \caption{DGP11. A sample of 200 simulated curves with 10\% of outliers (red).  }
    \end{figure}

    \item[DGP 12.] \textit{Main model}: $X_i(t)=e_i(t)$. 
    
    The outlying curves are generates as \\ $X_i(t)=\Tilde{e}_i(t),$ where $\Tilde{e}_i(t)$ is a different \\ zero-mean Gaussian process with covariance \\ function $\Tilde{c}(s,t)=6\exp\{-|s-t|^{0.1}\}$.
    
    \textit{Type of outliers:} \textbf{shape outliers}.

    \begin{figure}[htbp]
    \centering
    \includegraphics[width=.5\textwidth]{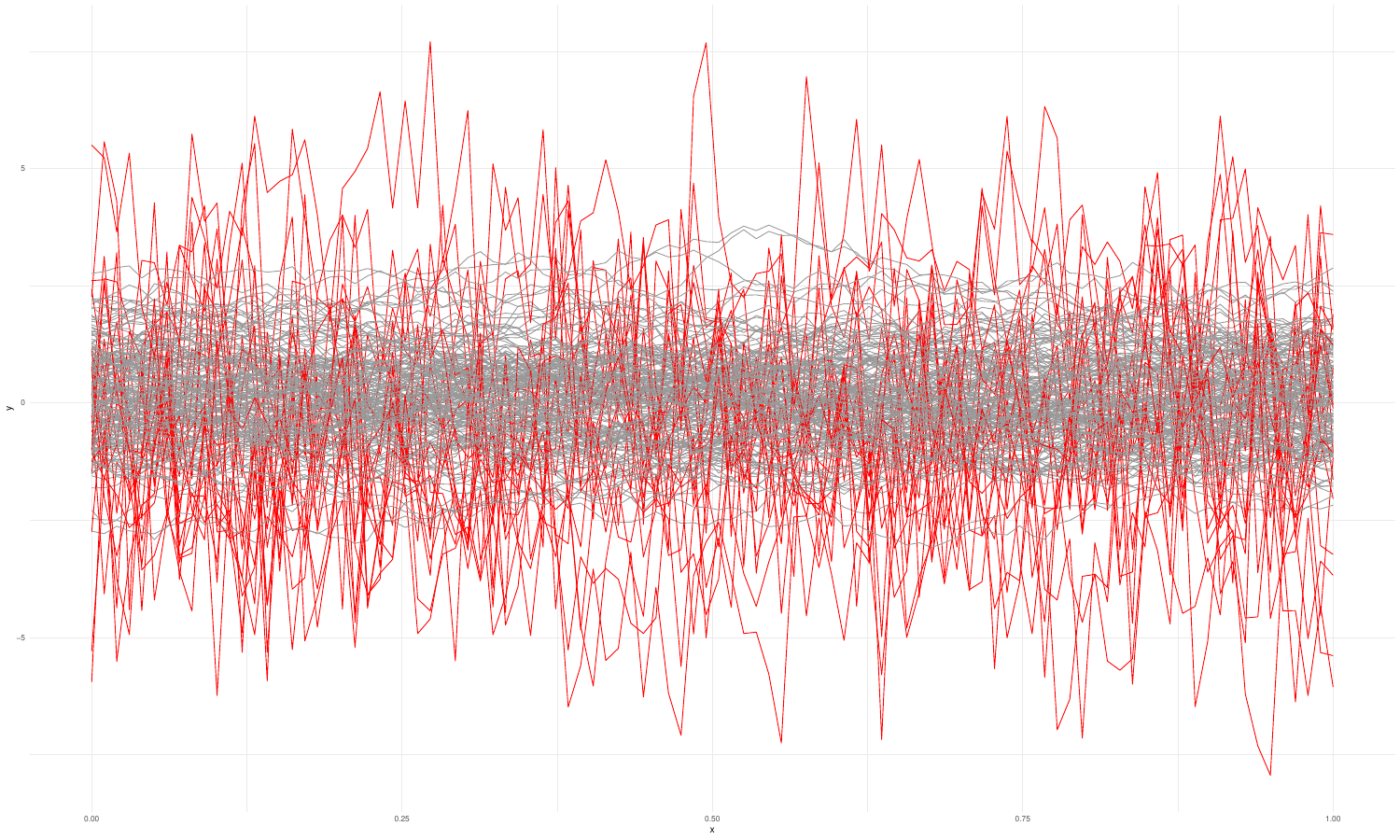}
    \caption{DGP12. A sample of 200 simulated curves with 10\% of outliers (red).  }
    \end{figure}
    
    \item[DGP 13.] \textit{Main model}: $X_i(t)=2\sin(15\pi t)+e_i(t)$.
    
    \textit{Outlier model}: $X_i(t)=2\sin(15\pi t +4)+e_i(t)$. 
    
    \textit{Type of outliers:} \textbf{shape outliers} in phase.

    \begin{figure}[htbp]
    \centering
    \includegraphics[width=.5\textwidth]{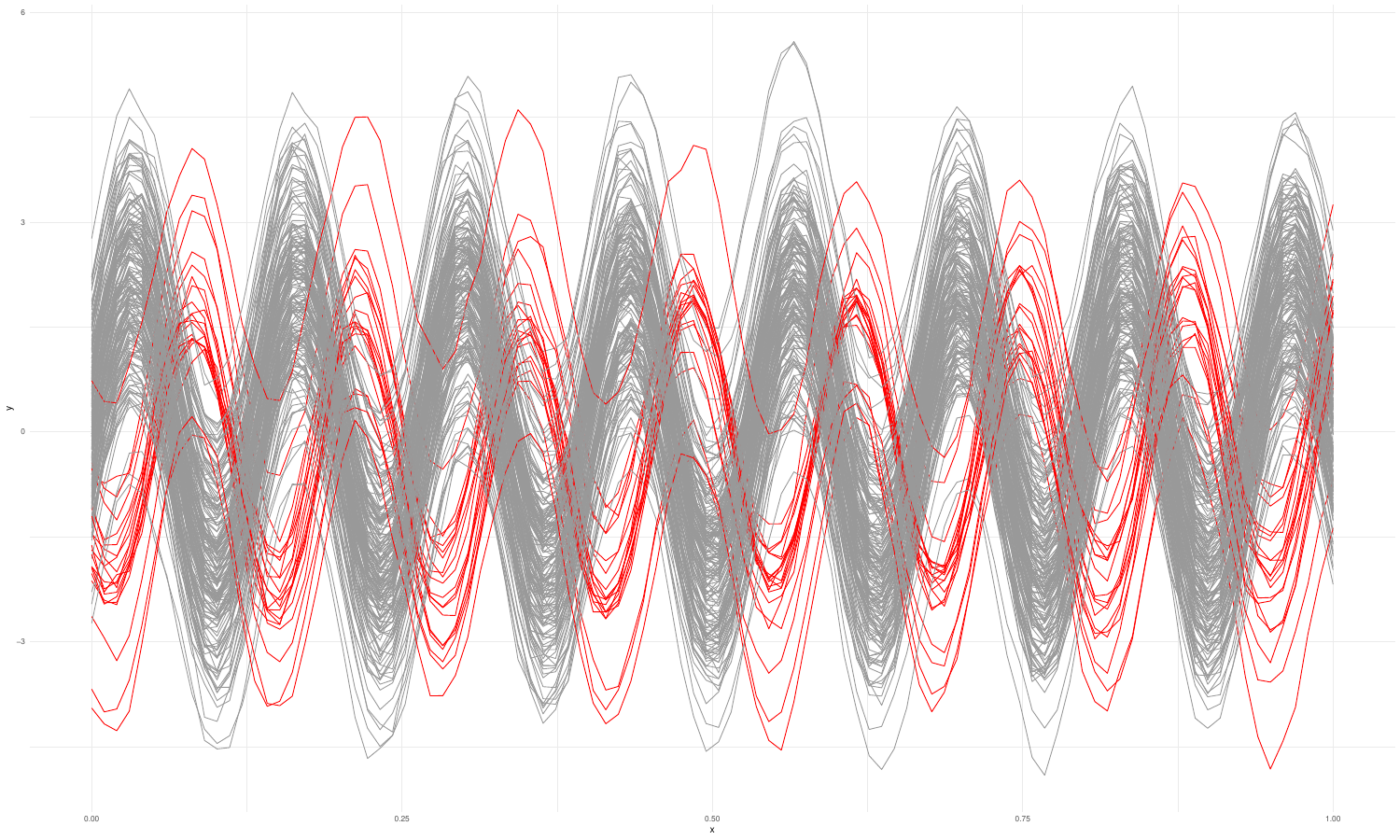}
    \caption{DGP13. A sample of 200 simulated curves with 10\% of outliers (red).  }
    \end{figure}

    \item[DGP 14.] \textit{Main model}: $X_i(t)=0.1+\arctan(t)+e_i(t)$.
    
    \textit{Outlier model}: $X_i(t)=0.1+\arctan(t)+e^*_i(t),$ where $e^*_i$ is another zero-mean Gaussian process with covariance function \\ $c^*(s,t)=0.1\exp\{-\frac{|s-t|^{0.1}}{4}\}.$ 
    
    \textit{Type of outliers:} both \textbf{shape and amplitude outliers}.

    \begin{figure}[htbp]
    \centering
    \includegraphics[width=.5\textwidth]{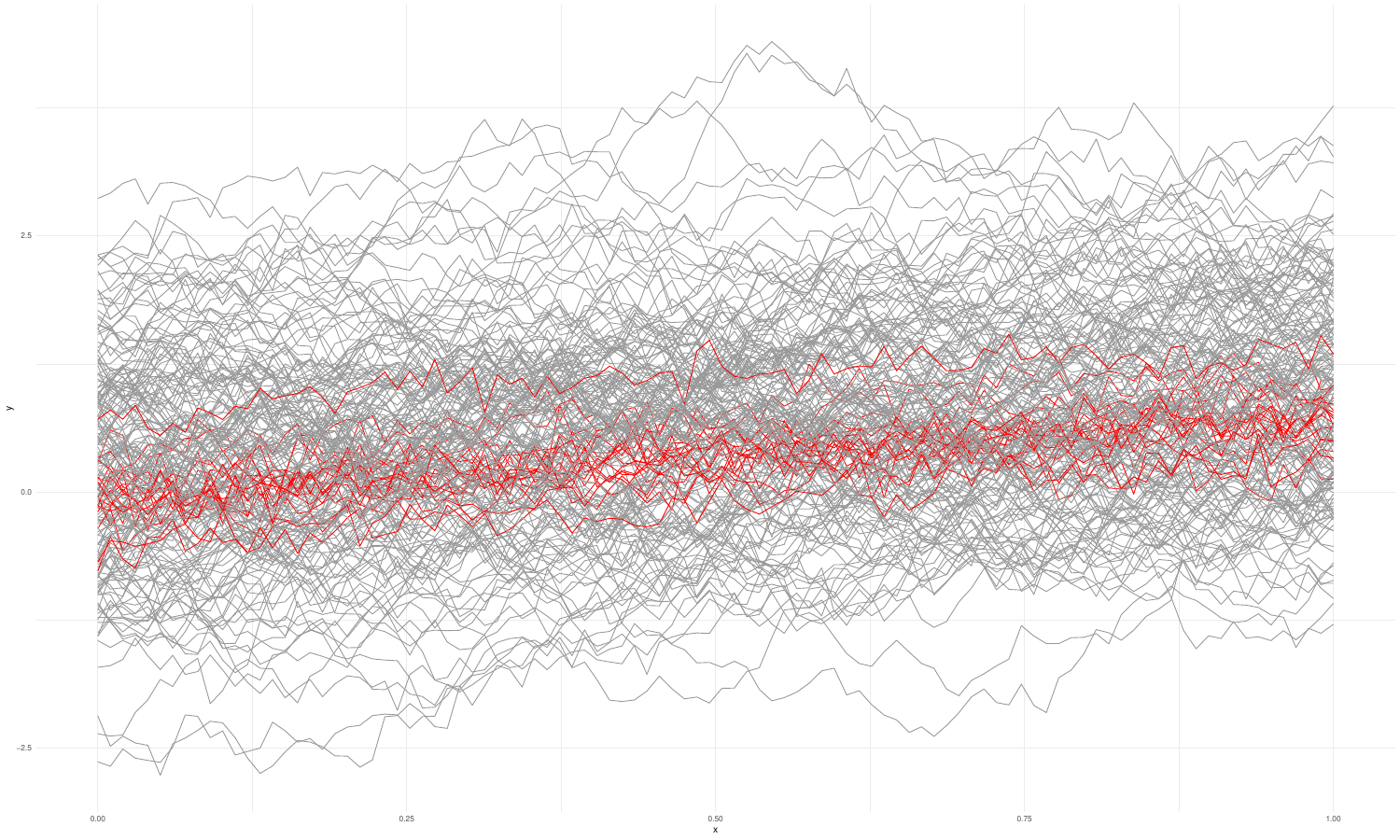}
    \caption{DGP14. A sample of 200 simulated curves with 10\% of outliers (red).  }
    \end{figure}

    \item[DGP 15.] \textit{Main model}: $X_i(t)=30t(1-t)^{3/2}+e_i(t)$.
    
    \textit{Outlier model}: $X_i(t)=30t(1-t)^{3/2}+e^*_i(t)$.
    
    \textit{Type of outliers:} both \textbf{shape and amplitude outliers}.

    \begin{figure}[htbp]
    \centering
    \includegraphics[width=.5\textwidth]{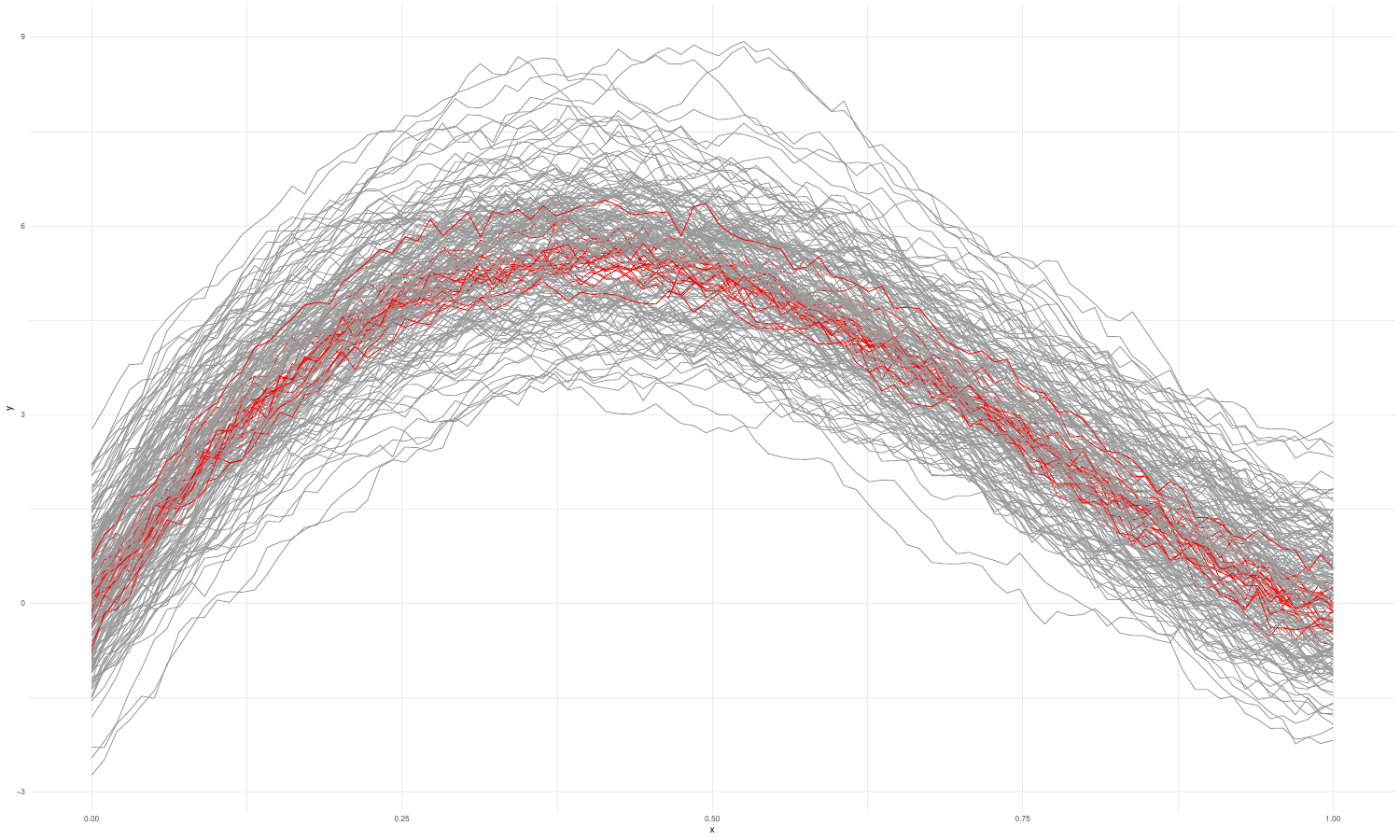}
    \caption{DGP15. A sample of 200 simulated curves with 10\% of outliers (red).  }
    \end{figure}
    
    \item[DGP 16.] \textit{Main model}: $X_i(t)=e_i(t)$.
    
    \textit{Outlier model}: $X_i(t)=0.1\sin(40(t+\theta)\pi)+e^*_i(t),$ where $\theta \sim \mathcal{U}(0.25,0.5).$ 
    
    \textit{Type of outliers:} both \textbf{shape and amplitude outliers}.

    \begin{figure}[htbp]
    \centering
    \includegraphics[width=.5\textwidth]{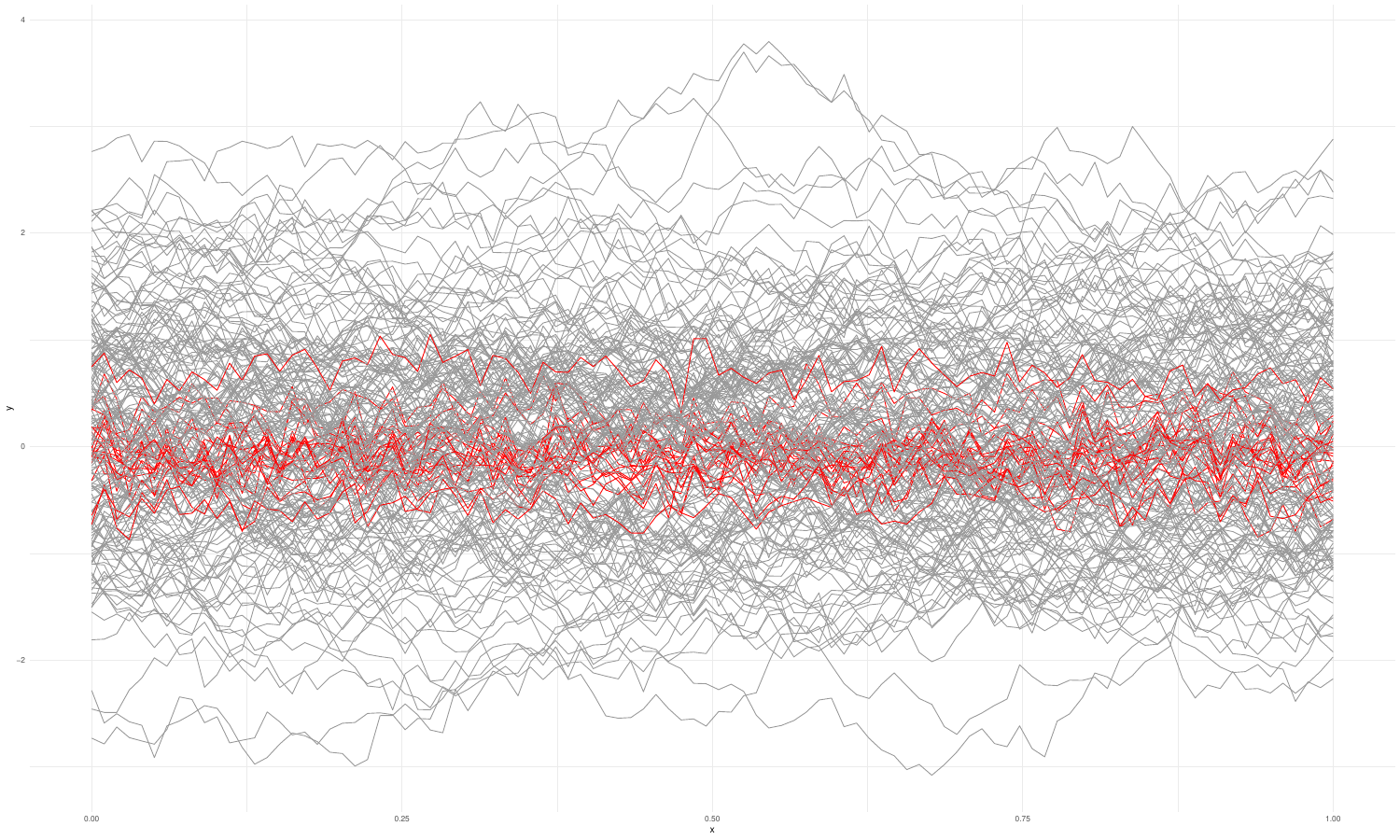}
    \caption{DGP16. A sample of 200 simulated curves with 10\% of outliers (red).  }
    \end{figure}

    \item[DGP 17.] \textit{Main model}: $X_i(t) = A+B \arctan(t)+G(t),$ where $A \sim \mathcal{N}(\mu=0,\,\sigma=2)$, $B \sim Exp(1)$, and $G:[0,1] \to \mathbb{R}$ is an independent Gaussian process with $Cov(G(s),G(t))=0.2 exp(\frac{-|s-t|}{0.3})$ for $s,t \in \mathcal{I}.$ 
    
    \textit{Outlier model}: $X_i(t)=1- 2 \arctan(t)+\Tilde{G}(t),$ where $\Tilde{G}$ is defined in the same way as $G$, but it is independent of it. 
    
    \textit{Type of outliers:} \textbf{shape outliers}, where the outlier curve tends to increase when the inliers decrease.

    \begin{figure}[htbp]
    \centering
    \includegraphics[width=.5\textwidth]{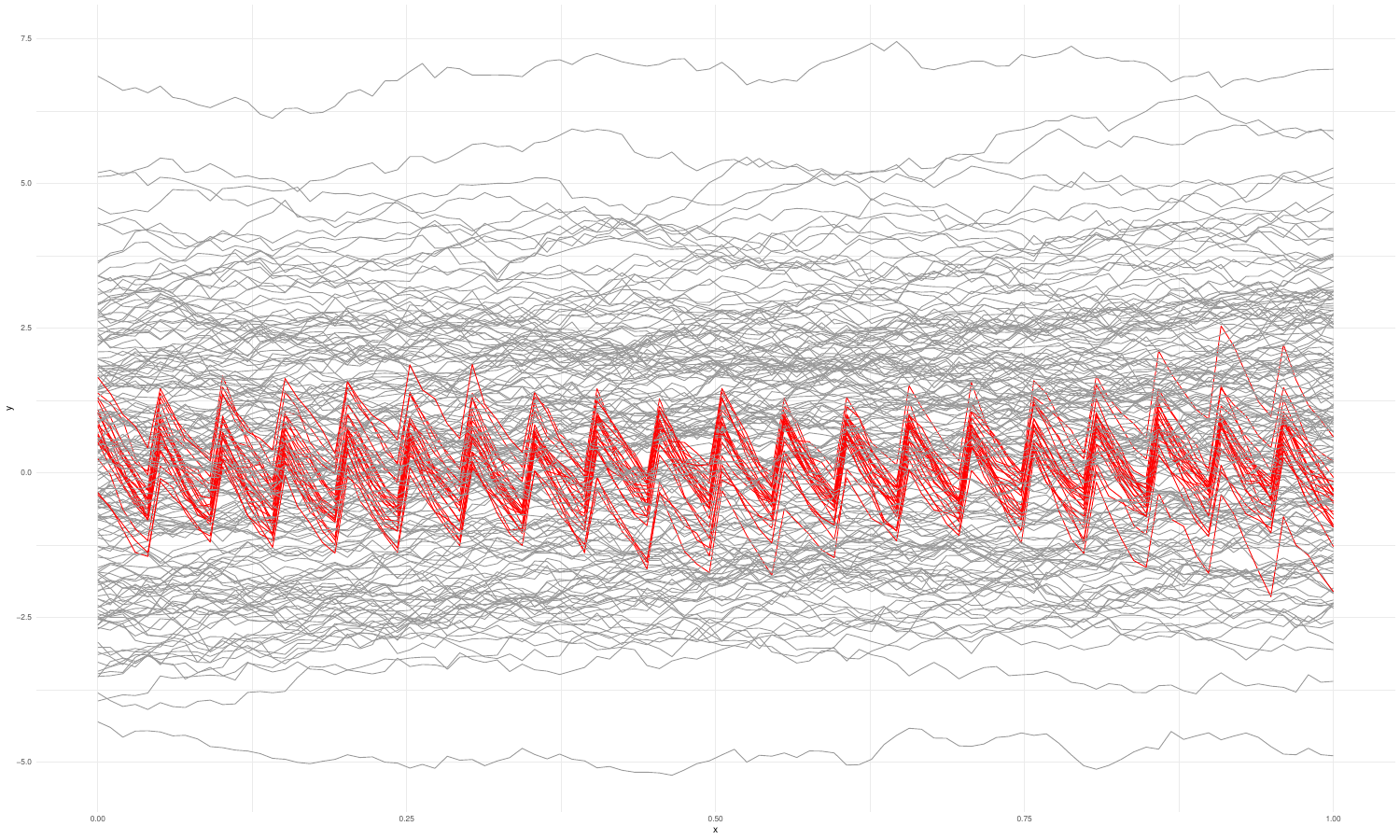}
    \caption{DGP17. A sample of 200 simulated curves with 10\% of outliers (red).  }
    \end{figure}

    \item[DGP 18.] \textit{Main model}: same as DGP 12. 
    
    \textit{Outlier model}: $X_i(t)=y_i(t)+ \Tilde{G}(t),$ where $$y_i(t) = \left\{ 
    \begin{array}{lll}
		 0.5+\log(2)\arctan(t)  & for & t \in [0,0.5], \\
		 -0.5+\log(2)\arctan(t) & for & t \in \left(0.5,1].\right.
	       \end{array}
        \right.$$ 
        
        \textit{Type of outliers:} \textbf{shape outliers}.

    \begin{figure}[htbp]
    \centering
    \includegraphics[width=.5\textwidth]{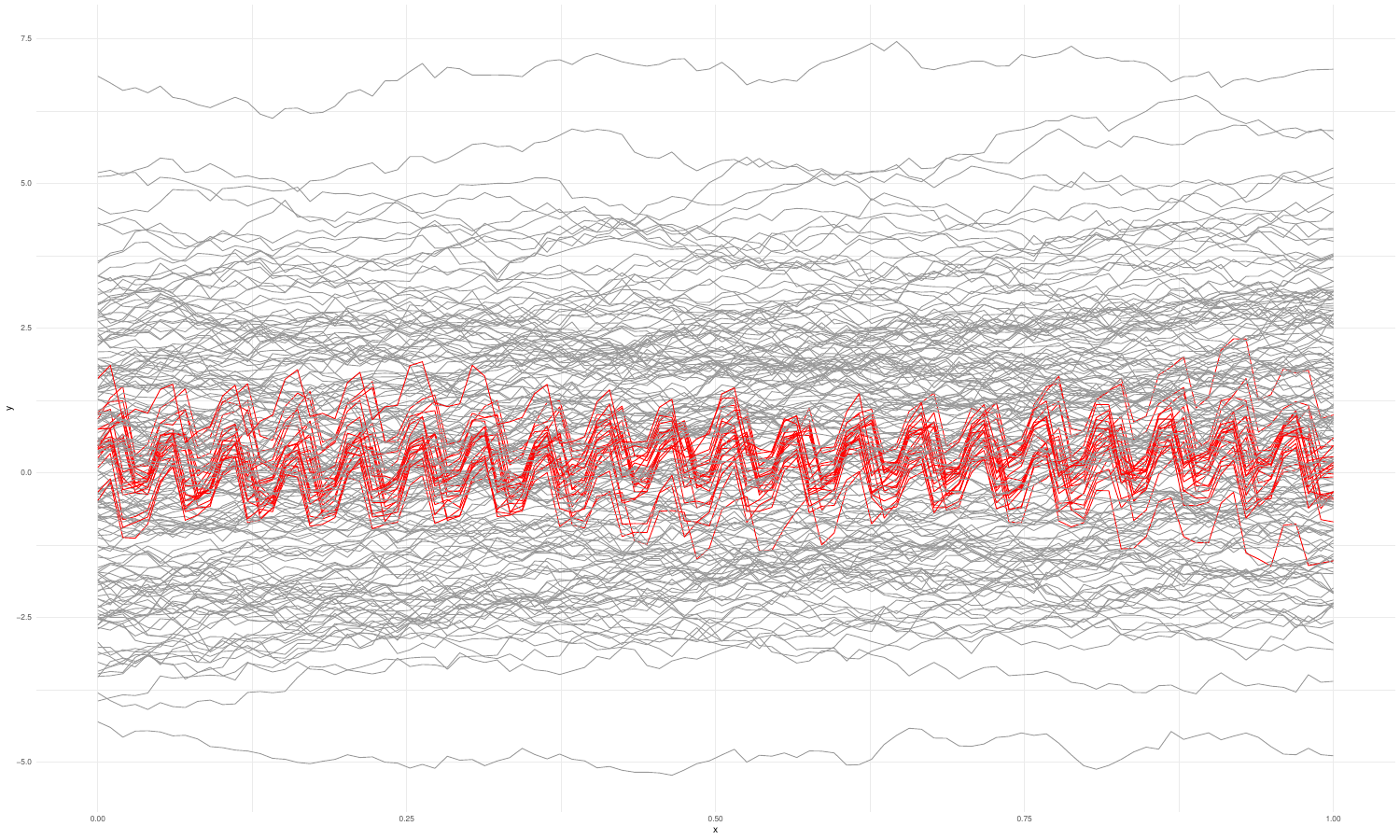}
    \caption{DGP18. A sample of 200 simulated curves with 10\% of outliers (red).  }
    \end{figure}

    \item[DGP 19.] \textit{Main model}: These curves are generated by first creating a B-splines basis with 25 basis functions. Coefficients for these curves are generated from a unit sphere by drawing from a standard normal distribution and normalizing them to lie on the unit sphere in the plane defined by the first two dimensions. These coefficients are then rotated into the basis-dimensional space using a random orthonormal rotation matrix obtained via QR decomposition.
    
    \textit{Outlier model}: it is similar to the main one. However, an additional step modifies the rotation matrix. Specifically, a new random orthonormal matrix is generated, scaled by a factor of $1.1$, and combined with the original rotation matrix using matrix multiplication and QR decomposition to ensure orthonormality. This combined matrix is then used to rotate the outlier coefficients. Finally, white noise with $\sigma = 0.3$ is added to the resulting curves. 
    
    \textit{Type of outliers:} \textbf{shape outliers}.

    \begin{figure}[htbp]
    \centering
    \includegraphics[width=.5\textwidth]{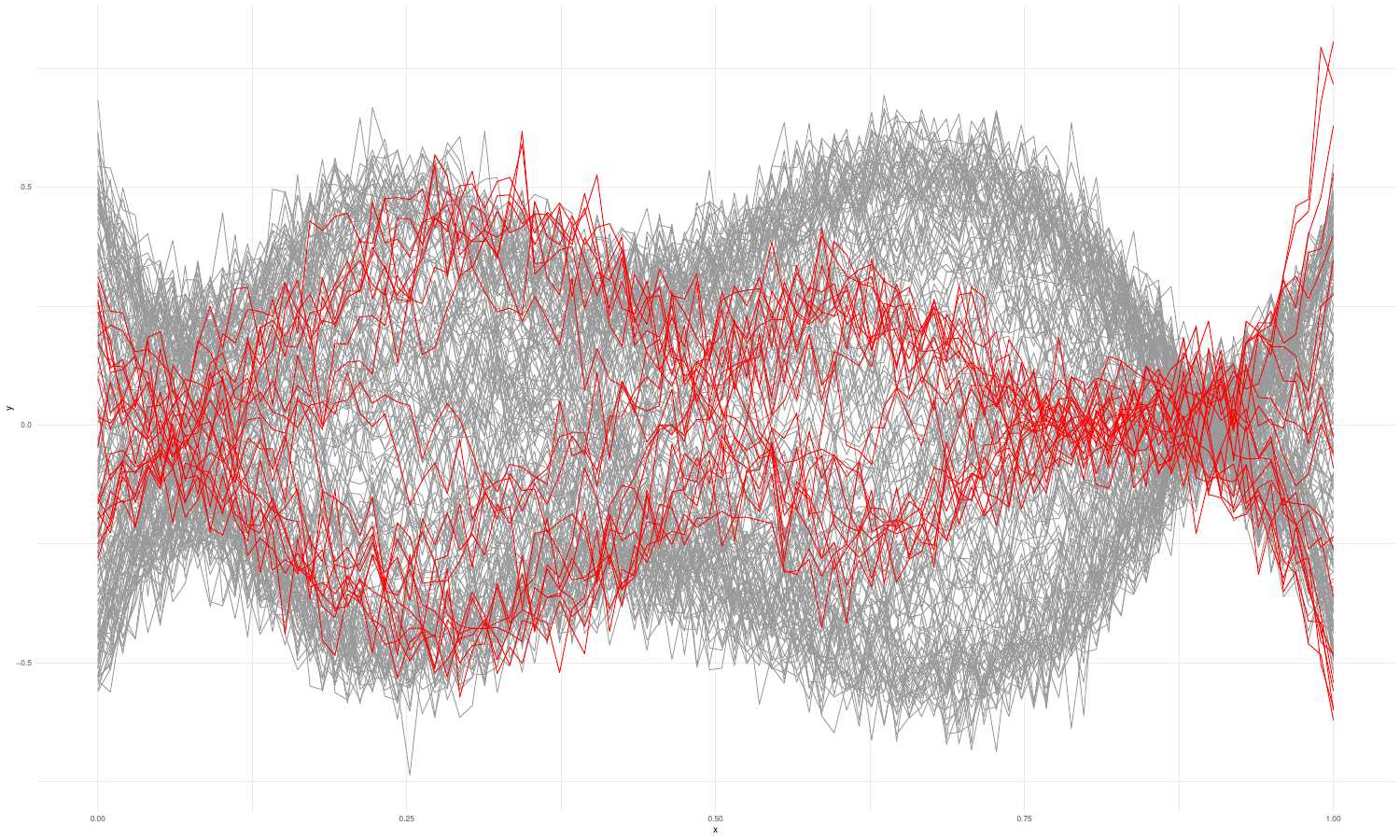}
    \caption{DGP19. A sample of 200 simulated curves with 10\% of outliers (red).  }
    \end{figure}
\end{itemize}

\FloatBarrier

{\onecolumn

\section{TPR, FPR and  for different $\alpha$ values and DGPs} \label{appC}

This section provides TPR, FPR and AUC for DGP1, \ldots, DGP19 for  $\alpha = 0.01, 0.05, \text{ and } 0.1$. Note that methods that do not threshold a single continuous "outlyingness" score to flag outliers (FASTMUOD, MSPLOT, TVD, and BP-PWD) appear as ``-'' in the AUC column.

\begin{table*}[htbp]

\caption{Results of outlier detection methodologies for DGP1 and different outliers proportions ($\alpha$).The values correspond to the median (sd) of TPR, FPR and AUC for 100 simulations.}
\centering
\scriptsize 
\begin{tabular}[t]{lccccccccc}
\toprule
\multicolumn{1}{c}{ } & \multicolumn{3}{c}{$\alpha = 0.01$} & \multicolumn{3}{c}{$\alpha = 0.05$} & \multicolumn{3}{c}{$\alpha = 0.1$} \\
\cmidrule(l{3pt}r{3pt}){2-4} \cmidrule(l{3pt}r{3pt}){5-7} \cmidrule(l{3pt}r{3pt}){8-10}
Method & TPR & FPR & AUC & TPR & FPR & AUC & TPR & FPR & AUC\\
\midrule
FASTMUOD & 1.000 (0.000) & 0.093 (0.019) & - & 1.000 (0.000) & 0.089 (0.019) & - & 1.000 (0.000) & 0.083 (0.018) & -\\
EHyOut & 1.000 (0.000) & 0.056 (0.021) & 1.000 (0.000) & 1.000 (0.000) & 0.042 (0.020) & 1.000 (0.000) & 1.000 (0.000) & 0.033 (0.017) & 1.000 (0.000)\\
OG & 1.000 (0.000) & 0.040 (0.016) & 1.000 (0.003) & 1.000 (0.000) & 0.042 (0.017) & 1.000 (0.002) & 1.000 (0.000) & 0.044 (0.016) & 0.998 (0.003)\\
MSPLOT & 1.000 (0.000) & 0.035 (0.016) & - & 1.000 (0.017) & 0.032 (0.015) & - & 1.000 (0.005) & 0.022 (0.014) & -\\
TVD & 1.000 (0.000) & 0.000 (0.000) & - & 1.000 (0.000) & 0.000 (0.000) & - & 1.000 (0.000) & 0.000 (0.000) & -\\
MBD & 1.000 (0.000) & 0.000 (0.000) & 1.000 (0.001) & 1.000 (0.000) & 0.000 (0.000) & 1.000 (0.001) & 1.000 (0.000) & 0.000 (0.000) & 1.000 (0.002)\\
MDS5LOF & 1.000 (0.000) & 0.091 (0.000) & 1.000 (0.000) & 1.000 (0.000) & 0.053 (0.000) & 1.000 (0.000) & 1.000 (0.000) & 0.000 (0.000) & 1.000 (0.000)\\
BP-PWD & 1.000 (0.000) & 0.045 (0.013) & - & 1.000 (0.000) & 0.047 (0.013) & - & 1.000 (0.000) & 0.044 (0.014) & -\\
\bottomrule
\end{tabular}
\end{table*}

\begin{table*}[htbp]

\caption{Results of outlier detection methodologies for DGP2 and different outliers proportions ($\alpha$). 
                    The values correspond to the median (sd) of TPR, FPR and AUC for 100 simulations.}
\centering
\scriptsize 
\begin{tabular}[t]{lccccccccc}
\toprule
\multicolumn{1}{c}{ } & \multicolumn{3}{c}{$\alpha = 0.01$} & \multicolumn{3}{c}{$\alpha = 0.05$} & \multicolumn{3}{c}{$\alpha = 0.1$} \\
\cmidrule(l{3pt}r{3pt}){2-4} \cmidrule(l{3pt}r{3pt}){5-7} \cmidrule(l{3pt}r{3pt}){8-10}
Method & TPR & FPR & AUC & TPR & FPR & AUC & TPR & FPR & AUC\\
\midrule
FASTMUOD & 1.000 (0.000) & 0.091 (0.017) & - & 1.000 (0.000) & 0.074 (0.017) & - & 1.000 (0.015) & 0.056 (0.017) & -\\
EHyOut & 1.000 (0.000) & 0.056 (0.021) & 1.000 (0.000) & 1.000 (0.000) & 0.047 (0.020) & 1.000 (0.000) & 1.000 (0.000) & 0.039 (0.019) & 1.000 (0.000)\\
OG & 0.500 (0.346) & 0.045 (0.016) & 0.797 (0.140) & 0.600 (0.154) & 0.037 (0.016) & 0.708 (0.088) & 0.500 (0.127) & 0.033 (0.015) & 0.715 (0.061)\\
MSPLOT & 1.000 (0.000) & 0.035 (0.016) & - & 1.000 (0.000) & 0.032 (0.014) & - & 1.000 (0.000) & 0.025 (0.013) & -\\
TVD & 1.000 (0.000) & 0.000 (0.001) & - & 1.000 (0.000) & 0.000 (0.001) & - & 1.000 (0.000) & 0.000 (0.000) & -\\
MBD & 0.500 (0.344) & 0.000 (0.001) & 0.631 (0.114) & 0.500 (0.163) & 0.000 (0.001) & 0.585 (0.055) & 0.500 (0.129) & 0.000 (0.001) & 0.569 (0.052)\\
MDS5LOF & 0.500 (0.362) & 0.096 (0.004) & 0.871 (0.086) & 0.300 (0.152) & 0.089 (0.008) & 0.752 (0.081) & 0.250 (0.088) & 0.083 (0.010) & 0.703 (0.065)\\
BP-PWD & 0.500 (0.346) & 0.045 (0.013) & - & 0.600 (0.161) & 0.042 (0.013) & - & 0.500 (0.131) & 0.039 (0.012) & -\\
\bottomrule
\end{tabular}
\end{table*}

\begin{table*}[htbp]
\caption{Results of outlier detection methodologies for DGP3 and different outliers proportions ($\alpha$). 
                    The values correspond to the median (sd) of TPR, FPR and AUC for 100 simulations.}
\centering
\scriptsize 
\begin{tabular}[t]{lccccccccc}
\toprule
\multicolumn{1}{c}{ } & \multicolumn{3}{c}{$\alpha = 0.01$} & \multicolumn{3}{c}{$\alpha = 0.05$} & \multicolumn{3}{c}{$\alpha = 0.1$} \\
\cmidrule(l{3pt}r{3pt}){2-4} \cmidrule(l{3pt}r{3pt}){5-7} \cmidrule(l{3pt}r{3pt}){8-10}
Method & TPR & FPR & AUC & TPR & FPR & AUC & TPR & FPR & AUC\\
\midrule
FASTMUOD & 1.000 (0.000) & 0.088 (0.018) & - & 1.000 (0.010) & 0.068 (0.017) & - & 1.000 (0.027) & 0.056 (0.016) & -\\
EHyOut & 1.000 (0.000) & 0.056 (0.021) & 1.000 (0.001) & 1.000 (0.000) & 0.047 (0.020) & 1.000 (0.000) & 1.000 (0.000) & 0.039 (0.020) & 1.000 (0.000)\\
OG & 1.000 (0.098) & 0.040 (0.016) & 0.543 (0.240) & 1.000 (0.089) & 0.037 (0.015) & 0.611 (0.099) & 0.950 (0.069) & 0.028 (0.013) & 0.664 (0.073)\\
MSPLOT & 1.000 (0.000) & 0.030 (0.016) & - & 1.000 (0.000) & 0.032 (0.015) & - & 1.000 (0.000) & 0.022 (0.013) & -\\
TVD & 1.000 (0.000) & 0.000 (0.001) & - & 1.000 (0.000) & 0.000 (0.000) & - & 1.000 (0.000) & 0.000 (0.000) & -\\
MBD & 1.000 (0.239) & 0.000 (0.001) & 0.876 (0.093) & 0.900 (0.138) & 0.000 (0.000) & 0.861 (0.047) & 0.850 (0.114) & 0.000 (0.000) & 0.875 (0.038)\\
MDS5LOF & 1.000 (0.086) & 0.091 (0.001) & 1.000 (0.012) & 1.000 (0.039) & 0.053 (0.002) & 0.999 (0.007) & 0.900 (0.049) & 0.011 (0.005) & 0.997 (0.006)\\
BP-PWD & 1.000 (0.239) & 0.045 (0.013) & - & 0.900 (0.132) & 0.047 (0.013) & - & 0.850 (0.114) & 0.044 (0.015) & -\\
\bottomrule
\end{tabular}
\end{table*}

\begin{table*}[htbp]

\caption{Results of outlier detection methodologies for DGP4 and different outliers proportions ($\alpha$). 
                    The values correspond to the median (sd) of TPR, FPR and AUC for 100 simulations.}
\centering
\scriptsize 
\begin{tabular}[t]{lccccccccc}
\toprule
\multicolumn{1}{c}{ } & \multicolumn{3}{c}{$\alpha = 0.01$} & \multicolumn{3}{c}{$\alpha = 0.05$} & \multicolumn{3}{c}{$\alpha = 0.1$} \\
\cmidrule(l{3pt}r{3pt}){2-4} \cmidrule(l{3pt}r{3pt}){5-7} \cmidrule(l{3pt}r{3pt}){8-10}
Method & TPR & FPR & AUC & TPR & FPR & AUC & TPR & FPR & AUC\\
\midrule
FASTMUOD & 1.000 (0.000) & 0.073 (0.020) & - & 1.000 (0.000) & 0.047 (0.016) & - & 1.000 (0.000) & 0.022 (0.013) & -\\
EHyOut & 1.000 (0.000) & 0.035 (0.016) & 1.000 (0.001) & 1.000 (0.024) & 0.026 (0.015) & 0.999 (0.002) & 1.000 (0.031) & 0.017 (0.014) & 0.999 (0.004)\\
OG & 1.000 (0.000) & 0.025 (0.014) & 1.000 (0.001) & 1.000 (0.010) & 0.016 (0.013) & 1.000 (0.024) & 1.000 (0.045) & 0.011 (0.009) & 0.999 (0.018)\\
MSPLOT & 1.000 (0.000) & 0.015 (0.011) & - & 1.000 (0.010) & 0.011 (0.011) & - & 1.000 (0.005) & 0.011 (0.009) & -\\
TVD & 0.000 (0.234) & 0.000 (0.000) & - & 0.000 (0.092) & 0.000 (0.000) & - & 0.000 (0.057) & 0.000 (0.000) & -\\
MBD & 0.000 (0.163) & 0.000 (0.000) & 0.997 (0.006) & 0.000 (0.083) & 0.000 (0.000) & 0.994 (0.006) & 0.000 (0.051) & 0.000 (0.000) & 0.987 (0.007)\\
MDS5LOF & 1.000 (0.000) & 0.091 (0.000) & 1.000 (0.002) & 1.000 (0.010) & 0.053 (0.001) & 1.000 (0.004) & 1.000 (0.027) & 0.000 (0.003) & 1.000 (0.001)\\
BP-PWD & 0.000 (0.163) & 0.035 (0.015) & - & 0.000 (0.083) & 0.042 (0.014) & - & 0.000 (0.051) & 0.044 (0.016) & -\\
\bottomrule
\end{tabular}
\end{table*}

\begin{table*}[htbp]

\caption{Results of outlier detection methodologies for DGP5 and different outliers proportions ($\alpha$). 
                    The values correspond to the median (sd) of TPR, FPR and AUC for 100 simulations.}
\centering
\scriptsize 
\begin{tabular}[t]{lccccccccc}
\toprule
\multicolumn{1}{c}{ } & \multicolumn{3}{c}{$\alpha = 0.01$} & \multicolumn{3}{c}{$\alpha = 0.05$} & \multicolumn{3}{c}{$\alpha = 0.1$} \\
\cmidrule(l{3pt}r{3pt}){2-4} \cmidrule(l{3pt}r{3pt}){5-7} \cmidrule(l{3pt}r{3pt}){8-10}
Method & TPR & FPR & AUC & TPR & FPR & AUC & TPR & FPR & AUC\\
\midrule
FASTMUOD & 1.000 (0.070) & 0.091 (0.017) & - & 1.000 (0.022) & 0.074 (0.016) & - & 0.950 (0.054) & 0.050 (0.016) & -\\
EHyOut & 1.000 (0.000) & 0.056 (0.021) & 1.000 (0.000) & 1.000 (0.000) & 0.047 (0.022) & 1.000 (0.000) & 1.000 (0.000) & 0.039 (0.019) & 1.000 (0.000)\\
OG & 1.000 (0.050) & 0.040 (0.015) & 0.995 (0.205) & 1.000 (0.017) & 0.032 (0.014) & 0.799 (0.127) & 1.000 (0.025) & 0.017 (0.011) & 0.799 (0.088)\\
MSPLOT & 1.000 (0.000) & 0.035 (0.016) & - & 1.000 (0.000) & 0.026 (0.014) & - & 1.000 (0.000) & 0.028 (0.014) & -\\
TVD & 1.000 (0.000) & 0.000 (0.001) & - & 1.000 (0.000) & 0.000 (0.001) & - & 1.000 (0.000) & 0.000 (0.001) & -\\
MBD & 0.500 (0.338) & 0.000 (0.001) & 0.876 (0.047) & 0.400 (0.189) & 0.000 (0.001) & 0.880 (0.026) & 0.350 (0.134) & 0.000 (0.000) & 0.883 (0.021)\\
MDS5LOF & 0.500 (0.352) & 0.096 (0.004) & 0.910 (0.121) & 0.500 (0.155) & 0.079 (0.008) & 0.837 (0.070) & 0.425 (0.086) & 0.064 (0.010) & 0.807 (0.056)\\
BP-PWD & 1.000 (0.000) & 0.045 (0.013) & - & 1.000 (0.000) & 0.037 (0.012) & - & 1.000 (0.000) & 0.028 (0.012) & -\\
\bottomrule
\end{tabular}
\end{table*}

\begin{table*}[htbp]

\caption{Results of outlier detection methodologies for DGP6 and different outliers proportions ($\alpha$). 
                    The values correspond to the median (sd) of TPR, FPR and AUC for 100 simulations.}
\centering
\scriptsize 
\begin{tabular}[t]{lccccccccc}
\toprule
\multicolumn{1}{c}{ } & \multicolumn{3}{c}{$\alpha = 0.01$} & \multicolumn{3}{c}{$\alpha = 0.05$} & \multicolumn{3}{c}{$\alpha = 0.1$} \\
\cmidrule(l{3pt}r{3pt}){2-4} \cmidrule(l{3pt}r{3pt}){5-7} \cmidrule(l{3pt}r{3pt}){8-10}
Method & TPR & FPR & AUC & TPR & FPR & AUC & TPR & FPR & AUC\\
\midrule
FASTMUOD & 1.000 (0.154) & 0.091 (0.017) & - & 1.000 (0.097) & 0.074 (0.017) & - & 0.900 (0.098) & 0.053 (0.017) & -\\
EHyOut & 1.000 (0.110) & 0.056 (0.021) & 0.995 (0.014) & 1.000 (0.051) & 0.047 (0.021) & 0.994 (0.007) & 1.000 (0.046) & 0.039 (0.018) & 0.995 (0.007)\\
OG & 1.000 (0.050) & 0.040 (0.016) & 0.995 (0.194) & 1.000 (0.051) & 0.032 (0.014) & 0.897 (0.096) & 0.950 (0.068) & 0.017 (0.011) & 0.893 (0.059)\\
MSPLOT & 1.000 (0.000) & 0.033 (0.015) & - & 1.000 (0.027) & 0.032 (0.014) & - & 1.000 (0.024) & 0.022 (0.014) & -\\
TVD & 0.000 (0.242) & 0.000 (0.001) & - & 0.050 (0.085) & 0.000 (0.001) & - & 0.050 (0.046) & 0.000 (0.001) & -\\
MBD & 0.000 (0.050) & 0.000 (0.001) & 0.886 (0.065) & 0.000 (0.038) & 0.000 (0.001) & 0.873 (0.034) & 0.000 (0.022) & 0.000 (0.001) & 0.879 (0.025)\\
\addlinespace
MDS5LOF & 0.500 (0.336) & 0.096 (0.003) & 0.889 (0.101) & 0.400 (0.157) & 0.084 (0.008) & 0.851 (0.056) & 0.400 (0.091) & 0.067 (0.010) & 0.864 (0.044)\\
BP-PWD & 0.000 (0.050) & 0.045 (0.013) & - & 0.000 (0.038) & 0.047 (0.014) & - & 0.000 (0.022) & 0.056 (0.016) & -\\
\bottomrule
\end{tabular}
\end{table*}

\begin{table*}[htbp]

\caption{Results of outlier detection methodologies for DGP7 and different outliers proportions ($\alpha$). The values correspond to the median (sd) of TPR, FPR and AUC for 100 simulations.}
\centering
\scriptsize 
\begin{tabular}[t]{lccccccccc}
\toprule
\multicolumn{1}{c}{ } & \multicolumn{3}{c}{$\alpha = 0.01$} & \multicolumn{3}{c}{$\alpha = 0.05$} & \multicolumn{3}{c}{$\alpha = 0.1$} \\
\cmidrule(l{3pt}r{3pt}){2-4} \cmidrule(l{3pt}r{3pt}){5-7} \cmidrule(l{3pt}r{3pt}){8-10}
Method & TPR & FPR & AUC & TPR & FPR & AUC & TPR & FPR & AUC\\
\midrule
FASTMUOD & 1.000 (0.050) & 0.091 (0.018) & - & 1.000 (0.050) & 0.074 (0.016) & - & 0.950 (0.090) & 0.056 (0.017) & -\\
EHyOut & 1.000 (0.000) & 0.056 (0.022) & 1.000 (0.000) & 1.000 (0.000) & 0.047 (0.021) & 1.000 (0.000) & 1.000 (0.000) & 0.039 (0.019) & 1.000 (0.000)\\
OG & 1.000 (0.000) & 0.040 (0.015) & 1.000 (0.136) & 1.000 (0.014) & 0.032 (0.015) & 0.900 (0.083) & 1.000 (0.023) & 0.017 (0.010) & 0.900 (0.071)\\
MSPLOT & 1.000 (0.000) & 0.035 (0.016) & - & 1.000 (0.000) & 0.029 (0.015) & - & 1.000 (0.000) & 0.022 (0.014) & -\\
TVD & 1.000 (0.334) & 0.000 (0.001) & - & 0.700 (0.191) & 0.000 (0.001) & - & 0.500 (0.194) & 0.000 (0.000) & -\\
MBD & 0.000 (0.070) & 0.000 (0.001) & 0.859 (0.032) & 0.000 (0.031) & 0.000 (0.000) & 0.861 (0.021) & 0.000 (0.024) & 0.000 (0.000) & 0.862 (0.023)\\
MDS5LOF & 0.500 (0.342) & 0.096 (0.003) & 0.852 (0.085) & 0.300 (0.149) & 0.089 (0.008) & 0.859 (0.053) & 0.300 (0.098) & 0.078 (0.011) & 0.860 (0.048)\\
BP-PWD & 0.000 (0.070) & 0.045 (0.014) & - & 0.000 (0.031) & 0.050 (0.014) & - & 0.000 (0.024) & 0.056 (0.016) & -\\
\bottomrule
\end{tabular}
\end{table*}

\begin{table*}[htbp]

\caption{Results of outlier detection methodologies for DGP8 and different outliers proportions ($\alpha$). 
                    The values correspond to the median (sd) of TPR, FPR and AUC for 100 simulations.}
\centering
\scriptsize 
\begin{tabular}[t]{lccccccccc}
\toprule
\multicolumn{1}{c}{ } & \multicolumn{3}{c}{$\alpha = 0.01$} & \multicolumn{3}{c}{$\alpha = 0.05$} & \multicolumn{3}{c}{$\alpha = 0.1$} \\
\cmidrule(l{3pt}r{3pt}){2-4} \cmidrule(l{3pt}r{3pt}){5-7} \cmidrule(l{3pt}r{3pt}){8-10}
Method & TPR & FPR & AUC & TPR & FPR & AUC & TPR & FPR & AUC\\
\midrule
FASTMUOD & 1.000 (0.000) & 0.061 (0.019) & - & 1.000 (0.000) & 0.047 (0.016) & - & 1.000 (0.000) & 0.033 (0.015) & -\\
EHyOut & 1.000 (0.000) & 0.056 (0.022) & 1.000 (0.000) & 1.000 (0.000) & 0.047 (0.020) & 1.000 (0.000) & 1.000 (0.000) & 0.039 (0.018) & 1.000 (0.000)\\
OG & 1.000 (0.000) & 0.043 (0.015) & 1.000 (0.000) & 1.000 (0.000) & 0.032 (0.014) & 1.000 (0.014) & 1.000 (0.000) & 0.017 (0.011) & 1.000 (0.012)\\
MSPLOT & 1.000 (0.000) & 0.033 (0.015) & - & 1.000 (0.000) & 0.032 (0.015) & - & 1.000 (0.000) & 0.022 (0.013) & -\\
TVD & 1.000 (0.000) & 0.000 (0.001) & - & 1.000 (0.000) & 0.000 (0.001) & - & 1.000 (0.000) & 0.000 (0.000) & -\\
MBD & 0.500 (0.356) & 0.000 (0.001) & 0.947 (0.014) & 0.300 (0.180) & 0.000 (0.000) & 0.941 (0.015) & 0.200 (0.140) & 0.000 (0.001) & 0.930 (0.016)\\
MDS5LOF & 1.000 (0.000) & 0.091 (0.000) & 0.995 (0.007) & 1.000 (0.000) & 0.053 (0.000) & 0.995 (0.006) & 0.950 (0.059) & 0.006 (0.007) & 0.996 (0.005)\\
BP-PWD & 1.000 (0.000) & 0.045 (0.013) & - & 1.000 (0.000) & 0.032 (0.013) & - & 1.000 (0.000) & 0.022 (0.010) & -\\
\bottomrule
\end{tabular}
\end{table*}

\begin{table*}[htbp]

\caption{Results of outlier detection methodologies for DGP9 and different outliers proportions ($\alpha$). The values correspond to the median (sd) of TPR, FPR and AUC for 100 simulations.}
\centering
\scriptsize 
\begin{tabular}[t]{lccccccccc}
\toprule
\multicolumn{1}{c}{ } & \multicolumn{3}{c}{$\alpha = 0.01$} & \multicolumn{3}{c}{$\alpha = 0.05$} & \multicolumn{3}{c}{$\alpha = 0.1$} \\
\cmidrule(l{3pt}r{3pt}){2-4} \cmidrule(l{3pt}r{3pt}){5-7} \cmidrule(l{3pt}r{3pt}){8-10}
Method & TPR & FPR & AUC & TPR & FPR & AUC & TPR & FPR & AUC\\
\midrule
FASTMUOD & 1.000 (0.098) & 0.066 (0.025) & - & 0.900 (0.134) & 0.063 (0.024) & - & 0.700 (0.169) & 0.067 (0.024) & -\\
EHyOut & 1.000 (0.086) & 0.020 (0.009) & 1.000 (0.004) & 1.000 (0.039) & 0.016 (0.010) & 0.998 (0.003) & 1.000 (0.055) & 0.017 (0.009) & 0.998 (0.004)\\
OG & 1.000 (0.344) & 0.000 (0.001) & 1.000 (0.128) & 0.600 (0.262) & 0.000 (0.000) & 0.900 (0.072) & 0.350 (0.211) & 0.000 (0.000) & 0.950 (0.044)\\
MSPLOT & 1.000 (0.217) & 0.000 (0.001) & - & 0.900 (0.136) & 0.000 (0.001) & - & 0.800 (0.156) & 0.000 (0.001) & -\\
TVD & 0.500 (0.351) & 0.000 (0.000) & - & 0.400 (0.168) & 0.000 (0.000) & - & 0.450 (0.107) & 0.000 (0.000) & -\\
MBD & 0.000 (0.000) & 0.000 (0.000) & 1.000 (0.002) & 0.000 (0.000) & 0.000 (0.000) & 1.000 (0.002) & 0.000 (0.000) & 0.000 (0.000) & 1.000 (0.001)\\
MDS5LOF & 1.000 (0.000) & 0.091 (0.000) & 1.000 (0.002) & 1.000 (0.010) & 0.053 (0.001) & 1.000 (0.001) & 1.000 (0.031) & 0.000 (0.003) & 1.000 (0.001)\\
BP-PWD & 0.000 (0.000) & 0.091 (0.017) & - & 0.000 (0.000) & 0.095 (0.019) & - & 0.000 (0.000) & 0.111 (0.020) & -\\
\bottomrule
\end{tabular}
\end{table*}

\begin{table*}

\caption{Results of outlier detection methodologies for DGP10 and different outliers proportions ($\alpha$). The values correspond to the median (sd) of TPR, FPR and AUC for 100 simulations.}
\centering
\scriptsize 
\begin{tabular}[t]{lccccccccc}
\toprule
\multicolumn{1}{c}{ } & \multicolumn{3}{c}{$\alpha = 0.01$} & \multicolumn{3}{c}{$\alpha = 0.05$} & \multicolumn{3}{c}{$\alpha = 0.1$} \\
\cmidrule(l{3pt}r{3pt}){2-4} \cmidrule(l{3pt}r{3pt}){5-7} \cmidrule(l{3pt}r{3pt}){8-10}
Method & TPR & FPR & AUC & TPR & FPR & AUC & TPR & FPR & AUC\\
\midrule
FASTMUOD & 1.000 (0.000) & 0.091 (0.018) & - & 1.000 (0.029) & 0.079 (0.017) & - & 1.000 (0.036) & 0.061 (0.016) & -\\
EHyOut & 1.000 (0.000) & 0.056 (0.021) & 1.000 (0.000) & 1.000 (0.000) & 0.047 (0.022) & 1.000 (0.000) & 1.000 (0.000) & 0.039 (0.018) & 1.000 (0.000)\\
OG & 1.000 (0.239) & 0.043 (0.016) & 0.864 (0.208) & 0.900 (0.101) & 0.037 (0.015) & 0.649 (0.089) & 0.850 (0.089) & 0.028 (0.013) & 0.662 (0.085)\\
MSPLOT & 1.000 (0.000) & 0.030 (0.015) & - & 1.000 (0.000) & 0.029 (0.016) & - & 1.000 (0.005) & 0.022 (0.014) & -\\
TVD & 1.000 (0.147) & 0.000 (0.001) & - & 1.000 (0.088) & 0.000 (0.001) & - & 0.900 (0.093) & 0.000 (0.001) & -\\
MBD & 0.500 (0.346) & 0.000 (0.001) & 0.872 (0.147) & 0.500 (0.146) & 0.000 (0.001) & 0.827 (0.069) & 0.450 (0.118) & 0.000 (0.001) & 0.832 (0.051)\\
MDS5LOF & 0.500 (0.347) & 0.096 (0.004) & 0.918 (0.090) & 0.500 (0.153) & 0.079 (0.008) & 0.856 (0.077) & 0.450 (0.107) & 0.061 (0.012) & 0.820 (0.058)\\
BP-PWD & 0.500 (0.353) & 0.045 (0.013) & - & 0.700 (0.134) & 0.042 (0.013) & - & 0.650 (0.124) & 0.039 (0.014) & -\\
\bottomrule
\end{tabular}
\end{table*}

\begin{table*}[htbp]

\caption{Results of outlier detection methodologies for DGP11 and different outliers proportions ($\alpha$). 
                    The values correspond to the median (sd) of TPR, FPR and AUC for 100 simulations.}
\centering
\scriptsize 
\begin{tabular}[t]{lccccccccc}
\toprule
\multicolumn{1}{c}{ } & \multicolumn{3}{c}{$\alpha = 0.01$} & \multicolumn{3}{c}{$\alpha = 0.05$} & \multicolumn{3}{c}{$\alpha = 0.1$} \\
\cmidrule(l{3pt}r{3pt}){2-4} \cmidrule(l{3pt}r{3pt}){5-7} \cmidrule(l{3pt}r{3pt}){8-10}
Method & TPR & FPR & AUC & TPR & FPR & AUC & TPR & FPR & AUC\\
\midrule
FASTMUOD & 1.000 (0.000) & 0.025 (0.013) & - & 1.000 (0.000) & 0.021 (0.014) & - & 1.000 (0.000) & 0.017 (0.013) & -\\
EHyOut & 1.000 (0.000) & 0.066 (0.023) & 1.000 (0.000) & 1.000 (0.000) & 0.058 (0.022) & 1.000 (0.000) & 1.000 (0.000) & 0.056 (0.022) & 1.000 (0.000)\\
OG & 1.000 (0.000) & 0.030 (0.025) & 1.000 (0.050) & 1.000 (0.000) & 0.005 (0.016) & 1.000 (0.028) & 1.000 (0.033) & 0.006 (0.008) & 1.000 (0.024)\\
MSPLOT & 1.000 (0.000) & 0.162 (0.043) & - & 1.000 (0.000) & 0.147 (0.037) & - & 1.000 (0.000) & 0.128 (0.034) & -\\
TVD & 1.000 (0.000) & 0.005 (0.008) & - & 1.000 (0.000) & 0.005 (0.006) & - & 1.000 (0.000) & 0.000 (0.005) & -\\
MBD & 0.000 (0.086) & 0.005 (0.009) & 0.601 (0.089) & 0.000 (0.036) & 0.005 (0.009) & 0.549 (0.050) & 0.000 (0.022) & 0.006 (0.008) & 0.551 (0.037)\\
MDS5LOF & 0.000 (0.157) & 0.101 (0.002) & 0.630 (0.113) & 0.100 (0.079) & 0.100 (0.004) & 0.551 (0.052) & 0.050 (0.059) & 0.106 (0.007) & 0.552 (0.045)\\
BP-PWD & 0.000 (0.086) & 0.015 (0.009) & - & 0.000 (0.036) & 0.105 (0.048) & - & 0.000 (0.048) & 0.119 (0.037) & -\\
\bottomrule
\end{tabular}
\end{table*}

\begin{table*}[htbp]

\caption{Results of outlier detection methodologies for DGP12 and different outliers proportions ($\alpha$).  The values correspond to the median (sd) of TPR, FPR and AUC for 100 simulations.}
\centering
\scriptsize 
\begin{tabular}[t]{lccccccccc}
\toprule
\multicolumn{1}{c}{ } & \multicolumn{3}{c}{$\alpha = 0.01$} & \multicolumn{3}{c}{$\alpha = 0.05$} & \multicolumn{3}{c}{$\alpha = 0.1$} \\
\cmidrule(l{3pt}r{3pt}){2-4} \cmidrule(l{3pt}r{3pt}){5-7} \cmidrule(l{3pt}r{3pt}){8-10}
Method & TPR & FPR & AUC & TPR & FPR & AUC & TPR & FPR & AUC\\
\midrule
FASTMUOD & 0.000 (0.314) & 0.020 (0.010) & - & 0.200 (0.134) & 0.016 (0.010) & - & 0.200 (0.099) & 0.011 (0.011) & -\\
EHyOut & 1.000 (0.000) & 0.056 (0.020) & 1.000 (0.000) & 1.000 (0.000) & 0.047 (0.016) & 1.000 (0.000) & 1.000 (0.000) & 0.042 (0.018) & 1.000 (0.000)\\
OG & 1.000 (0.000) & 0.045 (0.017) & 0.995 (0.202) & 1.000 (0.000) & 0.037 (0.015) & 0.798 (0.132) & 1.000 (0.005) & 0.017 (0.012) & 0.795 (0.096)\\
MSPLOT & 1.000 (0.000) & 0.030 (0.016) & - & 1.000 (0.000) & 0.026 (0.014) & - & 1.000 (0.000) & 0.022 (0.016) & -\\
TVD & 1.000 (0.000) & 0.000 (0.001) & - & 1.000 (0.000) & 0.000 (0.001) & - & 1.000 (0.000) & 0.000 (0.000) & -\\
MBD & 1.000 (0.335) & 0.000 (0.001) & 0.891 (0.042) & 0.700 (0.161) & 0.000 (0.001) & 0.896 (0.022) & 0.650 (0.116) & 0.000 (0.000) & 0.899 (0.018)\\
MDS5LOF & 0.500 (0.328) & 0.096 (0.003) & 0.914 (0.134) & 0.500 (0.141) & 0.079 (0.007) & 0.794 (0.079) & 0.400 (0.085) & 0.067 (0.009) & 0.763 (0.061)\\
BP-PWD & 1.000 (0.000) & 0.045 (0.016) & - & 1.000 (0.000) & 0.042 (0.017) & - & 1.000 (0.000) & 0.033 (0.014) & -\\
\bottomrule
\end{tabular}
\end{table*}

\begin{table*}[htbp]

\caption{Results of outlier detection methodologies for DGP13 and different outliers proportions ($\alpha$). The values correspond to the median (sd) of TPR, FPR and AUC for 100 simulations.}
\centering
\scriptsize 
\begin{tabular}[t]{lccccccccc}
\toprule
\multicolumn{1}{c}{ } & \multicolumn{3}{c}{$\alpha = 0.01$} & \multicolumn{3}{c}{$\alpha = 0.05$} & \multicolumn{3}{c}{$\alpha = 0.1$} \\
\cmidrule(l{3pt}r{3pt}){2-4} \cmidrule(l{3pt}r{3pt}){5-7} \cmidrule(l{3pt}r{3pt}){8-10}
Method & TPR & FPR & AUC & TPR & FPR & AUC & TPR & FPR & AUC\\
\midrule
FASTMUOD & 1.000 (0.000) & 0.061 (0.018) & - & 1.000 (0.000) & 0.047 (0.016) & - & 1.000 (0.000) & 0.033 (0.016) & -\\
EHyOut & 1.000 (0.000) & 0.056 (0.020) & 1.000 (0.000) & 1.000 (0.000) & 0.047 (0.016) & 1.000 (0.000) & 1.000 (0.000) & 0.044 (0.017) & 1.000 (0.000)\\
OG & 1.000 (0.000) & 0.045 (0.016) & 1.000 (0.001) & 1.000 (0.000) & 0.032 (0.014) & 1.000 (0.000) & 1.000 (0.000) & 0.017 (0.011) & 1.000 (0.007)\\
MSPLOT & 1.000 (0.000) & 0.030 (0.016) & - & 1.000 (0.000) & 0.032 (0.014) & - & 1.000 (0.000) & 0.022 (0.015) & -\\
TVD & 1.000 (0.000) & 0.000 (0.001) & - & 1.000 (0.000) & 0.000 (0.001) & - & 1.000 (0.000) & 0.000 (0.000) & -\\
MBD & 0.500 (0.379) & 0.000 (0.001) & 0.956 (0.013) & 0.400 (0.183) & 0.000 (0.001) & 0.950 (0.012) & 0.350 (0.157) & 0.000 (0.000) & 0.939 (0.015)\\
MDS5LOF & 1.000 (0.000) & 0.091 (0.000) & 0.997 (0.004) & 1.000 (0.000) & 0.053 (0.000) & 0.999 (0.003) & 1.000 (0.034) & 0.000 (0.004) & 1.000 (0.003)\\
BP-PWD & 1.000 (0.000) & 0.045 (0.016) & - & 1.000 (0.000) & 0.037 (0.013) & - & 1.000 (0.000) & 0.022 (0.012) & -\\
\bottomrule
\end{tabular}
\end{table*}

\begin{table*}[htbp]

\caption{Results of outlier detection methodologies for DGP14 and different outliers proportions ($\alpha$). 
                    The values correspond to the median (sd) of TPR, FPR and AUC for 100 simulations.}
\centering
\scriptsize 
\begin{tabular}[t]{lccccccccc}
\toprule
\multicolumn{1}{c}{ } & \multicolumn{3}{c}{$\alpha = 0.01$} & \multicolumn{3}{c}{$\alpha = 0.05$} & \multicolumn{3}{c}{$\alpha = 0.1$} \\
\cmidrule(l{3pt}r{3pt}){2-4} \cmidrule(l{3pt}r{3pt}){5-7} \cmidrule(l{3pt}r{3pt}){8-10}
Method & TPR & FPR & AUC & TPR & FPR & AUC & TPR & FPR & AUC\\
\midrule
FASTMUOD & 0.000 (0.000) & 0.015 (0.012) & - & 0.000 (0.000) & 0.021 (0.013) & - & 0.000 (0.000) & 0.028 (0.022) & -\\
EHyOut & 1.000 (0.213) & 0.056 (0.020) & 0.977 (0.030) & 0.900 (0.121) & 0.053 (0.018) & 0.975 (0.021) & 0.850 (0.147) & 0.056 (0.022) & 0.966 (0.026)\\
OG & 0.000 (0.000) & 0.051 (0.017) & 0.923 (0.019) & 0.000 (0.000) & 0.047 (0.017) & 0.917 (0.019) & 0.000 (0.000) & 0.044 (0.018) & 0.890 (0.021)\\
MSPLOT & 0.000 (0.000) & 0.030 (0.016) & - & 0.000 (0.000) & 0.024 (0.012) & - & 0.000 (0.000) & 0.019 (0.014) & -\\
TVD & 0.000 (0.128) & 0.000 (0.001) & - & 0.000 (0.035) & 0.000 (0.002) & - & 0.000 (0.016) & 0.000 (0.001) & -\\
MBD & 0.000 (0.000) & 0.000 (0.001) & 0.986 (0.094) & 0.000 (0.000) & 0.000 (0.002) & 0.941 (0.043) & 0.000 (0.000) & 0.000 (0.001) & 0.936 (0.030)\\
MDS5LOF & 0.000 (0.000) & 0.101 (0.000) & 0.856 (0.108) & 0.000 (0.000) & 0.105 (0.000) & 0.826 (0.055) & 0.000 (0.000) & 0.111 (0.000) & 0.818 (0.039)\\
BP-PWD & 1.000 (0.000) & 0.040 (0.015) & - & 1.000 (0.000) & 0.032 (0.013) & - & 1.000 (0.009) & 0.022 (0.011) & -\\
\bottomrule
\end{tabular}
\end{table*}

\begin{table*}

\caption{Results of outlier detection methodologies for DGP15 and different outliers proportions ($\alpha$). The values correspond to the median (sd) of TPR, FPR and AUC for 100 simulations.}
\centering
\scriptsize 
\begin{tabular}[t]{lccccccccc}
\toprule
\multicolumn{1}{c}{ } & \multicolumn{3}{c}{$\alpha = 0.01$} & \multicolumn{3}{c}{$\alpha = 0.05$} & \multicolumn{3}{c}{$\alpha = 0.1$} \\
\cmidrule(l{3pt}r{3pt}){2-4} \cmidrule(l{3pt}r{3pt}){5-7} \cmidrule(l{3pt}r{3pt}){8-10}
Method & TPR & FPR & AUC & TPR & FPR & AUC & TPR & FPR & AUC\\
\midrule
FASTMUOD & 0.000 (0.000) & 0.086 (0.021) & - & 0.000 (0.000) & 0.089 (0.023) & - & 0.000 (0.000) & 0.106 (0.023) & -\\
EHyOut & 1.000 (0.213) & 0.061 (0.020) & 0.979 (0.030) & 0.900 (0.123) & 0.053 (0.019) & 0.976 (0.021) & 0.850 (0.148) & 0.050 (0.021) & 0.967 (0.026)\\
OG & 0.000 (0.000) & 0.051 (0.018) & 0.924 (0.019) & 0.000 (0.000) & 0.047 (0.018) & 0.916 (0.020) & 0.000 (0.000) & 0.044 (0.018) & 0.889 (0.021)\\
MSPLOT & 0.000 (0.000) & 0.030 (0.016) & - & 0.000 (0.000) & 0.026 (0.012) & - & 0.000 (0.000) & 0.022 (0.014) & -\\
TVD & 0.000 (0.119) & 0.000 (0.001) & - & 0.000 (0.039) & 0.000 (0.002) & - & 0.000 (0.023) & 0.000 (0.001) & -\\
MBD & 0.000 (0.000) & 0.000 (0.001) & 0.984 (0.084) & 0.000 (0.000) & 0.000 (0.001) & 0.946 (0.038) & 0.000 (0.000) & 0.000 (0.001) & 0.942 (0.030)\\
MDS5LOF & 0.000 (0.000) & 0.101 (0.000) & 0.841 (0.101) & 0.000 (0.000) & 0.105 (0.000) & 0.825 (0.051) & 0.000 (0.000) & 0.111 (0.000) & 0.831 (0.039)\\
BP-PWD & 1.000 (0.000) & 0.040 (0.015) & - & 1.000 (0.000) & 0.037 (0.013) & - & 1.000 (0.009) & 0.022 (0.011) & -\\
\bottomrule
\end{tabular}
\end{table*}

\begin{table*}

\caption{Results of outlier detection methodologies for DGP16 and different outliers proportions ($\alpha$). The values correspond to the median (sd) of TPR, FPR and AUC for 100 simulations.}
\centering
\scriptsize 
\begin{tabular}[t]{lccccccccc}
\toprule
\multicolumn{1}{c}{ } & \multicolumn{3}{c}{$\alpha = 0.01$} & \multicolumn{3}{c}{$\alpha = 0.05$} & \multicolumn{3}{c}{$\alpha = 0.1$} \\
\cmidrule(l{3pt}r{3pt}){2-4} \cmidrule(l{3pt}r{3pt}){5-7} \cmidrule(l{3pt}r{3pt}){8-10}
Method & TPR & FPR & AUC & TPR & FPR & AUC & TPR & FPR & AUC\\
\midrule
FASTMUOD & 0.000 (0.000) & 0.020 (0.011) & - & 0.000 (0.000) & 0.026 (0.013) & - & 0.000 (0.000) & 0.028 (0.015) & -\\
EHyOut & 1.000 (0.050) & 0.056 (0.020) & 0.997 (0.008) & 1.000 (0.038) & 0.053 (0.019) & 0.995 (0.008) & 1.000 (0.030) & 0.050 (0.022) & 0.994 (0.007)\\
OG & 0.000 (0.000) & 0.051 (0.018) & 0.907 (0.020) & 0.000 (0.000) & 0.047 (0.017) & 0.898 (0.021) & 0.000 (0.000) & 0.044 (0.018) & 0.877 (0.021)\\
MSPLOT & 0.000 (0.000) & 0.030 (0.016) & - & 0.000 (0.000) & 0.026 (0.012) & - & 0.000 (0.000) & 0.022 (0.014) & -\\
TVD & 0.000 (0.311) & 0.000 (0.001) & - & 0.100 (0.131) & 0.000 (0.002) & - & 0.000 (0.046) & 0.000 (0.001) & -\\
MBD & 0.000 (0.000) & 0.000 (0.001) & 0.980 (0.085) & 0.000 (0.000) & 0.000 (0.001) & 0.943 (0.038) & 0.000 (0.000) & 0.000 (0.001) & 0.941 (0.030)\\
MDS5LOF & 0.000 (0.000) & 0.101 (0.000) & 0.842 (0.101) & 0.000 (0.000) & 0.105 (0.000) & 0.823 (0.052) & 0.000 (0.000) & 0.111 (0.000) & 0.832 (0.040)\\
BP-PWD & 1.000 (0.000) & 0.040 (0.015) & - & 1.000 (0.000) & 0.032 (0.013) & - & 1.000 (0.000) & 0.022 (0.011) & -\\
\bottomrule
\end{tabular}
\end{table*}

\begin{table*}[htbp]

\caption{Results of outlier detection methodologies for DGP17 and different outliers proportions ($\alpha$). The values correspond to the median (sd) of TPR, FPR and AUC for 100 simulations.}
\centering \scriptsize 
\centering
\scriptsize 
\begin{tabular}[t]{lccccccccc}
\toprule
\multicolumn{1}{c}{ } & \multicolumn{3}{c}{$\alpha = 0.01$} & \multicolumn{3}{c}{$\alpha = 0.05$} & \multicolumn{3}{c}{$\alpha = 0.1$} \\
\cmidrule(l{3pt}r{3pt}){2-4} \cmidrule(l{3pt}r{3pt}){5-7} \cmidrule(l{3pt}r{3pt}){8-10}
Method & TPR & FPR & AUC & TPR & FPR & AUC & TPR & FPR & AUC\\
\midrule
FASTMUOD & 0.000 (0.184) & 0.020 (0.013) & - & 0.000 (0.020) & 0.021 (0.012) & - & 0.000 (0.016) & 0.022 (0.012) & -\\
EHyOut & 0.500 (0.360) & 0.081 (0.021) & 0.934 (0.037) & 1.000 (0.000) & 0.079 (0.024) & 1.000 (0.000) & 1.000 (0.000) & 0.072 (0.023) & 1.000 (0.000)\\
OG & 1.000 (0.247) & 0.066 (0.017) & 0.970 (0.031) & 0.600 (0.192) & 0.058 (0.014) & 0.954 (0.013) & 0.275 (0.146) & 0.044 (0.015) & 0.945 (0.016)\\
MSPLOT & 0.500 (0.355) & 0.051 (0.018) & - & 0.100 (0.103) & 0.053 (0.017) & - & 0.025 (0.044) & 0.039 (0.017) & -\\
TVD & 1.000 (0.167) & 0.005 (0.006) & - & 1.000 (0.000) & 0.000 (0.006) & - & 1.000 (0.000) & 0.000 (0.004) & -\\
MBD & 0.000 (0.000) & 0.005 (0.007) & 0.750 (0.060) & 0.000 (0.000) & 0.005 (0.008) & 0.787 (0.039) & 0.000 (0.000) & 0.011 (0.010) & 0.793 (0.040)\\
MDS5LOF & 0.000 (0.000) & 0.101 (0.000) & 0.860 (0.083) & 0.000 (0.000) & 0.105 (0.000) & 0.847 (0.058) & 0.000 (0.000) & 0.111 (0.000) & 0.837 (0.064)\\
BP-PWD & 0.000 (0.000) & 0.056 (0.017) & - & 0.800 (0.151) & 0.034 (0.018) & - & 0.975 (0.060) & 0.033 (0.013) & -\\
\bottomrule
\end{tabular}
\end{table*}

\begin{table*}[htbp]

\caption{Results of outlier detection methodologies for DGP18 and different outliers proportions ($\alpha$). 
                    The values correspond to the median (sd) of TPR, FPR and AUC for 100 simulations.}
\centering
\scriptsize 
\begin{tabular}[t]{lccccccccc}
\toprule
\multicolumn{1}{c}{ } & \multicolumn{3}{c}{$\alpha = 0.01$} & \multicolumn{3}{c}{$\alpha = 0.05$} & \multicolumn{3}{c}{$\alpha = 0.1$} \\
\cmidrule(l{3pt}r{3pt}){2-4} \cmidrule(l{3pt}r{3pt}){5-7} \cmidrule(l{3pt}r{3pt}){8-10}
Method & TPR & FPR & AUC & TPR & FPR & AUC & TPR & FPR & AUC\\
\midrule
FASTMUOD & 0.000 (0.119) & 0.025 (0.013) & - & 0.000 (0.017) & 0.021 (0.013) & - & 0.000 (0.015) & 0.022 (0.012) & -\\
EHyOut & 1.000 (0.297) & 0.081 (0.022) & 0.943 (0.030) & 1.000 (0.000) & 0.079 (0.023) & 1.000 (0.000) & 1.000 (0.000) & 0.067 (0.020) & 1.000 (0.000)\\
OG & 0.500 (0.336) & 0.066 (0.017) & 0.942 (0.050) & 0.350 (0.175) & 0.058 (0.014) & 0.922 (0.021) & 0.150 (0.099) & 0.044 (0.016) & 0.913 (0.021)\\
MSPLOT & 0.000 (0.286) & 0.051 (0.018) & - & 0.000 (0.064) & 0.047 (0.017) & - & 0.000 (0.031) & 0.039 (0.018) & -\\
TVD & 1.000 (0.265) & 0.005 (0.006) & - & 1.000 (0.000) & 0.000 (0.006) & - & 1.000 (0.000) & 0.000 (0.004) & -\\
MBD & 0.000 (0.000) & 0.005 (0.007) & 0.804 (0.056) & 0.000 (0.000) & 0.005 (0.009) & 0.834 (0.032) & 0.000 (0.000) & 0.011 (0.011) & 0.840 (0.033)\\
MDS5LOF & 0.000 (0.000) & 0.101 (0.000) & 0.875 (0.062) & 0.000 (0.000) & 0.105 (0.000) & 0.891 (0.041) & 0.000 (0.000) & 0.111 (0.000) & 0.882 (0.048)\\
BP-PWD & 0.000 (0.139) & 0.051 (0.017) & - & 0.700 (0.173) & 0.037 (0.017) & - & 0.950 (0.083) & 0.039 (0.015) & -\\
\bottomrule
\end{tabular}
\end{table*}

\begin{table*}

\caption{Results of outlier detection methodologies for DGP19 and different outliers proportions ($\alpha$). The values correspond to the median (sd) of TPR, FPR and AUC for 100 simulations.}
\centering
\scriptsize 
\begin{tabular}[t]{lccccccccc}
\toprule
\multicolumn{1}{c}{ } & \multicolumn{3}{c}{$\alpha = 0.01$} & \multicolumn{3}{c}{$\alpha = 0.05$} & \multicolumn{3}{c}{$\alpha = 0.1$} \\
\cmidrule(l{3pt}r{3pt}){2-4} \cmidrule(l{3pt}r{3pt}){5-7} \cmidrule(l{3pt}r{3pt}){8-10}
Method & TPR & FPR & AUC & TPR & FPR & AUC & TPR & FPR & AUC\\
\midrule
FASTMUOD & 0.000 (0.383) & 0.000 (0.003) & - & 0.000 (0.333) & 0.000 (0.003) & - & 0.125 (0.332) & 0.000 (0.001) & -\\
EHyOut & 1.000 (0.329) & 0.000 (0.005) & 1.000 (0.052) & 0.700 (0.243) & 0.000 (0.003) & 0.992 (0.055) & 0.625 (0.317) & 0.000 (0.003) & 0.988 (0.074)\\
OG & 0.000 (0.350) & 0.000 (0.003) & 0.783 (0.168) & 0.000 (0.234) & 0.000 (0.003) & 0.727 (0.146) & 0.000 (0.227) & 0.000 (0.004) & 0.725 (0.135)\\
MSPLOT & 1.000 (0.310) & 0.000 (0.051) & - & 1.000 (0.240) & 0.000 (0.056) & - & 0.950 (0.307) & 0.000 (0.044) & -\\
TVD & 0.000 (0.269) & 0.000 (0.000) & - & 0.000 (0.162) & 0.000 (0.000) & - & 0.000 (0.137) & 0.000 (0.000) & -\\
MBD & 0.000 (0.269) & 0.000 (0.000) & 0.766 (0.166) & 0.000 (0.168) & 0.000 (0.000) & 0.738 (0.138) & 0.000 (0.133) & 0.000 (0.000) & 0.739 (0.151)\\
MDS5LOF & 0.000 (0.175) & 0.101 (0.002) & 0.801 (0.174) & 0.000 (0.071) & 0.105 (0.004) & 0.871 (0.159) & 0.000 (0.068) & 0.111 (0.008) & 0.843 (0.166)\\
BP-PWD & 0.000 (0.269) & 0.010 (0.008) & - & 0.000 (0.168) & 0.011 (0.007) & - & 0.000 (0.133) & 0.011 (0.006) & -\\
\bottomrule
\end{tabular}
\end{table*}
}
\end{appendix}
\FloatBarrier
\twocolumn






\begin{funding}
The first author was partially supported by the Ministerio de Ciencia e Innovación, Gobierno de España, (grant PID2022-137243OB-I00 funded by MCIN/AEI/10.13039/501100011033); the ENIA 2022 programme for the creation of university-industry chairs (AI-AImpulsa: UC3M-Universia); and a mobility grant from Universidad Carlos III de Madrid "Programa Propio de Investigación" (Ayudas para la Movilidad).

The second author was supported in part by Ministerio de Ciencia e Innovación, Gobierno de España, grant number PID2022-137050NB-I00 funded by MCIN/AEI/10.13039/501100011033, and European Union Next Generation EU/PRTR and ERDF A way of making Europe.

The third author was partially supported by the Spanish Ministry of Science and Innovation (MCIN/AEI/10.13039/501100011033) under grant PID2022-137243OB-I00. This work also received support from the European Union’s Recovery, Transformation and Resilience Plan – NextGenerationEU, through the INCIBE ANTICIPA grant and the ENIA 2022 programme for university–industry AI chairs (AImpulsa: UC3M-Universia).

The work of the fourth author has been funded by the German Federal Ministry of Education and Research (BMBF) under Grant No. 01IS18036A. The author takes full responsibilities for its content.
\end{funding}

\bibliographystyle{imsart-nameyear} 
\bibliography{biblio}       





\end{document}